\documentclass[a4paper,fleqn,10pt]{article}
\usepackage{amsmath}
\usepackage{amssymb}
\usepackage{array}
\usepackage{calc}
\usepackage{longtable}
\usepackage{multirow}
\usepackage{axodraw4j}
\usepackage{pstricks}
\usepackage{graphicx}
\usepackage{xspace}
\usepackage{mciteplus}
\usepackage[a4paper,pdfborder={0 0 0}]{hyperref}
\usepackage[format=hang,labelfont=bf,hypcap=true]{caption}
\usepackage{subfig}
\usepackage{sectsty}
\allsectionsfont{\sffamily}
\subsubsectionfont{\mdseries\itshape\large}
\let\spreprint\empty
\newcommand{\preprint}[1]{\def\spreprint{\protect#1}}
\let\sinstitute\empty
\newcommand{\institute}[1]{\def\sinstitute{\protect#1}}
\makeatletter
\renewcommand{\maketitle}{\begingroup
  \null\thispagestyle{empty}%
    \ifx\spreprint\empty
      \vskip 5ex
    \else
      \flushright\large\spreprint\vskip 2ex
    \fi
    \vskip 5ex
    \flushleft
      {\sffamily\bfseries\huge\@title}\vskip 2ex
      \@author\vskip 2ex
      \ifx\sinstitute\empty
      \else
        {\small\sinstitute}
      \fi
    \vskip 5ex
  \endgroup
}
\makeatother
\renewenvironment{abstract}{\begin{center}
  {\large\sffamily\bfseries Abstract: }
  \begin{minipage}[t]{0.75\textwidth}
}{\end{minipage}\end{center}\vskip 10ex}

\newcommand{\mytextwidthfigure}[3]{
  \begin{figure}[#1]
    \begin{center}
      #2\\
      \parbox[t]{\textwidth}{\caption{#3}}
    \end{center}
  \end{figure}
}
\newcommand{\myfigure}[3]{
  \begin{figure}[#1]
    \begin{center}
      #2\\
      \parbox[t]{\widthof{#2}}{\caption{#3}}
    \end{center}
  \end{figure}
}
\newcommand{\mytable}[3]{
  \begin{table}[#1]
    \begin{center}
      #2\\
      \parbox[t]{\widthof{#2}}{\caption{#3}}
    \end{center}
  \end{table}
}
\numberwithin{equation}{section}
\newcommand{\MCatNLO}{M\protect\scalebox{0.8}{C}@N\protect\scalebox{0.8}{LO}\xspace}
\newcommand{\POWHEG}{P\protect\scalebox{0.8}{OWHEG}\xspace}
\newcommand{\Nlo}{NLO\xspace}

\newcommand{\Vegas}{V\protect\scalebox{0.8}{EGAS}\xspace}
\newcommand{\Herwig}{H\protect\scalebox{0.8}{ERWIG}\xspace}

\newcommand{\Pythia}{P\protect\scalebox{0.8}{YTHIA}\xspace}

\newcommand{\MCFM}{M\protect\scalebox{0.8}{CFM}\xspace}
\newcommand{\Sherpa}{S\protect\scalebox{0.8}{HERPA}\xspace}
\newcommand{\Comix}{C\protect\scalebox{0.8}{OMIX}\xspace}

\newcommand{\Amegic}{A\protect\scalebox{0.8}{MEGIC++}\xspace}
\newcommand{\CSS}{C\protect\scalebox{0.8}{SS}\xspace}

\newcommand{\LEP}{LEP\xspace}
\newcommand{\LHC}{LHC\xspace}
\newcommand{\Tevatron}{Tevatron\xspace}
\newcommand{\Hera}{HERA\xspace}
\newcommand{\Aleph}{ALEPH\xspace}

\newcommand{\DO}{D\O\ }

\long\def\symbolfootnote[#1]#2{\begingroup%
\def\thefootnote{\fnsymbol{footnote}}\footnote[#1]{#2}\endgroup}
\newcommand{\abs}[1]{\left| #1\right|}
\newcommand{\rbr}[1]{\left( #1\right)}
\newcommand{\abr}[1]{\langle #1\rangle}

\newcommand{\sbr}[1]{\left[ #1\right]}
\newcommand{\im}{\imath}
\newcommand{\jm}{\jmath}
\newcommand{\args}[1]{\{\vec{#1}\}}
\newcommand{\argc}[2]{\{\vec{#1}\}_{\rm #2}}
\newcommand{\done}{{\rm d}}
\newcommand{\order}{\mathcal{O}}
\newcommand{\mc}[1]{\mathcal{#1}}
\newcommand{\mr}[1]{\mathrm{#1}}
\newcommand{\dst}{\displaystyle}

\newcommand{\bmap}[3]{b_{#1,#2}(#3)}
\newcommand{\rmap}[3]{r_{\widetilde{#1},\tilde{#2}}(#3)}
\newcommand{\bea}{\begin{eqnarray}}
\newcommand{\eea}{\end{eqnarray}}
\newcommand{\bi}{\begin{itemize}}
\newcommand{\ei}{\end{itemize}}
\newcommand{\hl}{\vphantom{$\int_A^B$}}

\hypersetup{
  pdfauthor={Stefan Hoeche, Frank Krauss,
    Marek Schoenherr, Frank Siegert},
  pdftitle={Automating the POWHEG method in Sherpa},
  pdfkeywords={QCD, POWHEG, Higher-Order corrections,
    Matrix Element, Parton Shower, Sherpa}
}
\preprint{ZU-TH 12/10\\IPPP/10/72\\DCPT/10/144\\CERN-PH-TH/2010-194\\
  MCNET/10/16}
\author{Stefan H{\"o}che$^1$, Frank Krauss$^{2,3}$,
  Marek Sch{\"o}nherr$^4$, Frank Siegert$^{2,5}$}
\title{Automating the \POWHEG method in \Sherpa}
\institute{$^1$ Institut f{\"u}r Theoretische Physik, 
  Universit{\"a}t Z{\"u}rich, CH-8057 Zurich, Switzerland\\
  $^2$ Institute for Particle Physics Phenomenology,
  Durham University, Durham DH1 3LE, UK\\
  $^3$ PH-TH, CERN, CH-1211 Geneva 23, Switzerland\\
  $^4$ Institut f{\"u}r Kern- und Teilchenphysik,
  Technische Universit{\"a}t Dresden, D-01062, Dresden, Germany\\
  $^5$ Department of Physics \& Astronomy,
  University College London, London WC13 6BT, UK\\}
\begin{document}
\maketitle
\begin{abstract}
  A new implementation of the \POWHEG method
\cite{Nason:2004rx,*Frixione:2007vw} into the Monte-Carlo event generator 
\Sherpa~\cite{Gleisberg:2003xi,*Gleisberg:2008ta} is presented, focusing 
on processes with a simple colour structure.  Results for a variety of 
processes, namely $e^+e^-\to$ hadrons, deep-inelastic lepton-nucleon 
scattering, hadroproduction of single vector bosons and of vector boson pairs
as well as the production of Higgs bosons in gluon fusion serve as test cases
for the successful realisation.  The algorithm is fully automated such that 
for further processes only virtual matrix elements need to be included.

\end{abstract}
%= text ===========================================
\section{Introduction}

Higher-order QCD corrections by now form an important ingredient
to many phenomenological studies and experimental analyses at both
the Tevatron and the LHC.  The impact of these corrections has been 
similarly important for various studies of HERA and LEP data.
Calculations invoking such corrections, typically at next-to leading 
order (NLO) in the perturbative expansion in the strong coupling 
$\alpha_s$, and in very few cases also at next-to-next-to leading order
(NNLO) accuracy, have been used for a wide range of precision tests of 
our understanding of QCD and the Standard Model. They are also 
important for the subtraction of backgrounds in searches for new
physics.  When being compared to such calculations, experimental 
measurements are usually corrected for detector effects, while the 
perturbative result is corrected for hadronisation.  Only after 
performing these corrections, theoretical predictions and experimental 
data are on the same footing.  Hadronisation 
corrections are typically determined by using multi-purpose Monte-Carlo 
event generators, such as \Pythia/\Pythia~8~\cite{Sjostrand:2006za,
  *Sjostrand:2007gs}, \Herwig/\Herwig{}++ \cite{Corcella:2000bw,*Bahr:2008pv}, 
or, more recently, \Sherpa \cite{Gleisberg:2003xi,*Gleisberg:2008ta}.
In the past two decades, such event generators have been the workhorses
of particle physics phenomenology, including in their simulation 
many features in addition to the perturbative aspects of a collision, 
such as hadronisation, the underlying event, hadron decays etc..  
But, to describe perturbative QCD, they typically rely on leading-order 
(LO) matrix elements only, combined with parton showers, which in turn 
model QCD radiation effects in a leading logarithmic approximation.  

Improvements to this approximation can be obtained through merging 
methods, pioneered in \cite{Andre:1997vh,Catani:2001cc,*Krauss:2002up}, 
and further worked out in different varieties at different accuracies and for 
different parton showers, e.g.\ in \cite{Lonnblad:2001iq,*Mangano:2001xp,
  Mrenna:2003if,*Mangano:2006rw,*Alwall:2008qv}. 
In this merging approach, tree-level matrix elements for processes 
with different jet multiplicities are combined with parton showers,
avoiding problems related to double counting of emissions.
Lately, a new formulation has been proposed which can be proved 
to preserve the formal accuracy of the parton shower, independent of the 
process under consideration~\cite{Hoeche:2009rj}.\footnote{A similar procedure, 
  although restricted to the case of $e^+e^-$-annihilation into hadrons
  was also presented in~\cite{Hamilton:2009ne}.}
Despite varying degrees of formal accuracy amongst the various methods,
their respective predictions tend to agree on a level expected 
from such improved leading order perturbation theory~\cite{Hoche:2006ph,
*Alwall:2007fs}. In most cases the approach leads to a dramatic improvement 
in the description of hard QCD radiation, which makes it a state-of-the-art 
tool for many analyses, that depend on the {\em shape} of distributions
related to hard QCD radiation. 

However sophisticated at modelling multi-jet topologies, the above methods lack 
the precision of higher-order calculations regarding the overall normalisation, 
i.e.\ the cross section of the inclusive process under consideration.
To accurately estimate uncertainties intrinsic to perturbative calculations, 
which manifest themselves for instance in uncertainties related to choices
of renormalisation and factorisation scales, full next-to leading order 
corrections are a {\em sine qua non}.  

So far, two different methods to achieve the systematic inclusion of complete 
NLO corrections for a fixed multiplicity have not only been described but 
also {\em implemented}, asserting their practicality.  The first one has 
been dubbed \MCatNLO~\cite{Frixione:2002ik}. By now it has been applied
to a variety of processes using the framework of both the \Herwig~%
\cite{Frixione:2003ei,*Frixione:2005vw,*Frixione:2008yi,*Weydert:2009vr}
and the \Pythia~\cite{Torrielli:2010aw} event generators, proving its
versatility.  It effectively relies on using the parton shower to perform
the subtraction of infrared divergences in the real-emission part of NLO
corrections, resulting in a separation of the event generation 
into two parts, one starting from the Born kinematics of the hard
process, the other starting with real-emission kinematics, i.e.\ including an 
extra parton in the final state.  The weights for these two samples are 
adjusted accordingly and yield the total cross section and the hardest
emission, correct to first order in $\alpha_s$.  However, this technique may 
produce a number of events with a negative weight.  The second method to 
include NLO corrections into parton showers is known as the \POWHEG 
technique~\cite{Nason:2004rx,*Frixione:2007vw}.  In contrast to \MCatNLO,
it can ensure that only events with positive weights are generated.  To 
achieve this, the simulation starts with a Born-level event,
reweighted to include the full NLO correction, i.e.\ including 
virtual and real corrections, integrated over the 
real-emission subspace.  The first emission is then produced using the 
exact real-emission matrix element, thus providing not only the correct 
differential cross section, but also the correct radiation pattern to first 
order in $\alpha_s$.  The big advantage of this technique is that it can 
straightforwardly be implemented in a shower-independent way, which has
been used to some extent in \cite{Nason:2006hfa,*Frixione:2007nw,
  *Alioli:2008tz,*Alioli:2009je,*Nason:2009ai,Alioli:2008gx}, where
different processes have been treated in the framework of \Pythia and/or
\Herwig.  In addition, there have been some implementations in the
framework of \Herwig{}++~\cite{LatundeDada:2006gx,*LatundeDada:2008bv,
  *Hamilton:2008pd,*Papaefstathiou:2009sr,Hamilton:2009za}\footnote{
  It is worth stressing, that in order to ensure formal
  accuracy a {\em truncated shower} must be used, as pointed out in the
  original publications proposing the method.  This option is not
  available for all parton shower algorithms that have been used in the
  actual implementations.}.
In fact, because the \POWHEG method is very similar to traditional 
matrix-element reweighting~\cite{Seymour:1994df,*Seymour:1994we,
  *Corcella:1998rs,*Miu:1998ju,*Corcella:1999gs,*Norrbin:2000uu}, 
it is relatively simple to implement in a process-independent 
way and thus very well suited for the systematic inclusion of NLO QCD 
corrections to arbitrary processes. This is reflected by the fact that a 
corresponding toolbox has already been advertised in~\cite{Alioli:2010xd}.  

Obviously, the ultimate goal is to have a multi-jet merging prescription, accurate 
at NLO, and to simultaneously respect the logarithmic accuracy of the parton 
shower.  A first step into this direction has been presented for the case of 
$e^+e^-$ annihilation into hadrons in \cite{Lavesson:2008ah}.  An alternative 
approach was suggested in~\cite{Hamilton:2010wh}, extending the method
of~\cite{Hoeche:2009rj} to NLO accuracy for the core interaction.  This
second method relies on the \POWHEG technique to ensure the NLO cross section
of the core process.  Parallel development has brought about a similar 
algorithm, which will be reported shortly~\cite{Hoeche:2010xx} and which is 
based on the generic \POWHEG implementation presented here.  

The outline of this paper is as follows: 
In Sec.~\ref{SEC:ReviewPOWHEG} the connection between the \POWHEG method and 
parton showers will be detailed, introducing also the notation used in this
publication. The possibility to implement the \POWHEG technique on top of 
an existing parton shower will be discussed. Section~\ref{SEC:Sherpa} briefly 
introduces the \Sherpa event generator, which sets the framework for this 
study.  Those parts which are relevant for the \POWHEG implementation will be 
presented in more detail.  Due to the overall setup of 
\Sherpa, incorporating matrix element generators capable of generating
the real correction terms for a given process and the corresponding 
differential and integrated subtraction terms in the Catani-Seymour
dipole subtraction scheme~\cite{Catani:1996vz,*Catani:2002hc},
the only missing bit of the NLO calculation driving the \POWHEG method
are the virtual contributions, which have been made accessible through
interfaces to BlackHat~\cite{whitehat:2010aa,*Berger:2009zg,*Berger:2009ep,*Berger:2010vm} 
and \MCFM~\cite{MCFM}. It is the first time that the \POWHEG method has
been automated using dipole subtraction rather than the Frixione-Kunszt-Signer
method~\cite{Frixione:1995ms,*Frixione:1997np}. The overall quality of the
implementation will be exemplified in a number of characteristic processes in 
Sec.~\ref{SEC:Results}, including the hitherto unpublished case of $W$-pair
production in hadronic collisions. Sec.~\ref{sec:conclusions} contains 
our conclusions.

\section{Parton showers and the \POWHEG method}
\label{SEC:ReviewPOWHEG}

In this publication, the \POWHEG method is reinterpreted as an advanced 
reweighting technique for standard parton showers.  The following section introduces 
the necessary notation and outlines the parallels between 
the \POWHEG method and traditional matrix-element reweighting.  The starting 
point of the discussion is the factorisation theorem underlying the specific 
parton-shower model, like the DGLAP equation~\cite{Gribov:1972ri,
  *Lipatov:1974qm,*Dokshitzer:1977sg,*Altarelli:1977zs},
the colour-dipole model~\cite{Gustafson:1986db,
  *Gustafson:1987rq,*Andersson:1989ki},
Catani-Seymour factorisation~\cite{Catani:1996vz,*Catani:2002hc} 
or antenna factorisation~\cite{Kosower:1997zr,*Kosower:2003bh,
  *GehrmannDeRidder:2005cm,*Daleo:2006xa}.
Except in collinear factorisation, the splitting functions of the parton shower
depend on (at least) one additional parton, which is often referred to as the 
``spectator''. In order to make this connection explicit, the notation of a 
dipole-like factorisation is adopted, which is sufficiently general to discuss 
all relevant features of the \POWHEG method and its implementation into 
the \Sherpa event generator.

%%%%%%%%%%%%%%%%%%%%%%%%%%%%%%%%%%%%%%%%%%%%%%%%%%%%%
%%%%%%%%%%%%%%%%%%%%%%%%%%%%%%%%%%%%%%%%%%%%%%%%%%%%%
\subsection{Decomposition of real-emission cross sections}
%%%%%%%%%%%%%%%%%%%%%%%%%%%%%%%%%%%%%%%%%%%%%%%%%%%%%
%%%%%%%%%%%%%%%%%%%%%%%%%%%%%%%%%%%%%%%%%%%%%%%%%%%%%
\label{sec:real_decomposition}
In the following, sets of $n$ particles in a $2\to (n-2)$ process will 
summarily be denoted by $\args{a}=\{a_1,\ldots,a_n\}$, and the particles will 
be specified through their flavours $\args{f\,}=\{f_1,\ldots,f_n\}$ and momenta 
$\args{p\,}=\{p_1,\ldots,p_n\}$.  The generic expression for a fully differential 
Born-level cross section in a scattering process with $(n-2)$ final-state 
particles can be written as a sum over all contributing flavour combinations as
\begin{equation}
  \done\sigma_B(\args{p\,})\,=\;
  \sum_{\args{f\,}}\done\sigma_B(\args{a})\;,
  \qquad\text{where}\qquad
  \done\sigma_B(\args{a})\,=\;\done\Phi_B(\args{p\,})\,\mr{B}(\args{a})\;,
\end{equation}
The individual terms in the sum are given by
\begin{equation}
  \begin{split}
  \mr{B}(\args{a})\,=&\;\mc{L}(\args{a})\,\mc{B}(\args{a})\;,
  &\mc{B}(\args{a})\,=&\;
  \frac{1}{F(\args{a})}\,\frac{1}{S(\args{f\,})}\,
  \abs{\mc{M}_B}^2(\args{a})\;,\\
  \done\Phi_B(\args{p\,})\,=&\;\frac{\done x_1}{x_1}\frac{\done x_2}{x_2}\,
  \done{\it\Phi}_B(\args{p\,})\;,
  &\mc{L}(\args{a};\mu^2)\,=&\;x_1f_{f_1}(x_1,\mu^2)\;x_2f_{f_2}(x_2,\mu^2)\;.
  \end{split}
\end{equation}
Here, $\abs{\mc{M}_B}^2(\args{a})$ denotes the partonic matrix element 
squared, with all factors due to averaging over initial state quantum numbers 
such as spin or colours absorbed into it, and $\done{\it\Phi}_B(\args{a})$ is 
the corresponding differential $n$-particle partonic phase-space element; 
$S(\args{f\,})$ is the symmetry factor due to identical flavours associated to 
the partonic subprocess, while $F(\args{a})$ denotes the flux factor
and $\mc{L}$ is the parton luminosity given by the corresponding
parton distribution functions (PDFs).  In the case of leptonic initial
states, ignoring QED initial state radiation, the parton distribution 
functions $f(x,\mu^2)$ are replaced by $\delta(1-x)$.

In a similar fashion, the real-emission part of the QCD next-to-leading order
cross section can be written as a sum, this time over parton configurations
$\{a_1,\ldots,a_{n+1}\}$, i.e.\ including one additional parton.  
A corresponding subprocess cross section reads
\begin{equation}\label{eq:real_xs}
  \done\sigma_R(\args{a})\,=\;
    \done\Phi_R(\args{p\,})\,\mr{R}(\args{a})\;.
\end{equation}
%Note that the expressions $\mr{B}(\args{a})$ and 
%$\mr{R}(\args{a})$ appearing in the discussion here coincide
%with the terms $\mr{B}$ and $\mr{R}$ in the original \POWHEG 
%publications~\cite{Nason:2004rx,*Frixione:2007vw}.

At this point, it is helpful to introduce a notation for mappings from 
real-emission parton configurations to Born parton configurations.
Such mappings combine the partons $a_i$ and $a_j$ into a common ``mother'' 
parton $a_{\widetilde{\im\jm}}$, in the presence of the spectator $a_k$
by defining a new flavour $f_{\widetilde{\im\jm}}$ and by redefining
the particle momenta. To be specific,
\begin{equation}\label{eq:parton_map_born}
  \bmap{ij}{k}{\args{a}}\,=\;\left\{\begin{array}{c}
    \args{f\,}\setminus\{f_i,f_j\}\cup\{f_{\widetilde{\im\jm}}\}\\
    \args{p\,}\to\args{\tilde{p\;}}
    \end{array}\right.
\end{equation}
The flavour of the ``mother'' parton, $f_{\widetilde{\im\jm}}$, is thereby
fixed unambiguously by the QCD interactions, while the flavour of the
spectator, $f_k$, remains unaltered. The momentum map guarantees that all
partons are kept on their mass shell.

Conversely, any Born parton configuration and a related branching process 
$\widetilde{\im\jm},\tilde{k}\to ij,k$ determine the parton configuration of a 
real-emission subprocess as
\begin{equation}\label{eq:parton_map_real}
  \rmap{\im\jm}{k}{f_i,\Phi_{R|B}\,;\args{a}}\,=\;
  \left\{\begin{array}{c}
    \args{f\,}\setminus\{f_{\widetilde{\im\jm}}\}\cup\{f_i,f_j\}\\
    \args{\tilde{p\;}}\to\args{p\,}
    \end{array}\right.\;.
\end{equation}
The radiative variables $\Phi_{R|B}$ are thereby employed to turn the 
$n$-parton momentum configuration into an n+1-parton momentum configuration
using the inverse of the phase-space map defined by 
Eq.~\eqref{eq:parton_map_born}.  The flavour $f_j$ is again determined
unambiguously by the QCD interactions. Here, also two obvious generalisations of 
Eq.~\eqref{eq:parton_map_born} shall be defined, $\bmap{ij}{k}{\args{f\,}}$ and 
$\bmap{ij}{k}{\args{p\,}}$, which act on the parton flavours and on the parton 
momenta only.  Correspondingly, such generalisations exist for 
Eq.~\eqref{eq:parton_map_real}.

In the soft and collinear limits, the partonic matrix element squared, 
$\mc{R}(\args{a})$, can be decomposed as a sum of terms 
$\mc{D}_{ij,k}(\args{a})$, 
\begin{equation}\label{eq:decomposition_real_me}
  \mc{R}(\args{a})\,\to\;
    \sum_{\{i,j\}}\sum_{k\neq i,j}\mc{D}_{ij,k}(\args{a})\;.
\end{equation} 
These terms factorise into a Born-level term and a universal splitting 
kernel, encoding the transition of $a_{\widetilde{\im\jm}}$ to 
$a_i$ and $a_j$~\cite{Catani:1996vz,*Catani:2002hc}.
The splitting is associated with a universal procedure for factorising
the phase space integral into a Born level part and a one-particle 
radiative phase space.
\begin{equation}
  \done\Phi_R(\args{p\,})\,=\;
  \done\Phi_B(\bmap{ij}{k}{\args{p\,}})\,
  \done\Phi_{R|B}^{ij,k}(\args{p\,})\;.
\end{equation}
The existence of universal decompositions like in Eq.~\eqref{eq:decomposition_real_me}
forms the basis of subtraction methods like the Catani-Seymour dipole 
subtraction~\cite{Catani:1996vz,*Catani:2002hc}, 
antenna subtraction~\cite{Kosower:1997zr,*Kosower:2003bh,
  *GehrmannDeRidder:2005cm,*Daleo:2006xa}, 
or the subtraction method of Frixione, Kunszt, and 
Signer~\cite{Frixione:1995ms,*Frixione:1997np}.  
It also serves as starting point for the construction of parton shower 
algorithms~\cite{Schumann:2007mg,Dinsdale:2007mf,*Giele:2007di}, 
which aim at approximating parton emissions in the collinear and soft limits
of the radiation phase space, to resum the associated large logarithms, cf.\
Sec.~\ref{sec:ps_construction}.

However, it is important to stress that, also away from the infrared limits, 
$\mc{R}(\args{a})$ can be decomposed into a number of terms $\mc{R}_{ij,k}$
analogous to $\mc{D}_{ij,k}$ through
\begin{equation}\label{eq:def_rho}
  \mc{R}_{ij,k}(\args{a})\,:=\;\rho_{ij,k}(\args{a})\,\mc{R}(\args{a})\;,
  \quad\text{where}\quad
  \rho_{ij,k}(\args{a})\,=\;\frac{\mc{D}_{ij,k}(\args{a})}{
    \sum_{\{m,n\}}\sum_{l\neq m,n}\mc{D}_{mn,l}(\args{a})}\;.
\end{equation}
Equation~\eqref{eq:real_xs} can now be rewritten as a sum of trivially 
factorised contributions
\begin{equation}\label{eq:split_real_xs}
  \done\sigma_R(\args{a})\,=\;\sum_{\{i,j\}}\sum_{k\neq i,j}
    \done\sigma_B(\bmap{ij}{k}{\args{a}})\,
    \done\sigma_{R|B}^{ij,k}(\args{a})\;,
\end{equation}
where
\begin{equation}\label{eq:def_real_dipole}
  \done\sigma_{R|B}^{ij,k}(\args{a})\,=\;
    \done\Phi_{R|B}^{ij,k}(\args{p\,})\,
    \frac{\mr{R}_{ij,k}(\args{a})}{\mr{B}(\bmap{ij}{k}{\args{a}})}\;.
\end{equation}
and $\mr{R}_{ij,k}(\args{a})=\mc{L}(\args{a})\mc{R}_{ij,k}(\args{a})$.
These equations are key ingredients to understanding and implementing the 
\POWHEG method.

%%%%%%%%%%%%%%%%%%%%%%%%%%%%%%%%%%%%%%%%%%%%%%%%%%%%%
%%%%%%%%%%%%%%%%%%%%%%%%%%%%%%%%%%%%%%%%%%%%%%%%%%%%%
\subsection{Construction of the parton shower}
%%%%%%%%%%%%%%%%%%%%%%%%%%%%%%%%%%%%%%%%%%%%%%%%%%%%%
%%%%%%%%%%%%%%%%%%%%%%%%%%%%%%%%%%%%%%%%%%%%%%%%%%%%%
\label{sec:ps_construction}
Due to the non-Abelian nature of QCD, the terms $\mc{D}_{ij,k}$ in 
Eq.~\eqref{eq:decomposition_real_me} in general do not factorise on the level 
of squared matrix elements, including all colour contributions.  To arrive at 
a practical model for a parton shower, sub-leading colour configurations are 
therefore neglected, which leads to an {\it assumed} factorisation 
on the level of squared matrix elements.
In the infrared limits one can then write
\begin{equation}\label{eq:factorisation_ps}
  \mc{D}_{ij,k}(\args{a})\,\to\;\mc{B}(\bmap{ij}{k}{\args{a}})\,
    \frac{S(\bmap{ij}{k}{\args{f\,}})}{S(\args{f\,})}\,
    \frac{1}{2\,p_ip_j}\,8\pi\,\alpha_s\,\mc{K}_{ij,k}(p_i,p_j,p_k)\;,
\end{equation}
where the set of momenta $\bmap{ij}{k}{\args{p\,}}$ is determined 
by the phase space map of the parton shower model.\footnote{
  Note that here only parton showers with local energy-momentum conservation
  are considered. Therefore, the phase-space maps $\argc{p\,}{R}\to\argc{p\,}{B}$ exist.}
The quantities $\mc{K}_{ij,k}$ are the parton shower evolution kernels, which depend
on the parton flavours $f_i$, $f_j$ and $f_k$ and on the radiative phase space.
The denominator factor $2\,p_ip_j$ or any linearly dependent quantity is 
usually used to define the parton shower evolution variable,
in the following denoted by $t$.

Using the above model, the parton-shower approximation of 
Eq.~\eqref{eq:def_real_dipole} can be derived as
\begin{equation}\label{eq:real_dipole_ps}
  \done\sigma_{R|B}^{{\rm(PS)}\,ij,k}(\args{a})\,=\;
    \done\Phi_{R|B}^{ij,k}(\args{p\,})\,
    \frac{S(\bmap{ij}{k}{\args{f\,}})}{S(\args{f\,})}\,
    \frac{1}{2\,p_ip_j}\,8\pi\,\alpha_s\,\mc{K}_{ij,k}(p_i,p_j,p_k)\,
    \frac{\mc{L}(\args{a})}{\mc{L}(\bmap{ij}{k}{\args{a}})}\;.
\end{equation}

Particles produced in the parton shower are resolved at a certain evolution 
scale and can therefore be distinguished from particles at higher and lower 
scales.  At most the two particles $a_i$ and $a_j$, emerging from the same 
splitting process, can be seen as identical.  
Hence, the ratio of symmetry factors in Eq.~\eqref{eq:def_real_dipole} 
changes to 
\begin{equation}\label{eq:ps_symfac}
  \frac{S(\bmap{ij}{k}{\args{f\,}})}{S(\args{f\,})}\,\to\;
  \frac{1}{S_{ij}}\,=\;\left\{\begin{array}{cc}
    1/2 & \text{if $\; i,j> 2$ and $b_i=b_j$}\\
    1 & \text{else}
    \end{array}\right.\;.
\end{equation}
The integral over the radiative phase space can be written as
\begin{equation}\label{eq:decomposition_radiative_ps}
  \done\Phi_{R|B}^{ij,k}(\args{p\,})\,=\;\frac{1}{16\pi^2}\,
    \done t\,\done z\,\frac{\done\phi}{2\pi}\; J_{ij,k}(t,z,\phi)\;,
\end{equation}
with $t$ the evolution variable, $z$ the splitting variable, and $\phi$
an azimuthal angle.  Here, $J$ denotes the Jacobian factor, that potentially 
arises due to the transformation of variables. 
Equation~\eqref{eq:real_dipole_ps} thus becomes
\begin{equation}\label{eq:gamma_one}
  \done\sigma_{R|B}^{{\rm(PS)}\,ij,k}(\args{a})\,=\;
    \frac{\done t}{t}\,\done z\,\frac{\done\phi}{2\pi}\,\frac{\alpha_s}{2\pi}\,
    \frac{1}{S_{ij}}\,J_{ij,k}(t,z,\phi)\,\mc{K}_{ij,k}(t,z,\phi)\,
    \frac{\mc{L}(\args{a};t)}
         {\mc{L}(\bmap{ij}{k}{\args{a}};t)}\;.
\end{equation}
The assignment of the mother parton, the spectator and the underlying Born process 
can now be assumed to be fixed.  Then, the sum runs over all possible real-emission 
configurations originating from this particular Born state instead.
Furthermore, assuming independence of the individual emissions, i.e.\ 
Poissonian statistics, this leads to the constrained no-branching probability
of the parton-shower model~\cite{Field:1989uq,*Ellis:1991qj} between the two 
scales $t''$ and $t'$
\begin{equation}\label{eq:nbp_constrained}
  \begin{split}
  \Delta^{\rm(PS)}_{\widetilde{\im\jm},\tilde{k}}(t',t'';\argc{a}{B})\,=&\;
  \exp\left\{-\sum_{f_i=q,g}
    \int_{t'}^{t''}\frac{\done t}{t}\,\int_{z_{\rm min}}^{z_{\rm max}}\done z\,
    \int_0^{2\pi}\frac{\done\phi}{2\pi}\,J_{ij,k}(t,z,\phi)\,
  \right.\\&\qquad\qquad\qquad\times\left.
    \frac{1}{S_{ij}}\,\frac{\alpha_s}{2\pi}\,
    \mc{K}_{ij,k}(t,z,\phi)\,
    \frac{\mc{L}(\rmap{\im\jm}{k}{f_i,t,z,\phi;\args{a}};t)}
         {\mc{L}(\args{a\,};t)}\,\right\}\;.
  \end{split}
\end{equation}
It is worth noting that Eq.~\eqref{eq:nbp_constrained} depends on the 
underlying Born process, since the flavour and momentum of the spectator 
enter as arguments of $J_{ij,k}$ and $\mc{K}_{ij,k}$.  The ratio of $\mc{L}$ in 
Eq.~\eqref{eq:nbp_constrained} accounts for a potential change of the parton 
luminosity when integrating over the initial-state phase space\footnote{
  Note that, depending on the parton shower model, the $x_i$ do not necessarily 
  fulfil the relation $x_i=\tilde{x}_i/z$~\cite{Platzer:2009jq,Hoeche:2009xc}.}.
Note that the partons $\argc{a}{B}$ denote a 
Born-level set, while in \eqref{eq:real_dipole_ps} and \eqref{eq:gamma_one} 
$\args{a}$ denote a set of partons at real-emission level.  This is also 
indicated by the subscript B in Eq.~\eqref{eq:nbp_constrained}.  

Using the definition
\begin{equation}\label{eq:nbp_global_prod}
  \Delta(t_0,\mu^2;\args{a}) = \prod\limits_{\{\widetilde{\im\jm},\tilde{k}\}}
    \Delta_{\widetilde{\im\jm},\tilde{k}}(t_0,\mu^2;\args{a})
\end{equation}
the total cross section in the parton shower approximation reads
\begin{equation}\label{eq:xsec_psapproximate}
  \begin{split}
  \sigma_B \,=&\;\sum_{\args{f\,}}\int\done\Phi_B(\args{p\,})\,{\mr{B}}(\args{a})
  \left[\,\vphantom{\int_A^B}\,
    \Delta^{\rm(PS)}(t_0,\mu^2;\args{a})\;
  \right.\\&\qquad\qquad+\,
    \sum_{\{\widetilde{\im\jm},\tilde{k}\}}
    \sum_{f_i=q,g}
    \int_{t_0}^{\mu^2}\frac{\done t}{t}\,
    \int_{z_{\rm min}}^{z_{\rm max}}\done z\,
    \int_0^{2\pi}\frac{\done\phi}{2\pi}\,
    J_{ij,k}(t,z,\phi)
  \\&\qquad\qquad\qquad\left.\times\,
    \frac{1}{S_{ij}}\,\frac{\alpha_s}{2\pi}\,
    \mc{K}_{ij,k}(t,z,\phi)\,
    \frac{\mc{L}(\rmap{\im\jm}{k}{f_i,t,z,\phi;\args{a}};t)}
         {\mc{L}(\args{a\,};t)}\,
    \Delta^{\rm(PS)}(t,\mu^2;\args{a})\,\right]\;.
  \end{split}
\end{equation}
The scale $t_0$ acts as the infrared cutoff of the parton shower.  
Simple inspection shows that the sum in the square bracket equals unity,
since the second term can be written as
\begin{equation}
  \int_{t_0}^{\mu^2}\done t\;
  \frac{\done\Delta^{\rm(PS)}(t,\mu^2;\args{a})}{\done t}\;.
\end{equation}
This makes the probabilistic properties of the parton shower explicit.  At the
same time it also shows that this unitarity leads to the cross section in
standard parton-shower Monte Carlos to be exactly the respective leading-order
cross section.  In order to evaluate the formal accuracy of the description
of the radiation pattern, induced by the second term in the square bracket --
the first term encodes the probability that there is no resolvable emission
off the Born-level configuration -- a corresponding observable must be 
introduced.  This complicates the discussion somewhat and is therefore
postponed to Secs.~\ref{sec:approx_nlo_xs} and~\ref{sec:accuracy}.

%%%%%%%%%%%%%%%%%%%%%%%%%%%%%%%%%%%%%%%%%%%%%%%%%%%%%
%%%%%%%%%%%%%%%%%%%%%%%%%%%%%%%%%%%%%%%%%%%%%%%%%%%%%
\subsection{Correcting parton showers with matrix elements}
%%%%%%%%%%%%%%%%%%%%%%%%%%%%%%%%%%%%%%%%%%%%%%%%%%%%%
%%%%%%%%%%%%%%%%%%%%%%%%%%%%%%%%%%%%%%%%%%%%%%%%%%%%%
\label{sec:ps_mecorrection}
The aim of this section is to devise a simple method for reinstating
$\order(\alpha_s)$ accuracy in the emission pattern of the parton shower,
i.e.~the hardest emission in the parton shower should
follow the distribution given by the corresponding real-emission matrix 
element.  Loosely speaking, the key idea is to 
replace the splitting kernels $\mc{K}$ with the ratio of real-emission and 
Born-level matrix elements.  Thus, instead of the 
splitting kernels, this ratio is exponentiated in the Sudakov form
factor and employed in simulating the splitting.

Comparing Eqs.~\eqref{eq:def_real_dipole} and~\eqref{eq:real_dipole_ps}, a 
corresponding factor correcting the latter to the former can be easily 
identified.  Using Eq.~\eqref{eq:ps_symfac} it reads
\begin{equation}\label{eq:def_me_correction}
  \begin{split}
  w_{ij,k}(\args{a})\,=&\;
    \frac{\done\sigma_{R|B}^{ij,k}(\args{a})}{
      \done\sigma_{R|B}^{{\rm(PS)}\,ij,k}(\args{a})}\,
  =\;\frac{2\,p_ip_j}{8\pi\,\alpha_s}\,
    \frac{S(\args{f\,})}{S(\bmap{ij}{k}{\args{f\,}})}\,
    \frac{\rho_{ij,k}(\args{a})\,\mc{R}(\args{a})}{
      \mc{B}(\bmap{ij}{k}{\args{a}})\;
      \mc{K}_{ij,k}(\args{a})}\;.
  \end{split}
\end{equation}
Employing the parton shower approximation, Eq.~\eqref{eq:factorisation_ps},
to replace $\rho_{ij,k}$ yields
\begin{equation}\label{eq:me_correction}
  w(\args{a})\,=\;\sbr{\,\dst\sum_{\{m,n\}}\sum_{l\neq m,n}
    \frac{S(\bmap{mn}{l}{\args{f\,}})}{S(\args{f\,})}\,
    \frac{\mc{B}(\bmap{mn}{l}{\args{a}})}{\mc{R}(\args{a})}\;
    \frac{8\pi\,\alpha_s}{2\,p_mp_n}\,
    \mc{K}_{mn,l}(\args{a})\,}^{-1}\;.
\end{equation}
Note that this implies a corrective weight, which is actually 
{\it splitter-spectator independent}.\\Correcting the parton shower 
to the full matrix element can thus be achieved through the following 
algorithm:
\begin{enumerate}
\item Determine an overestimate for Eq.~\eqref{eq:me_correction}, 
  i.e.\ find a set of $W_{\widetilde{\im\jm},f_i}(\args{a})$, such that\\ 
  $w(\rmap{\im\jm}{k}{f_i,\Phi_{R|B};\args{a}})\leq W_{\widetilde{\im\jm},f_i}(\args{a})$ 
  for all $\tilde{k}$ and throughout the real-emission phase space.
\item Replace the parton shower splitting kernels $\mc{K}_{ij,k}$ by
  $W_{\widetilde{\im\jm},f_i}(\args{a})\,\mc{K}_{ij,k}$.
\item Accept parton-shower branchings with probability 
$w(\rmap{\im\jm}{k}{f_i,\Phi_{R|B};\args{a}})/W_{\widetilde{\im\jm},f_i}(\args{a})$. 
\end{enumerate}
It is straightforward to show that the constrained no-branching probability 
of such a matrix-element corrected parton shower reads
\begin{equation}\label{eq:nbp_mecorr_constrained}
  \begin{split}
  \Delta^{\rm(ME)}_{\widetilde{\im\jm},\tilde{k}}(t',t'';\args{a})\,=&\;
    \exp\left\{-\sum_{f_i=q,g}
    \frac{1}{16\pi^2}\,
    \int_{t'}^{t''}\done t\,\int_{z_{\rm min}}^{z_{\rm max}}\done z\,
    \int_0^{2\pi}\frac{\done\phi}{2\pi}\,J_{ij,k}(t,z,\phi)
    \right.\\&\qquad\qquad\qquad\times\left.
    \frac{1}{S_{ij}}\,
    \frac{S(\rmap{\im\jm}{k}{f_i;\args{f\,}})}{S(\args{f\,})}\,
    \frac{\mr{R}_{ij,k}(\rmap{\im\jm}{k}{f_i,t,z,\phi;\args{a}})}{
      \mr{B}(\args{a})}\,\right\}\;.
  \end{split}
\end{equation}
The ratio $\mr{R}/\mr{B}$ in Eq.~\eqref{eq:nbp_mecorr_constrained} coincides with the ratio
in the original publications presenting the \POWHEG method.  In the relatively
simple cases treated so far~\cite{Nason:2004rx,*Frixione:2007vw,Nason:2006hfa,
  *Frixione:2007nw,*Alioli:2008tz,*Alioli:2009je,*Nason:2009ai,Alioli:2008gx}, 
the various symmetry factors in the equation above cancel and can be
neglected.  For more complicated flavour structures this factor may 
differ from one and therefore must be retained.

Employing again the definition of Eq.~\eqref{eq:nbp_global_prod}, but this
time for the Sudakov form factor constructed from the ratio 
${\mr R}/\mr{B}$ yields the cross section in the matrix element improved 
parton shower approximation.  It reads
\begin{equation}\label{eq:xsec_me_improved}
  \begin{split}
  \sigma_B \,=&\;\sum_{\args{f\,}}\int\done\Phi_B(\args{p\,})\,{\mr{B}}(\args{a})
  \left[\,\vphantom{\int_A^B}\,
    \Delta^{\rm(ME)}(t_0,\mu^2;\args{a})\;
  \right.\\&\qquad\qquad+\,
    \sum_{\{\widetilde{\im\jm},\tilde{k}\}}
    \sum_{f_i=q,g}
    \frac{1}{16\pi^2}\,\int_{t_0}^{\mu^2}\done t\,
    \int_{z_{\rm min}}^{z_{\rm max}}\done z\,
    \int_0^{2\pi}\frac{\done\phi}{2\pi}\,
    J_{ij,k}(t,z,\phi)
  \\&\qquad\qquad\qquad\qquad\left.\times\,\frac{1}{S_{ij}}\,
    \frac{S(\rmap{\im\jm}{k}{f_i;\args{f\,}})}{S(\args{f\,})}\,
    \frac{\mr{R}_{ij,k}(\rmap{\im\jm}{k}{f_i,t,z,\phi;\args{a}})}{
      \mr{B}(\args{a})}\,
    \Delta^{\rm(ME)}(t,\mu^2;\args{a})\,\right]\;.
  \end{split}
\end{equation}
Again, the term in the square bracket equals one and thus reflects the
probabilistic nature of this approach.  Consequently, in the
matrix-element improved parton-shower approximation the total cross section
is given by the Born cross section, although the radiation 
pattern has improved. For a detailed discussion of the real-emission
term see Sec.~\ref{sec:accuracy}.

%%%%%%%%%%%%%%%%%%%%%%%%%%%%%%%%%%%%%%%%%%%%%%%%%%%%%
%%%%%%%%%%%%%%%%%%%%%%%%%%%%%%%%%%%%%%%%%%%%%%%%%%%%%
\subsection{Approximate NLO cross sections}
\label{sec:approx_nlo_xs}
%%%%%%%%%%%%%%%%%%%%%%%%%%%%%%%%%%%%%%%%%%%%%%%%%%%%%
%%%%%%%%%%%%%%%%%%%%%%%%%%%%%%%%%%%%%%%%%%%%%%%%%%%%%

In the previous two sections it has become clear that the total cross section
of events simulated in a parton-shower Monte-Carlo is determined by the ``seed''
cross section, typically computed at Born level.  While matrix-element improvement
of the naive parton shower picture will lead to radiation patterns which are 
accurate to $\order(\alpha_s)$, the total cross section of the event sample 
and any observable that can be defined at Born level will still be 
given by the respective leading-order expression.   
To allow for a simulation with next-to-leading order accuracy, including the
cross section of the event sample, a prescription to assign a corresponding
weight and multiplicity of the seed event must be found.

The solution is to replace the original Born-level matrix element with a modified
one~\cite{Nason:2004rx,*Frixione:2007vw}, denoted by $\bar{\mr{B}}$,
\begin{equation}\label{eq:def_fdep_bbar}
  \done\sigma_B(\args{a})\,\to\;
  \done\sigma_{\bar{B}}(\args{a})\,:=\;
  \done\Phi_B(\args{p\,})\,\bar{\mr{B}}(\args{a})
\end{equation}
such that the ``seed'' cross section, $\dst\done\sigma_{\bar{B}}$, 
integrates to the full NLO result. When constructing such an 
{\em NLO-weighted differential cross section for the Born configuration}, 
certain approximations must be made, since NLO cross sections have two contributions, 
one with Born-like kinematics and one with real-emission like kinematics,
both of which exhibit divergent structures. The value of a given infrared 
and collinear safe observable, $O$, computed at NLO, is given in terms of 
the Born term $\mr{B}$, the real emission term $\mr{R}$, and the virtual 
contribution (including the collinear counter-terms), denoted by 
$\tilde{\mr{V}}$, as
\begin{equation}\label{eq:obs_xs_nlo}
  \begin{split}
  \abr{O}^{(\rm NLO)}\,=&\; 
  \sum_{\args{f\,}}\int\done\Phi_B(\args{p\,})
  \sbr{\vphantom{\int}\,\mr{B}(\args{a})+\tilde{\mr{V}}(\args{a})\,}O(\args{p\,})\\ 
  &\qquad +\sum_{\args{f\,}}\int\done\Phi_R(\args{p\,})\,
    \mr{R}(\args{a})\,O(\args{p\,})\;.
  \end{split}
\end{equation}
It is obvious that the real-emission contribution cannot be simply combined 
with the Born and virtual terms, as it depends on different kinematics.  In 
the following, the solution of this problem in the framework of the \POWHEG 
method is outlined.

In order to compute Eq.~\eqref{eq:obs_xs_nlo} in a Monte-Carlo 
approach, subtraction terms, rendering the real emission finite in $D=4$ 
space-time dimensions are introduced. Corresponding integrated subtraction
terms regularise the infrared divergences of the virtual 
terms.  In the dipole subtraction method~\cite{Catani:1996vz,*Catani:2002hc}, 
the equation above can then be written as
\begin{equation}\label{eq:obs_xs_nlo_sub}
  \begin{split}
  \abr{O}^{(\rm NLO)}\,=&\; 
  \sum_{\args{f\,}} \int\done\Phi_B(\args{p\,})\sbr{\vphantom{\int}\,
    \mr{B}(\args{a})+\tilde{\mr{V}}(\args{a})+\mr{I}(\args{a})\,}
  O(\args{p\,})\\
  &\quad+
  \sum_{\args{f\,}}\int\done\Phi_R(\args{p\,})
  \sbr{\,\mr{R}(\args{a})\,O(\args{p\,})
    -\sum_{\{i,j\}}\sum_{k\neq i,j}\mr{S}_{ij,k}(\args{a})\,
    O(\bmap{ij}{k}{\args{p\,}})\,}\;.
  \end{split}
\end{equation}
Note that each $\mr{S}_{ij,k}$ defines a separate phase space map and that 
the observable $O$ in the last term depends on $\bmap{ij}{k}{\args{p\,}}$, 
rather than $\args{p\,}$, which is a crucial feature of the subtraction 
procedure. The real and integrated subtraction terms $\mr{S}_{ij,k}(\args{a})$
and $\mr{I}(\args{a})$ fulfil the relation
\bea\label{eq:sub_term_cancelation}
  \begin{split}
  \sum_{\args{f\,}}\int\done\Phi_B(\args{p\,})\,\mr{I}(\args{a})
 \,=&\;
  \sum_{\args{f\,}}\sum_{\{i,j\}}\sum_{k\neq i,j}
    \int\done\Phi_R(\args{p\,})\,
    \mr{S}_{ij,k}(\args{a}) \;.
  \end{split}
\eea

Identifying $\mr{D}_{ij,k}$ with $\mr{S}_{ij,k}$, the term with real-emission 
kinematics in Eq.~\eqref{eq:obs_xs_nlo_sub} can then be decomposed 
according to Eq.~\eqref{eq:def_rho}, resulting in
\begin{equation}
  \sum_{\args{f\,}}\done\Phi_R(\args{p\,})
    \sum_{\{i,j\}}\sum_{k\neq i,j}
    \sbr{\vphantom{\int}\,\mr{R}_{ij,k}(\args{a})\,O(\args{p\,})
      -\mr{S}_{ij,k}(\args{a})\,O(\bmap{ij}{k}{\args{p\,}})\,}\;.
\end{equation}
In the \POWHEG method, this term is {\em approximated} as
\begin{equation}
  \begin{split}
  &\sum_{\args{f\,}}\done\Phi_R(\args{p\,})
    \sum_{\{i,j\}}\sum_{k\neq i,j}
    \sbr{\vphantom{\int}\,\mr{R}_{ij,k}(\args{a})
      -\mr{S}_{ij,k}(\args{a})\,}O(\bmap{ij}{k}{\args{p\,}})\\
  &\quad=\;\sum_{\args{f\,}}\done\Phi_B(\args{p\,})
    \sum_{\{\widetilde{\im\jm},\tilde{k}\}}\sum_{f_i=q,g}
    \done\Phi_{R|B}^{ij,k}\,
    \sbr{\vphantom{\int}\,\mr{R}_{ij,k}(\rmap{\im\jm}{k}{\args{a}})
    -\mr{S}_{ij,k}(\rmap{\im\jm}{k}{\args{a}})\,}
  O(\args{p\,})\;.
  \end{split}
\end{equation}
This allows the recombination of all contributions to the NLO cross section.
Employing Eq.~\eqref{eq:def_fdep_bbar}, therefore
\begin{equation}\label{eq:def_bbar}
  \begin{split}
  \bar{\mr{B}}(\args{a})\,=&\;
    \mr{B}(\args{a})+\tilde{\mr{V}}(\args{a})+\mr{I}(\args{a})\\
  &\quad+
    \sum_{\{\widetilde{\im\jm},\tilde{k}\}}
    \sum_{f_i=q,g}
    \int\done\Phi_{R|B}^{ij,k}
  \sbr{\vphantom{\int}\,\mr{R}_{ij,k}(\rmap{\im\jm}{k}{\args{a}})
    -\mr{S}_{ij,k}(\rmap{\im\jm}{k}{\args{a}})\,}\;.
  \end{split}
\end{equation}
In the next section it will be shown that the combination of this term,
$\bar{\mr{B}}$ with a matrix-element reweighted parton shower yields the 
attempted $\mc{O}(\alpha_s)$ accuracy, 
not only of the total cross section, but also of the real-emission 
contribution.

%%%%%%%%%%%%%%%%%%%%%%%%%%%%%%%%%%%%%%%%%%%%%%%%%%%%%
%%%%%%%%%%%%%%%%%%%%%%%%%%%%%%%%%%%%%%%%%%%%%%%%%%%%%
\subsection{The \POWHEG method and its accuracy}
\label{sec:accuracy}
%%%%%%%%%%%%%%%%%%%%%%%%%%%%%%%%%%%%%%%%%%%%%%%%%%%%%
%%%%%%%%%%%%%%%%%%%%%%%%%%%%%%%%%%%%%%%%%%%%%%%%%%%%%

The key point of the \POWHEG method is, to supplement Monte Carlo event samples
from matrix-element corrected parton showers with a next-to-leading order weight 
to arrive at full NLO accuracy.  This is achieved by combining the two methods 
discussed in Secs.~\ref{sec:ps_mecorrection} and~\ref{sec:approx_nlo_xs}.
To obtain the $\order(\alpha_s)$ approximation to the cross section in the 
\POWHEG method, the parton-shower expression of the real-emission probability 
is combined with the approximated initial cross section, 
$\dst\done\sigma_{\bar{B}}$.  This yields the following master formula for the
value of an infrared and collinear safe observable, $O$,
\begin{equation}\label{eq:powheg_masterformula}
  \begin{split}
  \abr{O}^{\rm (POWHEG)}\,=&\;\sum_{\args{f\,}}\int\done\Phi_B(\args{p\,})\,
  \bar{\mr{B}}(\args{a})\left[\,\vphantom{\int_A^B}
    \Delta^{\rm(ME)}(t_0,\mu^2;\args{a})\,O(\args{p\,})\right.\\
  &\qquad+\,
    \sum_{\{\widetilde{\im\jm},\tilde{k}\}}
    \sum_{f_i=q,g}\frac{1}{16\pi^2}
    \int_{t_0}^{\mu^2}\done t\,\int_{z_{\rm min}}^{z_{\rm max}}\done z\,
    \int_0^{2\pi}\frac{\done\phi}{2\pi}\,J_{ij,k}(t,z,\phi)\\
  &\qquad\qquad\left.\times\,
    \frac{1}{S_{ij}}\,
    \frac{S(\rmap{\im\jm}{k}{\args{f\,}})}{S(\args{f\,})}\,
    \frac{\mr{R}_{ij,k}(\rmap{\im\jm}{k}{\args{a}})}{
      \mr{B}(\args{a})}\;
    \Delta^{\rm(ME)}(t,\mu^2;\args{a})\,
    O(\rmap{\im\jm}{k}{\args{p\,}})\,\right]\;,
  \end{split}
\end{equation}
where obvious arguments of the parton maps $r_{\widetilde{\im\jm},\tilde{k}}$
have been suppressed.  Clearly, if the observable $\mc{O}$ on the right hand side 
of Eq.~\eqref{eq:powheg_masterformula} becomes one, the quantity computed is the
total cross section, as for the cases discussed in Secs.~\ref{sec:ps_construction} 
and~\ref{sec:ps_mecorrection}.  This particular case, however, 
is insensitive to the details of the radiation pattern.  
To continue the discussion, note that the second term in the 
square bracket of Eq.~\eqref{eq:powheg_masterformula} can be rewritten as
\begin{equation}
  \sum_{\{\widetilde{\im\jm},\tilde{k}\}}\int_{t_0}^{\mu^2}\done t\;
  \frac{\done\log\Delta^{\rm(ME)}_{\widetilde{\im\jm},\tilde{k}}
    (t,\mu^2;\args{a})}{\done t}\;\Delta^{\rm(ME)}(t,\mu^2;\args{a})\;
  O(\rmap{\im\jm}{k}{\args{p\,}})\;,
\end{equation}
which allows to rearrange the expressions in 
Eq.~\eqref{eq:powheg_masterformula} as
\begin{equation}\label{eq:powheg_rearranged}
  \begin{split}
  &\abr{O}^{\rm (POWHEG)}\,=\;
  \sum_{\args{f\,}}\int\done\Phi_B(\args{p\,})\,\bar{\mr{B}}(\args{a})
  \left\{\vphantom{\int_A^B}\,O(\args{p\,})
  \right.\\&\qquad\qquad\left.
  +\sum_{\{\widetilde{\im\jm},\tilde{k}\}}\int_{t_0}^{\mu^2}\done t\;
  \frac{\done\log\Delta^{\rm(ME)}_{\widetilde{\im\jm},\tilde{k}}
    (t,\mu^2;\args{a})}{\done t}\;\Delta^{\rm(ME)}(t,\mu^2;\args{a})\;
  \sbr{\,O(\rmap{\im\jm}{k}{\args{p\,}})-O(\args{p\,})\,}\,\right\}\;,
  \end{split}
\end{equation}
Two special cases should now be considered
\begin{itemize}
\item The infrared limit ($t\to 0$)\\
In this case, only the first term in Eq.~\eqref{eq:powheg_rearranged}
contributes, as any infrared safe observable maps the real-emission
kinematics for collinear (soft) emissions to the kinematics of 
the (any) underlying Born configuration
\begin{equation*}
  O(\rmap{\im\jm}{k}{\args{p\,}})\overset{t\to 0}{\to}O(\args{p\,})
\end{equation*}
The contribution to $\abr{O}^{\rm(POWHEG)}$ from this phase-space 
region is therefore correct to $\order(\alpha_s)$.
\item Hard emissions ($t\to\mu^2$)\\
In this case, a cancellation between the first and the last term 
in Eq.~\eqref{eq:powheg_rearranged} is achieved and only the second 
term remains. Also, $\Delta(t,\mu^2)\to1$ as $t\to\mu^2$.
To $\order(\alpha_s)$ one can then replace $\bar{\mr{B}}\to\mr{B}$,
leading to 
\begin{equation*}
  \sum_{\args{f\,}}\int\done\Phi_B(\args{p\,})
  \sum_{\{\widetilde{\im\jm},\tilde{k}\}}\sum_{f_i=q,g}
    \frac{1}{16\pi^2}\int^{\mu^2}\done t\,\int\done z\,
    \int\frac{\done\phi}{2\pi}\,J_{ij,k}(t,z,\phi)\,
    \frac{S(\rmap{\im\jm}{k}{\args{f\,}})}{S(\args{f\,})\,S_{ij}}\,
    \mr{R}_{ij,k}(\rmap{\im\jm}{k}{\args{a}})\;.
\end{equation*}
Comparing this result with the real-emission term in Eq.~\eqref{eq:def_bbar} 
reveals that both expressions differ by the factor
$S(\args{f\,})\,S_{ij}/S(\rmap{\im\jm}{k}{\args{f\,}})$,
which arises solely due to the way the real-emission phase space
is populated by the parton shower (cf.\ Sec.~\ref{sec:ps_construction}).\\
Therefore, the contribution to $\abr{O}^{\rm(POWHEG)}$ from this 
phase-space region is correct to $\order(\alpha_s)$.
\end{itemize}

In the phase-space regions ``between'' these limits, the \POWHEG method
interpolates smoothly between the two above results.

\section{Realisation of the \POWHEG method in the \Sherpa Monte Carlo}
\label{SEC:Sherpa}

\Sherpa is a multi-purpose Monte-Carlo event generator for collider 
experiments~\cite{Gleisberg:2003xi,*Gleisberg:2008ta}.  The goal of this
project is a complete simulation of all aspects of the collision.  Despite
being focused on developments improving the treatment of perturbative 
QCD, over the past years significant improvements have been achieved 
regarding the description of soft QCD and QED dynamics, like the process of
hadronisation, the decays of the produced hadrons, and the implementation of 
QED radiation in these decays~\cite{Schonherr:2008av}.  
One of the traditional key features of the \Sherpa program, however, is a 
consistent merging of multi-jet matrix elements at tree-level with the 
subsequent parton shower in the spirit of~\cite{Catani:2001cc,*Krauss:2002up}.
An improvement of this method and its consistent implementation have been 
presented in~\cite{Hoeche:2009rj} and extended to include hard QED 
radiation~\cite{Hoeche:2009xc}.  
To this end, \Sherpa uses its two internal tree-level matrix element 
generators \Amegic~\cite{Krauss:2001iv} and \Comix~\cite{Gleisberg:2008fv}, 
which are capable of calculating cross sections for processes in the Standard 
Model (\Amegic and \Comix) and beyond (\Amegic), involving final states with
high multiplicities.  Soft and collinear parton radiation is generated in 
\Sherpa by means of a parton shower based on Catani--Seymour dipole 
factorisation~\cite{Schumann:2007mg}. 
The program also allows to steer external modules for the computation of 
virtual corrections using a standardised interface~\cite{Binoth:2010xt}.  The 
corresponding real corrections and subtraction terms in the Catani-Seymour 
formalism~\cite{Catani:1996vz,*Catani:2002hc} are then provided automatically 
by \Amegic~\cite{Gleisberg:2007md}.

\Sherpa is therefore perfectly suited to implement the \POWHEG method as
all prerequisites outlined in Sec.~\ref{SEC:ReviewPOWHEG} are found within
a single, coherent framework. In this section the basic features of the 
corresponding modules of the generator are reviewed, as far as they are 
important in the context of this work.  A complete overview
of \Sherpa can be found in~\cite{Gleisberg:2003xi,*Gleisberg:2008ta}.

%%%%%%%%%%%%%%%%%%%%%%%%%%%%%%%%%%%%%%%%%%%%%%%%%%%%%
\subsection{Matrix elements and subtraction terms}
%%%%%%%%%%%%%%%%%%%%%%%%%%%%%%%%%%%%%%%%%%%%%%%%%%%%%

\mytextwidthfigure{t}{
  %% FF dipole
  \begin{picture}(160,120)(0,0)
    \Line( 80, 50)(110, 70)
    \Line(110, 70)(130,100)
    \Line( 80, 50)(120, 20)
    \Line(110, 70)(143, 70)
    \LongArrow( 90,30)(105, 18)
    \LongArrow(110,85)(120, 100)
    \LongArrow(120,60)(135, 60)
    \Vertex(110,70){2.5}
    \GCirc(80,50){10}{1}
    \put(-10,110){(1)\hspace*{3mm}FF}
    \put( 90, 70){$\widetilde{\im\jm}$}
    \put(140,100){$i$}
    \put(145, 60){$j$}
    \put(125, 15){$k$}
    %\put( 10,110){\underline{$\cD_{ij,k}$}:}
    %\put( 40, 50){$\bV_{ij,k}$}
    \put( 90, 14){$p_k$}
    \put(100, 99){$p_i$}
    \put(120, 50){$p_j$}
  \end{picture}
  \hspace*{20mm}
  %% FI dipole
  \begin{picture}(160,120)(0,0)
    \Line( 80, 50)(110, 70)
    \Line(110, 70)(130,100)
    \Line( 80, 50)( 40, 20)
    \Line(110, 70)(143, 70)
    \LongArrow( 52,18)( 67, 30)
    \LongArrow(110,85)(120, 100)
    \LongArrow(120,60)(135, 60)
    \Vertex(110,70){2.5}
    \GCirc(80,50){10}{1}
    \put(-10,110){(2)\hspace*{3mm}FI}
    \put( 90, 70){$\widetilde{\im\jm}$}
    \put(140,100){$i$}
    \put(145, 60){$j$}  
    \put( 28, 18){$a$}
    %\put( 10,110){\underline{$\cD_{ij}^a$}:}
    %\put( 40, 50){$\bV_{ij}^a$}
    \put( 65, 18){$p_a$}
    \put(100, 99){$p_i$}
    \put(120, 50){$p_j$}
  \end{picture}
  %% IF dipole
  \begin{picture}(160,120)(0,0)
    \Line( 80, 50)( 20, 90)
    \Line( 80, 50)(120, 20)
    \Line( 50, 70)( 80, 90)
    \LongArrow( 90,30)(105, 18)
    \LongArrow( 65,70)( 80, 80)
    \LongArrow( 20,80)( 35, 70)
    \Vertex( 50,70){2.5}
    \GCirc(80,50){10}{1}
    \put(-10,110){(3)\hspace*{3mm}IF}
    \put( 49, 46){$\widetilde{a\jm}$}
    \put( 10, 88){$a$}
    \put( 87, 90){$j$}
    \put(125, 15){$k$}
    %\put( 10,110){\underline{$\cD_k^{aj}$}:}
    %\put(100, 50){$\bV_k^{aj}$}
    \put( 90, 14){$p_k$}
    \put( 20, 63){$p_a$}
    \put( 76, 67){$p_j$}
  \end{picture}
  \hspace*{20mm}
  %% II dipole
  \begin{picture}(160,120)(0,0)
    \Line( 80, 50)( 20, 90)
    \Line( 80, 50)( 40, 20)
    \Line( 50, 70)( 80, 90)
    \LongArrow( 52,18)( 67, 30)
    \LongArrow( 65,70)( 80, 80)
    \LongArrow( 20,80)( 35, 70)
    \Vertex( 50,70){2.5}
    \GCirc(80,50){10}{1}
    \put(-10,110){(4)\hspace*{3mm}II}
    \put( 49, 46){$\widetilde{a\jm}$}
    \put( 10, 88){$a$}
    \put( 87, 90){$j$}
    \put( 28, 18){$b$}
    %\put( 10,110){\underline{$\cD^{aj,b}$}:}
    %\put(100, 50){$\bV^{aj,b}$}
    \put( 65, 18){$p_b$}
    \put( 20, 63){$p_a$}
    \put( 76, 67){$p_j$}
  \end{picture}
  }
  {Effective diagram for the splitting of (1) a final-state parton  
   connected to a final-state spectator, (2) a final-state parton  
   connected to an initial-state spectator, (3) an initial-state parton  
   connected to a final-state spectator and (4) an initial-state parton  
   connected to an initial-state spectator in the standard Catani-Seymour 
   notation.  The blob denotes the colour correlated leading order 
   matrix element, and the incoming and outgoing lines label the initial-state 
   and final-state partons participating in the splitting.
   \label{Fig:dip_configs}}

For this study, the matrix-element generator \Amegic~\cite{Krauss:2001iv}, is
employed.  It is based on the construction of Feynman diagrams, which are evaluated 
using the helicity methods introduced in~\cite{Kleiss:1985yh,*Ballestrero:1994jn}.
For the computation of NLO cross sections in QCD-associated processes, \Amegic 
provides the fully automated generation of dipole subtraction terms
\cite{Gleisberg:2007md}, implementing the Catani-Seymour formalism 
\cite{Catani:1996vz,*Catani:2002hc}. 
As outlined in Sec.~\ref{sec:approx_nlo_xs}, such a subtraction procedure is 
a necessary ingredient to be able and compute NLO QCD cross sections
with Monte Carlo methods.

In the Catani-Seymour method, the soft and collinear singularities of the 
real emission amplitude squared, $\mc{R}(\args{a})$, are removed by a local 
subtraction term (cf.\ Eq.~\eqref{eq:obs_xs_nlo_sub}) 
\bea
 \begin{split}
 \mc{S}(\args{a})
 \,=&\; \sum_{\{i,j\}}\sum_{k\neq i,j} \mc{S}_{ij,k}(\args{a})\\
 \,=&\;   \sum_{i,j}\sum_{k\neq i,j} \tilde{\mc{D}}_{ij,k}(\args{a})
        + \sum_{i,j}\sum_{a}         \tilde{\mc{D}}_{ij}^a(\args{a})
        + \sum_{a,j}\sum_{k\neq j}   \tilde{\mc{D}}_k^{aj}(\args{a})
        + \sum_{a,j}\sum_{b\neq a}   \tilde{\mc{D}}^{aj,b}(\args{a}) \;.
 \end{split}
\eea
On the first line, the notation of Sec.~\ref{sec:approx_nlo_xs} is adopted,
while on the second line the definitions of~\cite{Catani:1996vz,*Catani:2002hc}
are restored by defining $i,j,k>2$, $a,b=1,2$ and requiring all indices to 
be mutually distinct.  Along those lines, $\tilde{\mc{D}}_{ij,k}$,
$\tilde{\mc{D}}_{ij}^a$, $\tilde{\mc{D}}_k^{aj}$ and $\tilde{\mc{D}}^{aj,b}$ 
are the four types of Catani-Seymour dipole terms, as depicted in 
Fig.~\ref{Fig:dip_configs}, for final-state splittings with final-state 
spectators, final-state splittings with initial-state spectators, initial-state 
splittings with final-state spectators and initial-state splittings 
with initial-state spectators, respectively. This implies that for final state 
splittings, i.e. $i,j>2$,
\bea
  \mc{S}_{ij,k}=\mc{S}_{ji,k}=\frac{1}{2}\,\tilde{\mc{D}}_{ij,k}
  \qquad\mbox{and}\qquad
  \mc{S}_{ij,a}=\mc{S}_{ji,a}=\frac{1}{2}\,\tilde{\mc{D}}_{ij}^a \,,
\eea
while for 
initial state splittings
\bea
  \mc{S}_{aj,k}=\tilde{\mc{D}}_k^{aj}
  \qquad\mbox{and}\qquad
  \mc{S}_{aj,b}=\tilde{\mc{D}}^{aj,b}\,.
\eea
Due to its analytic integrability over the extra emission phase space 
$\done\Phi_{R|B}^{ij,k}$  in $D=4-2\epsilon$ dimensions, the subtraction term 
$\mc{S}$, in its integrated form $\mc{I}$ of Eq.~(\ref{eq:sub_term_cancelation}), 
as well as the collinear counter-terms can be added back to the virtual 
contributions to cancel their poles in $\epsilon$,
\bea
  \left\{\epsilon\left[
                        \tilde{\mc{V}}(\args{a})+\mc{I}(\args{a})
  \right]\right\}_{\epsilon=0}
 \,=\; 
  \left\{\epsilon\left[ \vphantom{\tilde{\mc{V}}}
                        \mc{V}(\args{a})+\mc{I}(\args{a})+\mc{C}(\args{a})
  \right]\right\}_{\epsilon=0}
 \,=\;
  0 \;,
\eea
wherein $\mc{V}$ is the one-loop matrix element convoluted with the Born 
amplitude and $\mc{C}$ is the collinear counter-term.

The implementation in \Sherpa's matrix element generator \Amegic, expanding 
upon its tree-level capabilities to generate $\mc{B}$ and $\mc{R}$, is able to 
generate both the subtraction terms $\mc{S}$ and their integrated counterparts 
$\mc{I}$ as well as the collinear counter-term $\mc{C}$ in an automated fashion.
The virtual contributions $\mc{V}$, however, are obtained from dedicated 
external codes interfaced using the Binoth-Les Houches Accord~\cite{Binoth:2010xt}.
Having all this at hand, the assembly of the 
$\bar{\mc{B}}$-function of Eq.~(\ref{eq:def_bbar}), integrable in $D=4$ 
dimensions, is feasible in an automated way. This
involves integrating over the real emission subspace of the phase space of 
the NLO real correction to the Born process, cf.\ Eq.~\eqref{eq:def_bbar} 
and adding the result to the terms with Born-level kinematics.  

\mytable{t}{
  \begin{tabular}{|c|c|c|c|c|c|c|}\cline{1-3}\cline{5-7}
    \vphantom{$\dst\int$}Type & \hspace{6mm}$\alpha$\hspace{6mm} & 
     \hspace{6mm}$z$\hspace{6mm} & & Type & \hspace{6mm}$\alpha$\hspace{6mm} & 
     \hspace{6mm}$z$\hspace{6mm} \\\cline{1-3}\cline{5-7}
    \vphantom{$\dst\int$}FF & $y_{ij,k}$ & $\tilde{z}_i$ & &
      IF & $u_i$ & $x_{ik,a}$ \\
    \vphantom{$\dst\int$}FI & $1-x_{ij,k}$ & $\tilde{z}_i$ & &
      II & $\tilde{v}_i$ & $x_{i,ab}$ \\
  \cline{1-3}\cline{5-7}\end{tabular}\vspace*{1ex}
  }{Definition of integration variables in Eq.~\eqref{eq:eeg_ps} 
    for the various dipole configurations of Fig.~\ref{Fig:dip_configs}.
    \label{tab:eeg_variables}}

In \Sherpa, this integration is performed in a Monte-Carlo fashion, 
by selecting a single point in the real-emission phase space. This technique 
potentially generates negative weights.  In the standard \POWHEG method, 
the emergence of such negative weights is suppressed by either performing 
the integration analytically or by sampling over sufficiently many 
real-emission phase space points.  Tests to decide which method is better for 
practical applications are beyond the scope of this publication and will be 
addressed in a future work; here it should suffice to state that, of course, 
sampling over more than one phase space point is neither a conceptual problem 
nor a practical obstacle.  Also, for all processes under study in this publication, 
no problems have been encountered by the possibility of having negative weighted events.
Loosely speaking, the problem can in no case be more severe than the possibility 
of having negative weights in a standard NLO calculation.
Therefore, the only remaining issue is to construct an 
integration method, which, starting from a given Born configuration, is able 
to fill the real-emission phase space in an efficient manner.  Having an 
implementation of the Catani-Seymour subtraction method at hand, the 
construction of an integrator for the real-emission subspace based on 
CS-subtraction terms is rendered a straightforward exercise.  The actual integration can 
be decomposed into three one-dimensional integrals
(cf.\ Eq.~\eqref{eq:decomposition_radiative_ps})~\cite{Catani:1996vz,*Catani:2002hc}
\begin{equation}\label{eq:eeg_ps}
  \done\Phi_{R|B}^{ij,k}\,=\;\frac{2\,p_{\widetilde{\im\jm}}p_{\tilde{k}}}{16\pi^2}\,
    \done \alpha\,\done z\,\frac{\done\phi}{2\pi}\; \tilde{J}_{ij,k}(\alpha,z,\phi)\;,
\end{equation}
with the two integration variables $\alpha$ and $z$ given in 
Tab.~\ref{tab:eeg_variables}, cf.~\cite{Gleisberg:2007md}. The azimuthal angle,
$\phi$, is common to all configurations.

Several different integration channels, each based on a a separate CS dipole,
can be combined to yield a multi-channel integrator~\cite{Kleiss:1994qy} for 
the real-emission phase space. The a-priori weights in the multi-channel can 
be employed to better adapt to the emission pattern of the process under 
consideration.  Additionally, every one-dimensional integrator can be 
individually improved using the \Vegas 
algorithm~\cite{Lepage:1977sw,*Lepage:1980dq}.

%%%%%%%%%%%%%%%%%%%%%%%%%%%%%%%%%%%%%%%%%%%%%%%%%%%%%
\subsection{The parton shower}
%%%%%%%%%%%%%%%%%%%%%%%%%%%%%%%%%%%%%%%%%%%%%%%%%%%%%
\label{sec:css}
\Sherpa implements a parton shower based on Catani-Seymour (CS)
dipole factorisation, which will be denoted by \CSS~\cite{Schumann:2007mg}. 
The model was originally proposed in~\cite{Nagy:2005aa,*Nagy:2006kb} 
and worked out and implemented in parallel 
in~\cite{Schumann:2007mg,Dinsdale:2007mf}. 
It relies on the factorisation of real-emission matrix elements in the 
CS subtraction framework~\cite{Catani:1996vz,*Catani:2002hc}.  Since the 
original algorithm has been improved in several ways, which have not yet
been compiled in a single reference, the parton shower model and its features 
are briefly reviewed here.

\subsubsection{Ordering parameters and splitting functions}
\mytable{t}{
  \begin{tabular}{|c|c|c|c|c|c|c|c|c|}
  \cline{1-4}\cline{6-9}
  \vphantom{$\dst\int$}Type &
    \hspace*{3mm}$z_{i,jk}$\hspace*{3mm} & 
    \hspace*{3mm}$\xi_{i,jk}$\hspace*{3mm} & 
    \hspace*{7mm}$\tilde{y}_{ij,k}$\hspace*{7mm} & &
    Type & 
    \hspace*{3mm}$z_{j,ak}$\hspace*{3mm} & 
    \hspace*{3mm}$\xi_{j,ak}$\hspace*{3mm} & 
    \hspace*{7mm}$\tilde{y}_{ja,k}$\hspace*{7mm} \\
  \cline{1-4}\cline{6-9}
  \vphantom{$\dst\int\limits_A^B$}FF & 
    $\tilde{z}_i$ & $\tilde{z}_i$ & $y_{ij,k}$ & &
  \vphantom{$\dst\int\limits_A^B$}IF & 
    $x_{jk,a}$ & $\abs{\,1-\dst\frac{1-u_j}{x_{jk,a}-u_j}\,}$ &
    $\dst\frac{u_j}{x_{jk,a}}$ \\
  \cline{1-4}\cline{6-9}
  \vphantom{$\dst\int\limits_A^B$}FI & 
    $\tilde{z}_i$ & $-\,\tilde{z}_i$ & 
    $\dst\frac{1-x_{ij,a}}{x_{ij,a}}$ & &
  \vphantom{$\dst\int\limits_A^B$}II & 
    $x_{j,ab}$ & $\dst1-\frac{1}{x_{j,ab}+\tilde{v}_j}$ &
    $\dst\frac{\tilde{v}_j}{x_{j,ab}}$ \\
  \cline{1-4}\cline{6-9}
  \end{tabular}\vspace*{1ex}
  \vspace*{2mm}
  }{Mapping of variables for Eqs.~\eqref{eq:def_kt_fss},~\eqref{eq:def_kt_iss} 
  and~\eqref{eq:def_zi_kt}.  Note that the definitions for massless partons 
  in~\cite{Catani:1996vz,*Catani:2002hc} are employed.
  \label{tab:css_op_sp_kin}}

Consider the process depicted in Fig.~\ref{Fig:dip_configs}, where a parton
$\widetilde{\im\jm}$, accompanied by a spectator parton $\tilde{k}$, splits into 
partons $i$ and $j$, with the recoil absorbed by the spectator $k$. Conveniently
the combined momenta are identified as $p_{ij}=p_i+p_j$ and $Q=p_{ij}+p_k$.
A Lorentz--invariant transverse momentum, $\tilde{\rm k}_T^2$, which acts as 
ordering parameter in the parton shower algorithm, can now be defined as
\begin{equation}\label{eq:def_kt_fss}
  \tilde{\rm k}_T^{\rm (FS)\,2}\,=\;\abs{Q^2-m_i^2-m_j^2-m_k^2}\,
  \tilde{y}_{ij,k}\,z_{i,jk}(1-z_{i,jk})-
  (1-z_{i,jk})^2\,m_i^2-z_{i,jk}^2\,m_j^2\;,
\end{equation}
This relation holds, independent of whether the spectator parton is a final- 
or initial-state particle for all final-state splittings, while for 
initial-state splittings $a\to\widetilde{a\jm}\;j$, again independent of the 
type of spectator, the ordering parameter is given by
%\footnote{Note that these relations are pure kinematical definitions. 
%  They do not imply that an evolution for massive initial state particles 
%  is implemented in the shower algorithm.}
\begin{equation}\label{eq:def_kt_iss}
  \tilde{\rm k}_T^{\rm (IS)\,2}\,=\;\abs{Q^2-m_j^2-m_a^2-m_k^2}\,
    \tilde{y}_{ja,k}\,(1-z_{j,ak})-m_j^2-(1-z_{j,ak})^2\,m_a^2\;,
\end{equation}
The precise definition of $\tilde{y}$ and the splitting variables $z$ for 
the various dipole types are listed in Tab.~\ref{tab:css_op_sp_kin}.

Sudakov form factors for all branching types, taking into account finite 
masses of final-state partons and strictly relying on the Lorentz-invariant 
variables $z$ and $\tilde{\rm k}_T^2$, have been derived.  The corresponding 
evolution kernels, as defined in Eq.~\eqref{eq:factorisation_ps}, read
\begin{equation}\label{eq:kernel_css}
  \mc K_{ij,k}(\tilde{\rm k}_T^2,z)\,=\;\frac{1}{\bar{S}_{ij}}\,\frac{1}{\mc{N}_{\rm spec}}\,
    \abr{{\rm V}_{ij,k}(\tilde{\rm k}_T^2,z)}\;,
\end{equation}
where $\mc{N}_{\rm spec}$ is the number of spectator partons of the off-shell
particle $a_{\widetilde{\im\jm}}$ in the large $N_C$ limit.  The spin-averaged 
dipole functions $\abr{\rm V}$, taken in four dimensions, depend on the type 
of emitter and spectator parton and are listed in~\cite{Schumann:2007mg}.
Their infrared singularities are regularised through the parton shower 
cutoff, $t_0$, typically of the order of 1 GeV$^2$. The denominator factor
$\bar{S}_{ij}$ avoids double-counting identical final states in final-state
evolution and is given by
\begin{equation}\label{eq:symfac_ps}
  \frac{1}{\bar{S}_{ij}}\,=\;\left\{\begin{array}{cc}
    1/2 & \text{if $\; i,j> 2$}\\
    1 & \text{else}
    \end{array}\right.\;.
\end{equation}

A potential shortcoming of the original approach in~\cite{Schumann:2007mg} 
is that certain dipole functions connecting the initial and final state may 
acquire negative values in some non-singular regions of the phase space.  This 
prohibits their naive interpretation in terms of splitting probabilities and 
leaves the corresponding parts of the phase space unpopulated.  The problem 
was solved recently by altering the finite parts of the affected splitting 
functions such that they reproduce corresponding full matrix 
elements~\cite{Carli:2010cg}.

\subsubsection{Splitting kinematics}
\label{sec:css_kinematics}
All branchings in the \CSS formalism implement exact four-momentum 
conservation and the particles are kept on their mass-shell before and after 
every evolution step.  The phase-space maps from $n$- to n+1-particle 
final states are exact and cover the whole phase space. They are, however, 
not unambiguous.  One of the most prominent criticisms of this particular 
parton-shower model is the ``unphysical'' recoil strategy originally employed 
in configurations with initial-state splitter and final-state spectator.  This 
problem was addressed in~\cite{Platzer:2009jq} for the case of massless 
partons and has been implemented for prompt photon production and extended
to the fully massive case in~\cite{Hoeche:2009xc}.  The latter publication 
defined a general algorithm for constructing the emission kinematics, 
independent of the type of splitter and spectator parton, which is summarised 
as follows:
\begin{enumerate}
\item Determine the new momentum of the spectator parton 
  as~\cite{Catani:1996vz,*Catani:2002hc}
  \begin{equation}\label{eq:def_pk}
    \begin{split}
      p_k=&\;\left(\tilde{p}_k-\frac{Q^2+m_k^2-m_{ij}^2}{2\,Q^2}\,Q\right)\,
      \sqrt{\frac{\lambda(Q^2,s_{ij},m_k^2)}{\lambda(Q^2,m_{ij}^2,m_k^2)}}
      +\frac{Q^2+m_k^2-s_{ij}}{2\,Q^2}\,Q\;,
    \end{split}
  \end{equation}
  with $\lambda$ denoting the K{\"a}llen function 
  $\lambda(a,b,c)=\rbr{a-b-c}^2-4\,bc$ and
  \begin{equation}\label{eq:def_sij}
    s_{ij}\,=\;\tilde{y}_{ij,k}\,(Q^2-m_k^2)+(1-\tilde{y}_{ij,k})\,(m_i^2+m_j^2)\;.
  \end{equation}
\item Find the light-like helper vectors $l$ and $n$ as
  \begin{align}\label{eq:sudakov_momenta}
    l\,=&\;\frac{p_{ij}-\alpha_{ij}\, p_k}{1-\alpha_{ij}\alpha_k}\,\;,
    &n\,=&\;\frac{p_k-\alpha_k\, p_{ij}}{1-\alpha_{ij}\alpha_k}\,\;,
  \end{align}
  where $\alpha_{ij}=s_{ij}/\gamma_{ij,k}\;$, $\alpha_k=m_k^2/\gamma_{ij,k}$ and
  \begin{align}
    \gamma_{ij,k}\,=&\;2\,ln\,=\;\frac{1}{2}\sbr{\,\rbr{Q^2-s_{ij}-m_k^2}
      +{\rm sgn}\rbr{Q^2-s_{ij}-m_k^2}\sqrt{\lambda(Q^2,s_{ij},m_k^2)}\;}\;.
  \end{align}
\item Express the momenta $p_i$ and $p_j$ in terms of $l$, $n$ and a 
  transverse component $k_\perp$ as
  \begin{align}\label{eq:def_pi_pj_FF}
    p_i^\mu\,=&\;\bar{z}_i\,l^\mu+\frac{m_i^2+{\rm k}_\perp^2}{\bar{z}_i}\,
    \frac{n^\mu}{2\,ln}+k_\perp^\mu\;,
    &p_j^\mu\,=&\;(1-\bar{z}_i)\,l^\mu+
    \frac{m_j^2+{\rm k}_\perp^2}{1-\bar{z}_i}\,
    \frac{n^\mu}{2\,ln}-k_\perp^\mu\;,
  \end{align}
  The parameters $\bar{z}_i$ and ${\rm k}_\perp^2$ of this decomposition are 
  given by
  \begin{equation}\label{eq:def_zi_kt}
    \begin{split}
      \bar{z}_i\,=&\;\frac{Q^2-s_{ij}-m_k^2}{
        \sqrt{\lambda\rbr{Q^2,s_{ij},m_k^2}}}\,
      \sbr{\;\xi_{i,jk}\,-\,\frac{m_k^2}{\abs{\gamma_{ij,k}}}\rbr{\,
          \frac{\tilde{y}_{ij,k}}{1-\tilde{y}_{ij,k}}+
          \frac{2\,m_i^2}{Q^2-s_{ij}-m_k^2}}\,}\;,\\
          {\rm k}_\perp^2\,=&\;\rbr{Q^2-m_i^2-m_j^2-m_k^2}\,
          \tilde{y}_{ij,k}\;\bar{z}_i\,(1-\bar{z}_i)-
          (1-\bar{z}_i)^2 m_i^2-\bar{z}_i^2 m_j^2\;,
    \end{split}
  \end{equation}
  where $\xi_{i,jk}$ is defined in Tab.~\ref{tab:css_op_sp_kin}. 
  Equations~\eqref{eq:def_pk} and~\eqref{eq:def_zi_kt} are valid for all 
  dipole configurations, i.e.\ initial and final-state branchings with 
  the recoil partner being either in the initial or in the final state.
\end{enumerate}

To generate an emission using the reweighting technique presented 
in Sec.~\ref{sec:ps_mecorrection}, it is important to be able and access
matrix-element information during the parton-shower evolution, such that
Eq.~\eqref{eq:me_correction} can be implemented in a process-independent
manner.  \Sherpa provides an interface between its tree-level matrix-element 
generators and its parton shower, which allows for all the necessary 
interactions.  Together with an implementation of the phase-space maps
$\bmap{ij}{k}{\args{a}}$, that correspond to the inverse of the above 
splitting kinematics, Eq.~\eqref{eq:me_correction} can then be realised
easily.

\subsection{Automatic identification of Born zeros}
\label{sec:born_zeros}

It was noted in~\cite{Alioli:2008gx} that Eq.~\eqref{eq:me_correction}
can develop spurious singularities as the matrix element of the underlying 
Born process may be zero, while the real-emission matrix element is not. 
Such configurations do not exponentiate, as $\mc{R}$ is not singular 
when $\mc{B}\to 0$. This fact can be employed to formulate a general solution
to the problem~\cite{Alioli:2008gx}. One can split $\mc{R}$ into two parts, 
a singular one, $\mc{R}^{\rm(s)}$, and a regular one, $\mc{R}^{\rm(r)}$.
\begin{align}\label{eq:zh_split}
  \mc{R}^{\rm(s)}\,=&\;\mc{R}\,\frac{Z}{Z+H}\;,
  &&\text{and}
  &\mc{R}^{\rm(r)}\,=&\;\mc{R}\,\frac{H}{Z+H}\;,
\end{align}
where
\begin{align}\label{eq:zh_def}
  Z\,=&\;\frac{\mc{B}}{\mc{B}_{\rm max}}\;,
  &&\text{and}
  &H\,=\;\kappa^2_\text{res}\;\frac{t}{t_{\rm max}}\;.
\end{align}
Note that $\mc{B}_{\rm max}$ can be determined during the integration
of the seed cross section, while $t_{\rm max}$ is given as a universal 
function of the hadronic centre-of-mass energy, depending only on
the definition of $t$ in the parton shower model. The resolution factor 
$\kappa_\text{res}$ then determines the relative splitting between 
$\mc{R}^{\rm(s)}$ and $\mc{R}^{\rm(r)}$: the larger $\kappa_\text{res}$, 
the larger the fraction $\mc{R}^{\rm(r)}$ of $\mc{R}$.

The necessity of such a splitting of the real emission matrix element 
can be determined on an event-by-event basis by comparing the correction 
factor to the parton shower of Eq.~\eqref{eq:def_me_correction}, $w_{ij,k}$, 
to a predefined threshold $w_{ij,k}^\text{th}$. Thus, regular non-exponentiated 
$\mc{R}^{\rm(r)}$ events are only produced if $w_{ij,k}>w_{ij,k}^\text{th}$. 
Such a treatment ensures that both the exponentiation of the real-emission 
matrix element is as inclusive as possible and the parton shower correction 
factor does not get too large, rendering event generation too inefficient.

\section{Results}
\label{SEC:Results}

This section collects results obtained with the implementation of the \POWHEG{}
algorithm in the \Sherpa{} event generator.
We exemplify the performance in a variety of processes which are listed in
Sec.~\ref{sec:results:processes} with their cuts and relevant settings.
In Sec.~\ref{sec:results:consistency} the internal consistency of the
implementation is checked by performing scale variations, cross section comparisons with
ordinary NLO calculations, and variations of internal parameters of \POWHEG{}.
Comparisons of results from the new implementation with predictions
from tree-level matrix-element parton-shower merging (ME+PS) are presented in 
Sec.~\ref{sec:results:meps}. Finally, comparisons with experimental data
are made in Sec.~\ref{sec:results:data}.

\subsection{Process listing}
\label{sec:results:processes}

\subsubsection{Jet production in $e^+e^-$ collisions}
\label{sec:results_ee}

The annihilation of $e^+e^-$ into hadrons is studied at LEP Run 1 energies,
$E_\mathrm{CMS}=91.25$ GeV. This setup allows to validate the algorithms of 
Sec.~\ref{SEC:ReviewPOWHEG} in pure final-state QCD evolution, which is the 
simple-most testing ground. The parton shower cut-off scale has been set to
$k_{T,\,\mr{min}}^2=1.6~\text{GeV}^2$. Even though the improvements in this paper
are purely related to perturbative physics, the results are presented after
hadronisation with the Lund model~\cite{Sjostrand:2006za} to make comparison to
experimental results more meaningful. The ME+PS samples have been
generated with up to one additional jet in the matrix elements and the phase
space slicing parameter was set to $\log(y_\mathrm{cut})=-2.25$.
For the virtual matrix elements, the code provided by the BlackHat
collaboration~\cite{whitehat:2010aa,*Berger:2009zg,*Berger:2009ep,*Berger:2010vm} 
was used.

\subsubsection{Deep-inelastic lepton-nucleon scattering}
\label{sec:results_dis}

Hadronic final states in deep-inelastic lepton-nucleon scattering (DIS) are
studied at HERA Run 1 energies, $E_\mathrm{CMS}=300$ GeV. 
Just like $e^+e^-$-annihilation into hadrons, this process boasts a wealth of 
precise experimental data. From the theoretical perspective, it is invaluable, 
as it allows to test QCD factorisation in an extremely clean environment.
The associated scale, given by the virtuality of the exchanged $\gamma^*/Z$-boson 
is not fixed, but potentially varies by orders of magnitude, which allows to test 
perturbative QCD predictions in various kinematic limits. 
Our results are presented at the parton level only, as hadronisation corrections 
have little effect on the observables and the focus lies on the potential 
improvements on the perturbative part of the simulation.
The CTEQ6.6~\cite{Nadolsky:2008zw} parton distribution functions have been
employed and the strong coupling has been defined accordingly as
$\alpha_s(m_Z)=0.118$ with NLO running for both the matrix elements and the parton
shower. The remaining settings correspond to those in~\cite{Carli:2010cg}.
ME+PS samples have been generated with up to one additional jet in the 
matrix element and the phase space slicing parameters were set to $\bar{Q}_\mr{cut}=5$
and $S_\mr{DIS}=0.6$ (cf.~\cite{Carli:2010cg}).
Virtual matrix elements were provided by BlackHat~\cite{whitehat:2010aa,
  *Berger:2009zg,*Berger:2009ep,*Berger:2010vm}.

\subsubsection{Drell-Yan lepton pair production}
\label{sec:drellyan}

We investigate Drell-Yan lepton pair production at Tevatron Run 2 energies,
simulating $p\bar{p}$ collisions at $E_\mathrm{CMS}=1.96$~TeV.
The CTEQ6.6~\cite{Nadolsky:2008zw} parton distribution functions are
employed and the strong coupling is defined accordingly as
$\alpha_s(m_Z)=0.118$ with NLO running for both the matrix elements and the parton
shower. A cut on the invariant mass of the lepton pair of
$66<m_{\ell\ell}/\text{GeV}<116$ is applied at the matrix-element level.
For the ME+PS samples matrix elements with up to one additional jet were 
generated and a phase-space slicing cut of $Q_{\text{cut}}=20$~GeV was applied.
Virtual matrix elements were provided by BlackHat~\cite{whitehat:2010aa,
  *Berger:2009zg,*Berger:2009ep,*Berger:2010vm}.
The factorisation and renormalisation scales for the NLO matrix element were
chosen as $\mu^2_R=\mu^2_F=m^2_{\perp,\,\ell\ell}$. In all tree-level matrix elements
\Sherpa's default scale choice was employed: The matrix element is clustered
onto a core $2\to 2$ configuration using a $k_T$-type algorithm with
recombination into on-shell particles. Scales are defined as the lowest invariant
mass or negative virtuality in the core process.
Hadronisation and multiple parton interactions have been disabled to allow for
a study at the parton shower level. The $Z\to\ell\ell$ decay is corrected 
for QED next-to-leading order and soft-resummation effects in the 
Yennie-Frautschi-Suura (YFS) approach~\cite{Schonherr:2008av}.

The three reactions listed in Sec.~\ref{sec:results_ee}-\ref{sec:drellyan}
essentially amount to one and the same process at the
parton level, as they only differ by crossing of initial and final state legs. 
Their combination allows to validate the implementation of the matrix-element 
reweighting in Sec.~\ref{sec:ps_mecorrection} for all possible dipole configurations 
with quark splitters.

\subsubsection{$W$ boson production}

Production of $W$ bosons is presented in $p\bar{p}$ collisions at
$E_\mathrm{CMS}=1.8$~TeV.
Although in principle similar to the Drell-Yan case, this process
is of special interest to validate the automatic 
decomposition of the real-emission term into singular and non-singular pieces,
as outlined in Sec.~\ref{sec:born_zeros}. If not stated otherwise, the
parameters for this decomposition are set to $\kappa_\text{res}=4$ and 
$w_{ij,k}^\text{th}=100$.
The CTEQ6.6~\cite{Nadolsky:2008zw} parton distribution functions have been
employed and the strong coupling has been defined accordingly as
$\alpha_s(m_Z)=0.118$ with NLO running for both the matrix elements and the parton
shower. A cut on the invariant mass of the lepton-neutrino
pair of $m_{\ell\nu}>10$~GeV was applied at the matrix-element level.
For the ME+PS samples matrix elements with up to one additional jet 
were used and a phase space slicing cut of $Q_{\text{cut}}=20$~GeV 
was applied.
Virtual matrix elements were provided by BlackHat~\cite{whitehat:2010aa,
  *Berger:2009zg,*Berger:2009ep,*Berger:2010vm}.
The factorisation and renormalisation scales for the NLO matrix element were
chosen as $\mu^2_R=\mu^2_F=m^2_{\perp,\,\ell\nu}$. In all tree-level matrix elements
\Sherpa's default scale choice was employed, cf. Sec.~\ref{sec:drellyan}.
Hadronisation and multiple parton interactions have been disabled. 
The $W\to\ell\nu$ decay is corrected for QED next-to-leading order and 
soft-resummation effects in the YFS approach~\cite{Schonherr:2008av}.

\subsubsection{Higgs boson production through gluon-gluon fusion}

The production of Higgs bosons through gluon-gluon fusion is
simulated for proton-proton collisions at $E_{\text{CMS}}=14$~TeV.
The coupling to gluons is mediated by a top-quark loop and modeled through 
an effective Lagrangian \cite{Dawson:1990zj}.
Again, this process is technically very similar to the Drell-Yan case, but it also
allows to validate matrix-element corrections to the remaining initial state splitting 
functions. Next-to-leading order corrections are rather large at nominal LHC energies, 
with a ratio of $K\approx 2$ between the NLO and the LO result for the total cross section. 
This fact has spurred tremendous efforts to perform fully differential calculations at 
NNLO~\cite{Anastasiou:2005qj,*Anastasiou:2007mz,*Anastasiou:2008tj} and several predictions 
have been presented which merged such fixed-order results with resummation at next-to-next-to-leading 
logarithmic accuracy~\cite{Catani:2003zt,*Bozzi:2005wk}, as the process is expected to have high 
phenomenological relevance at LHC energies. However, this publication
centres on the behaviour of the theory at NLO only, as a prediction
beyond this level of accuracy is clearly not within the capabilities of the \POWHEG method.
The CTEQ6.6~\cite{Nadolsky:2008zw} parton distribution functions have been
employed and the strong coupling has been defined accordingly as
$\alpha_s(m_Z)=0.118$ with NLO running for both the matrix elements and the parton
shower. A cut for the invariant mass of the $\tau$ pair of 
$115<m_{\tau\tau}/\text{GeV}<125$ was applied at the matrix-element level.
For the ME+PS merged samples matrix elements with up to one additional jet 
were used and a phase-space slicing cut of $Q_{\text{cut}}=20$~GeV 
was applied.
The virtual matrix elements have been implemented according to~\cite{Hamilton:2009za}.
The factorisation and renormalisation scales for the NLO matrix element were
chosen as $\mu^2_R=\mu^2_F=m^2_{\perp,\,\tau\tau}$. In all tree-level matrix elements
\Sherpa's default scale choice was employed, cf. Sec.~\ref{sec:drellyan}.
Hadronisation and multiple parton interactions have been disabled.
The $h\to\tau\tau$ decay is corrected for QED soft-resummation and 
approximate next-to-leading order effects in the YFS approach~\cite{Schonherr:2008av}.

\subsubsection{$Z$--pair production}

The production of pairs of $Z$ bosons is studied for proton-proton collisions at
$E_{\text{CMS}}=14$~TeV.  This is an important background for the golden-plated
Higgs boson discovery mode at the LHC.  Detailed studies of the decay properties
of the $Z$ bosons and their correlations are known to allow for a determination
of some properties of the Higgs boson, when found.  Among these correlations
are, e.g.\ the relative orientation of the decay planes of the bosons.  

The CTEQ6.6~\cite{Nadolsky:2008zw} parton distribution functions have been
employed and the strong coupling has been defined accordingly as
$\alpha_s(m_Z)=0.118$ with NLO running for both the matrix elements and the 
parton shower. A cut on the invariant mass of each lepton pair of
$66<m_{\ell\ell}/\text{GeV}<116$ was applied at the matrix-element level.
For the ME+PS samples matrix elements with up to one additional jet were 
used and a phase-space slicing cut of $Q_{\text{cut}}=20$~GeV was applied.
Virtual matrix elements were provided by \MCFM~\cite{MCFM,Campbell:1999ah,*Dixon:1998py}.
The factorisation and renormalisation scales were chosen as $\mu^2_R=\mu^2_F=m^2_{ZZ}$.
Hadronisation and multiple parton interactions have been disabled to allow for
a study at the parton shower level. Each $Z\to\ell\ell$ decay is corrected 
for QED next-to-leading order and soft-resummation effects in the YFS 
approach~\cite{Schonherr:2008av}.

\subsubsection{$W^+ W^-$--production}

$W^+W^-$--production is also studied for proton-proton collisions at
$E_{\text{CMS}}=14$~TeV.  It is worth noting that this process hitherto has not
been treated in the \POWHEG approach.  Similar to the $Z$--pair production,
it is an important background to the search channel for the Standard Model 
Higgs boson, at masses around and above 130 GeV.  Again, in order to suppress
this background, distributions which depend on correlations of decay products
of the $W$'s in phase space are heavily employed.   

In the simulation here, again the CTEQ6.6~\cite{Nadolsky:2008zw} parton 
distribution functions have been employed and the strong coupling has been 
defined accordingly as $\alpha_s(m_Z)=0.118$ with NLO running for both the 
matrix elements and the parton shower. A cut on the invariant mass of each 
lepton-neutrino pair of $m_{\ell\nu}>10$~GeV was applied at the matrix-element 
level.  For the ME+PS samples matrix elements with up to one additional jet 
were used and a phase-space slicing cut of $Q_{\text{cut}}=20$~GeV was applied.
Virtual matrix elements were provided by \MCFM~\cite{MCFM,Campbell:1999ah,*Dixon:1998py}.  
The factorisation and renormalisation scales were chosen as 
$\mu^2_R=\mu^2_F=m^2_{WW}$.  Hadronisation and multiple parton interactions 
have been disabled to allow for a study at the parton shower level. Each 
$W\to\ell\nu$ decay is corrected for QED next-to-leading order and 
soft-resummation effects in the YFS approach~\cite{Schonherr:2008av}.

\subsection{Tests of internal consistency}
\label{sec:results:consistency}

%%                %    ee    %    DY    %    W-    %    DIS   %    ggh   %
%%%%%%%%%%%%%%%%%%%%%%%%%%%%%%%%%%%%%%%%%%%%%%%%%%%%%%%%%%%%%%%%%%%%%%%%%%
%%                %  LEP I   %  TeV II  %  TeV I   %   HERA   %    LHC   %
%%%%%%%%%%%%%%%%%%%%%%%%%%%%%%%%%%%%%%%%%%%%%%%%%%%%%%%%%%%%%%%%%%%%%%%%%%
%% Scale % Factor % PH % NLO % PH % NLO % PH % NLO % PH % NLO % PH % NLO %
%%%%%%%%%%%%%%%%%%%%%%%%%%%%%%%%%%%%%%%%%%%%%%%%%%%%%%%%%%%%%%%%%%%%%%%%%%
%%       %  1/2   %    %     %    %     %    %     %    %     %    %     %
%%  mV   %   1    %    %     %    %     %    %     %    %     %    %     %
%%       %   2    %    %     %    %     %    %     %    %     %    %     %
%%%%%%%%%%%%%%%%%%%%%%%%%%%%%%%%%%%%%%%%%%%%%%%%%%%%%%%%%%%%%%%%%%%%%%%%%%
%%       %  1/2   %    %     %    %     %    %     %    %     %    %     %
%% mperp %   1    %    %     %    %     %    %     %    %     %    %     %
%%       %   2    %    %     %    %     %    %     %    %     %    %     %
%%%%%%%%%%%%%%%%%%%%%%%%%%%%%%%%%%%%%%%%%%%%%%%%%%%%%%%%%%%%%%%%%%%%%%%%%%

\mytable{t}{
\begin{tabular}{|c|c|c|c|c|c|}
 \hline
 %%%%%
 \multicolumn{2}{|c|}{\hphantom{lalala}} 
 & \multicolumn{2}{|c|}{\hl $e^+e^-\to$ hadrons} 
 & \multicolumn{2}{|c|}{$e^+p\to e^+\!+j+X$} \\\cline{3-6}
 \multicolumn{2}{|c|}{\hphantom{lalala}} 
 & \multicolumn{2}{|c|}{\multirow{2}{*}{$E_{\rm cms}$ = 91.2 GeV}} 
 & \multicolumn{2}{|c|}{$E_{\rm cms}$ = 300 GeV} \\
 \multicolumn{2}{|c|}{\hphantom{lalala}} 
 & \multicolumn{2}{|c|}{} 
 & \multicolumn{2}{|c|}{$Q^2\,>\,$150 GeV$^2$} \\\hline
 %%%%%
 \hl $\mu=\mu_R=\mu_F$ & Factor 
  & \POWHEG & \Nlo 
  & \POWHEG & \Nlo \\\hline\hline
 %%%%%
 \multirow{3}{*}{$\sqrt{Q^2}$} & $1/2\vphantom{\frac{|}{|}}$ 
  & 30179(18) & 30195(20)
  & 3906(9)& 3908(10) \\\cline{2-6}
 %%%%%
  & $1\vphantom{\frac{|}{|}}$
  & 29411(17) & 29416(18)
  & 4047(10) & 4050(11) \\\cline{2-6}
 %%%%%
  & $2\vphantom{\frac{|}{|}}$
  & 28680(16) & 28697(18)
  &  4180(10)& 4188(11) \\\hline
\end{tabular}\vspace*{1mm}
}{Cross sections in pb for $e^+\!$-$\,e^-\!$ annihilation into hadrons at \LEP and 
  deep-inelastic positron-proton scattering at \Hera as calculated in the 
  \POWHEG framework and in a conventional fixed order \Nlo calculation 
  \cite{Gleisberg:2007md}.\label{Tab:xsec_consistency_eeDIS}}

\mytable{t}{
\begin{tabular}{|c|c|c|c|c|c|c|c|}
 \hline
 %%%%%
 \multicolumn{2}{|c|}{\hphantom{lalala}} 
 & \multicolumn{2}{|c|}{\hl $p\bar{p}\to W^+\!+X$}
 & \multicolumn{2}{|c|}{$p\bar{p}\to Z+X$}
 & \multicolumn{2}{|c|}{$pp\to h+X$} \\\cline{3-8}
 \multicolumn{2}{|c|}{\hphantom{lalala}} 
 & \multicolumn{2}{|c|}{$E_{\rm cms}$ = 1.8 TeV}
 & \multicolumn{2}{|c|}{$E_{\rm cms}$ = 1.96 TeV}
 & \multicolumn{2}{|c|}{$E_{\rm cms}$ = 14 TeV} \\
 \multicolumn{2}{|c|}{\hphantom{lalala}} 
 & \multicolumn{2}{|c|}{$m_{\ell\nu}\,>\,10$ GeV}
 & \multicolumn{2}{|c|}{$66\,<\,m_{\ell\ell}\,<\,116$ GeV}
 & \multicolumn{2}{|c|}{$115\,<\,m_{\tau\tau}\,<\,125$ GeV} \\\hline
 %%%%%
 \hl $\mu=\mu_R=\mu_F$ & Factor 
  & \POWHEG & \Nlo
  & \POWHEG & \Nlo 
  & \POWHEG & \Nlo \\\hline\hline
 %%%%%
 \multirow{3}{*}{$m_{\ell\nu}/m_{\ell\ell}$} & $1/2\vphantom{\frac{|}{|}}$ 
  & \hspace*{1mm}1235.4(5)\hspace*{1mm} & 1235.1(1.0)
  & 243.96(14) & 243.84(16)
  & 2.3153(13) & 2.3130(13)\\\cline{2-8}
 %%%%%
  & $1\vphantom{\frac{|}{|}}$
  & 1215.0(5) & 1214.9(9)
  & 239.70(13) & 239.59(16)
  & 2.4487(12)& 2.4474(13) \\\cline{2-8}
 %%%%%
  & $2\vphantom{\frac{|}{|}}$
  & 1201.4(5) & 1202.0(9)
  & 236.72(13) & 236.77(15)
  & 2.5811(13) & 2.5786(13) \\\hline\hline
 %%%%%
 \multirow{3}{*}{$m_\perp$} & $1/2\vphantom{\frac{|}{|}}$ 
  & 1231.0(5) & 1230.3(1.0)
  & 243.00(14) & 243.06(16)
  & 2.2873(13) & 2.2869(14) \\\cline{2-8}
 %%%%%
  & $1\vphantom{\frac{|}{|}}$
  & 1211.8(5) & 1211.7(9)
  & 239.01(13) & 238.96(15)
  & 2.4255(12) & 2.4231(19) \\\cline{2-8}
 %%%%%
  & $2\vphantom{\frac{|}{|}}$
  & 1198.8(5) & 1199.3(9)
  & 236.23(13) & 236.13(14)
  & 2.5623(13) & 2.5620(14) \\\hline
\end{tabular}\vspace*{1mm}
}{Cross sections in pb for inclusive $W^+[\to e^+\nu_e]$ and $Z[\to e^+e^-]$ 
  production at the \Tevatron and $h[\to\tau^+\tau^-]$ production via a 
  top-quark loop at the \LHC as calculated in the \POWHEG framework and in a 
  conventional fixed order \Nlo calculation 
  \cite{Gleisberg:2007md}.\label{Tab:xsec_consistency_WZh}}

The aim of this section is to provide consistency checks on the different 
aspects of the \POWHEG implementation in \Sherpa. At first, total cross sections 
as obtained from \POWHEG are compared with the corresponding results from
a standard NLO calculation. In this case, the public release 
\Sherpa-1.2.2~\footnote{See \href{http://www.sherpa-mc.de}{http://www.sherpa-mc.de}.}
serves as the reference, which includes an implementation of~\cite{Gleisberg:2007md}.
Results for $e^+\!$-$\,e^-\!$ annihilation into hadrons and deep-inelastic 
positron-proton scattering are presented in Tab.~\ref{Tab:xsec_consistency_eeDIS}.
Numbers for inclusive $Z$-boson production with decay to an electron-positron pair, 
for inclusive $W$-boson production with decay to an electron-neutrino pair, 
and for Higgs-boson production via a top-quark loop with decay into $\tau$
are listed in Tab.~\ref{Tab:xsec_consistency_WZh}. The agreement between the
\POWHEG results and those of the standard integration method typically is within
a $1\sigma$ range as given by the respective Monte-Carlo errors.

To examine differences between \POWHEG and a parton-shower Monte Carlo regarding the
exponentiation of the real-emission matrix elements in \POWHEG, $\mc{R}$ can be 
approximated by $\mc{R}^\text{(PS)}$ in Eq.~\eqref{eq:me_correction}. Performing this 
replacement does not only constitute a mandatory cross-check, whether the parton-shower 
approximation is retained, but it also estimates the size of corrections that can be 
expected at all when switching to NLO accuracy in the event simulation.
Apart from the overall normalisation, in processes with no additional phase space 
dependence introduced by the loop matrix element, the emission pattern in \POWHEG 
should be identical to the parton-shower result. This is verified in inclusive $Z$-boson 
production at \Tevatron energies as displayed in Fig.~\ref{Fig:approxme_Z}. 
For low transverse momentum (low jet resolution) $p_\perp\ll\mu_F$ both distributions 
coincide within statistical errors. For large values the emission phase space is 
severely restricted in the parton-shower approach, as $t<\mu_F^2\approx m_Z^2$ and
$p_\perp\lesssim t$. Any contribution to this phase-space region must therefore originate 
from configurations where more than one hard parton recoils against the lepton pair.
Such configurations are suppressed by higher orders of $\alpha_s$, and therefore
the emission rate is gravely underestimated by the parton shower. As a direct 
consequence, all deviations are then manifestations of the 
exponentiation of non-logarithmic terms, which can be sizeable in the hard 
wide-angle emission region.

The automatic splitting of the real-emission matrix element into singular and
regular contributions as presented in Sec.~\ref{sec:born_zeros} contains 
two unphysical parameters: $\kappa_\text{res}$, which governs the relative sizes
of the exponentiated, singular part $\mc{R}^{\rm(s)}$ and the non-exponentiated, 
regular part $\mc{R}^{\rm(r)}$, and $w_{ij,k}^\text{th}$, which determines 
when the above separation is actually employed. The effect of 
$\kappa_\text{res}$ on the central parton shower reweighting factor $w_{ij,k}$ 
is detailed Fig.~\ref{Fig:ZH_RBs}. There, it can be seen that for values of 
$\kappa_\text{res}$ chosen neither too low, such that the maximum of the 
reweighting factor rises beyond reasonable bounds rendering the reweighting of 
the parton shower inoperable, nor too high, such that parts of leading 
logarithmic structure of $\mc{R}$ are not exponentiated, event generation with 
the accuracy aimed at by the \POWHEG algorithm is feasible. 
Hence, the results of the Monte-Carlo simulation should be fairly 
independent of $\kappa_\text{res}$ and $w_{ij,k}^\text{th}$, if varied within a reasonable 
range. Figure~\ref{Fig:ZH_splitting} displays predictions for transverse momentum 
spectra in $W$-boson production for several values of $\kappa_\text{res}$. As 
expected, no significant variations of the emission pattern can be observed. 
The small differences that can be seen when changing the resolution scale 
$\kappa_\text{res}$ are entirely within the logarithmic accuracy of the 
parton-shower approach and therefore also within the logarithmic accuracy of 
the real-emission contribution in \POWHEG. Variations 
in $w_{ij,k}^\text{th}$ only have very little influence on physical
distributions. %% maybe also add plots to the latter statement

\myfigure{t}{
  \includegraphics[width=0.42\textwidth]{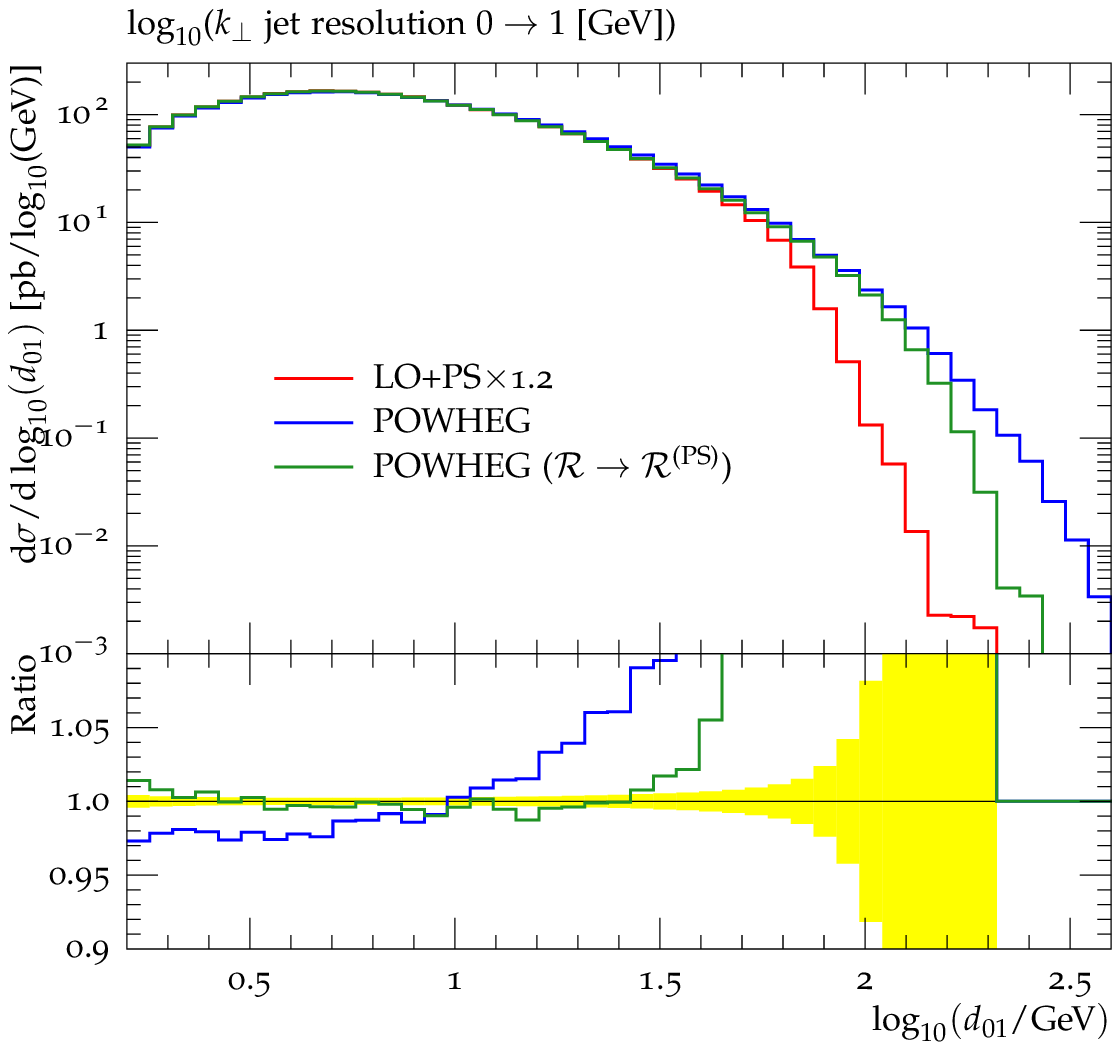}
  \hspace*{0.05\textwidth}
  \includegraphics[width=0.42\textwidth]{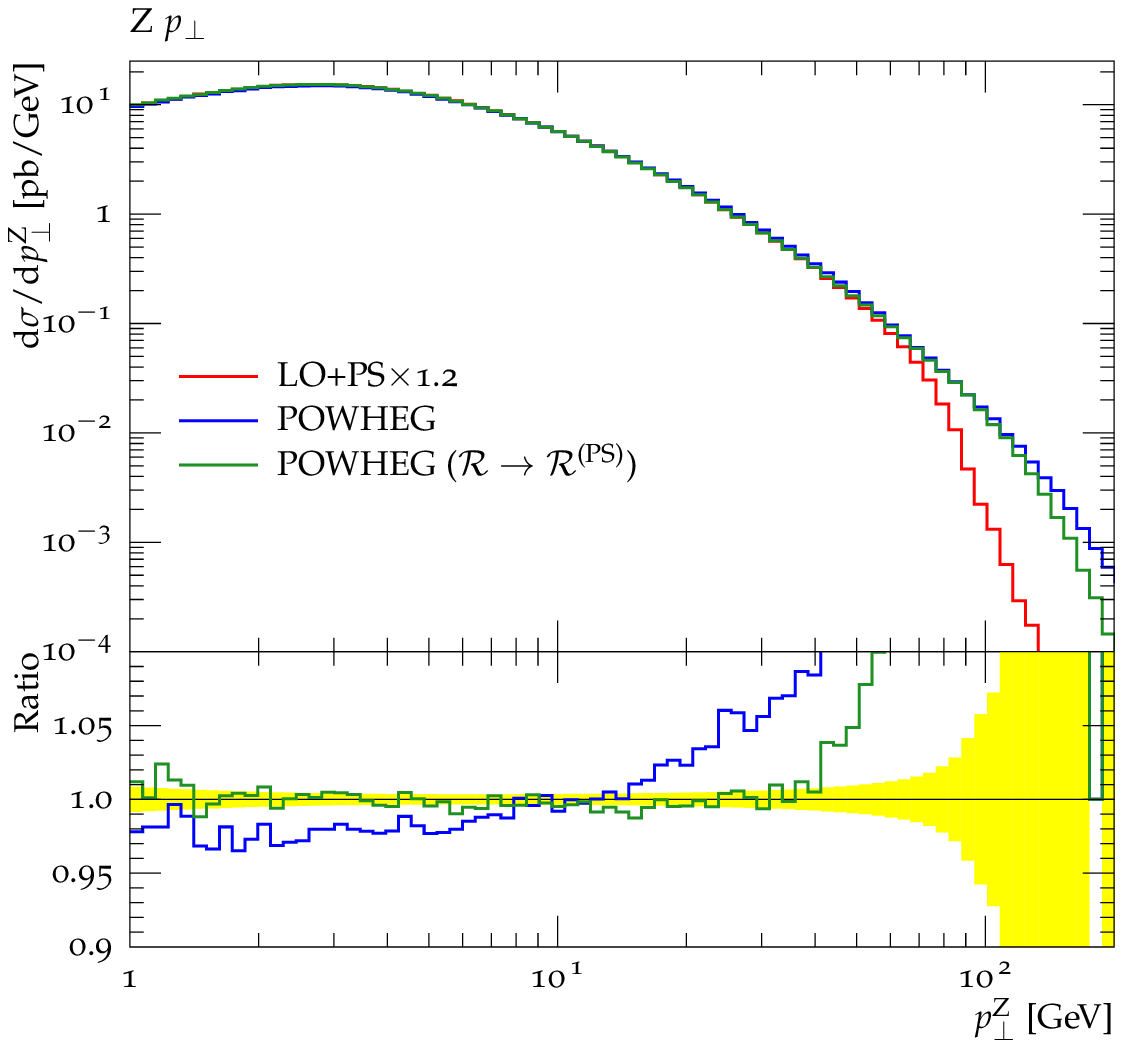}
}
{
  $0\to 1$ jet resolution in $k_T$ clustered jets and transverse momentum of 
  the $e^+e^-\!$-pair in $Z/\gamma^*$ boson production at the \Tevatron.
  The standard parton shower effected on the leading order matrix elements 
  (red) is compared to the \POWHEG formulation (blue) and to \POWHEG with 
  the real emission matrix element $\mc{R}$ replaced by its parton-shower approximation 
  $\mc{R}^\text{(PS)}$ (green).\label{Fig:approxme_Z}
}

\clearpage

\myfigure{p}{
  \includegraphics[width=0.42\textwidth]{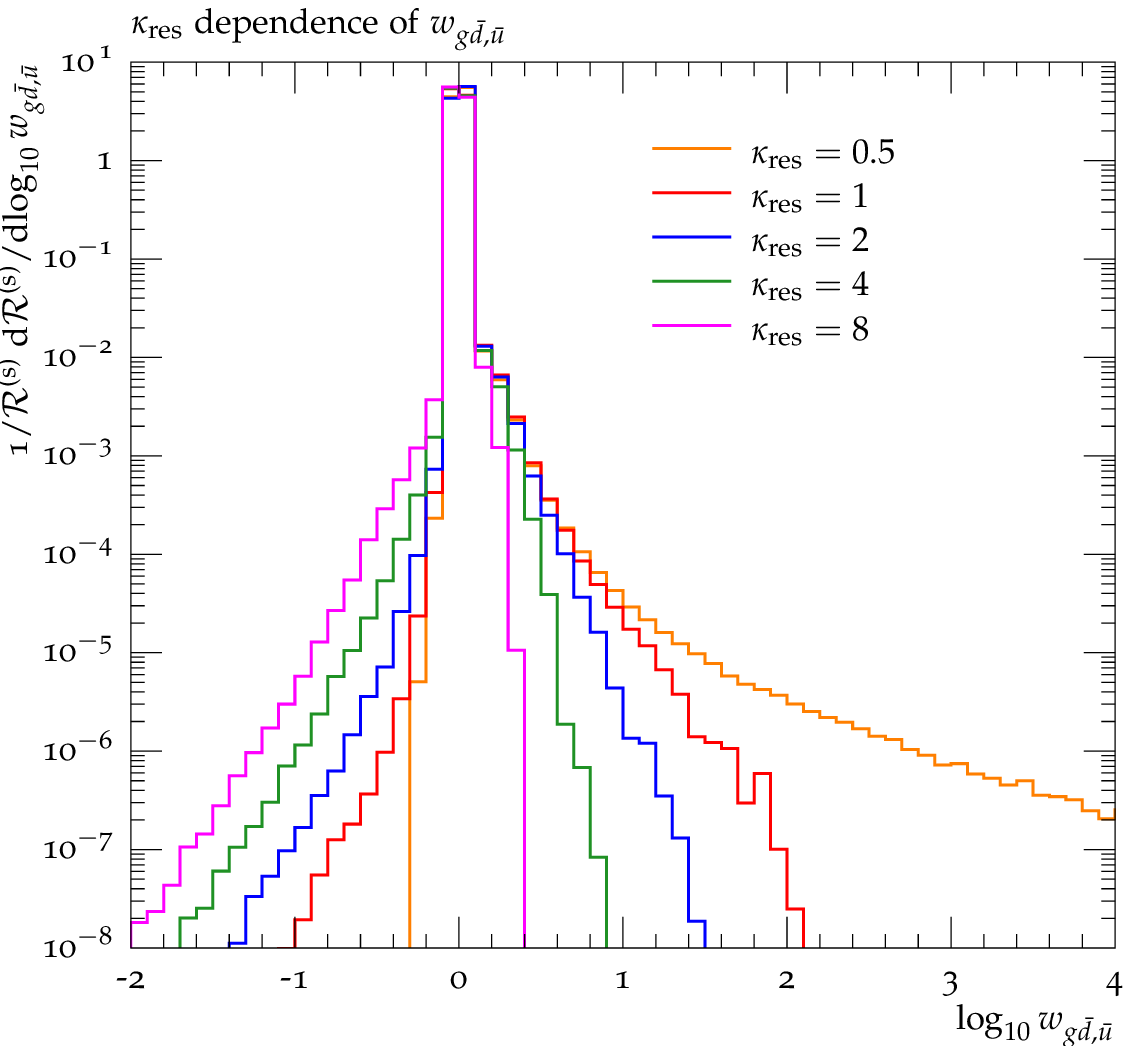}
  \hspace*{0.05\textwidth}
  \includegraphics[width=0.42\textwidth]{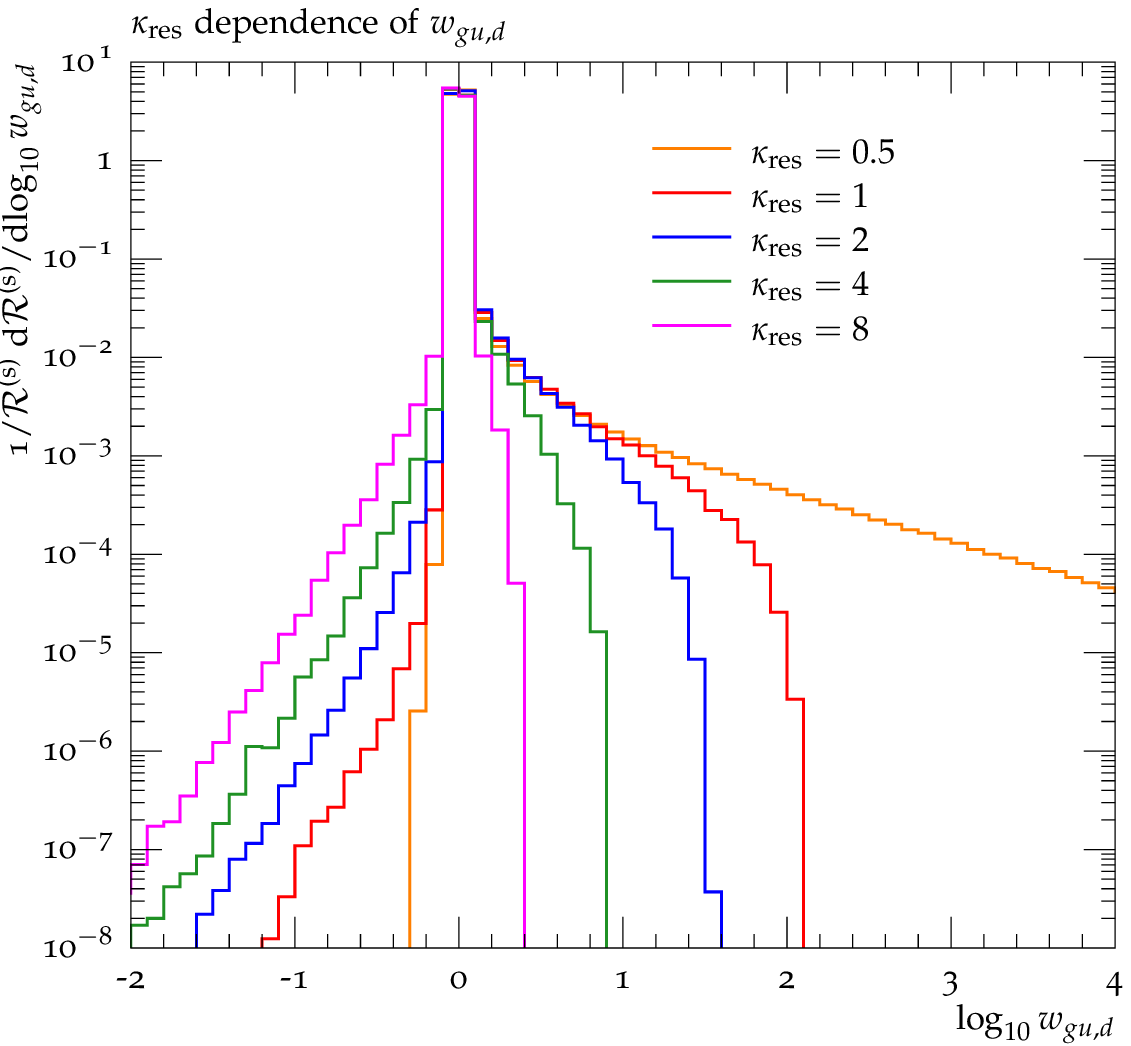}
}
{
  Dependence of the parton shower correction factor $w_{ij,k}$ on 
  the $Z$-$H$-splitting parameter $\kappa_\text{res}$ for $W^-$ production 
  at the \Tevatron. % $w_{ij,k}^\text{th}=100$ was kept fixed.
  \label{Fig:ZH_RBs}
}

\myfigure{p}{
  \includegraphics[width=0.42\textwidth]{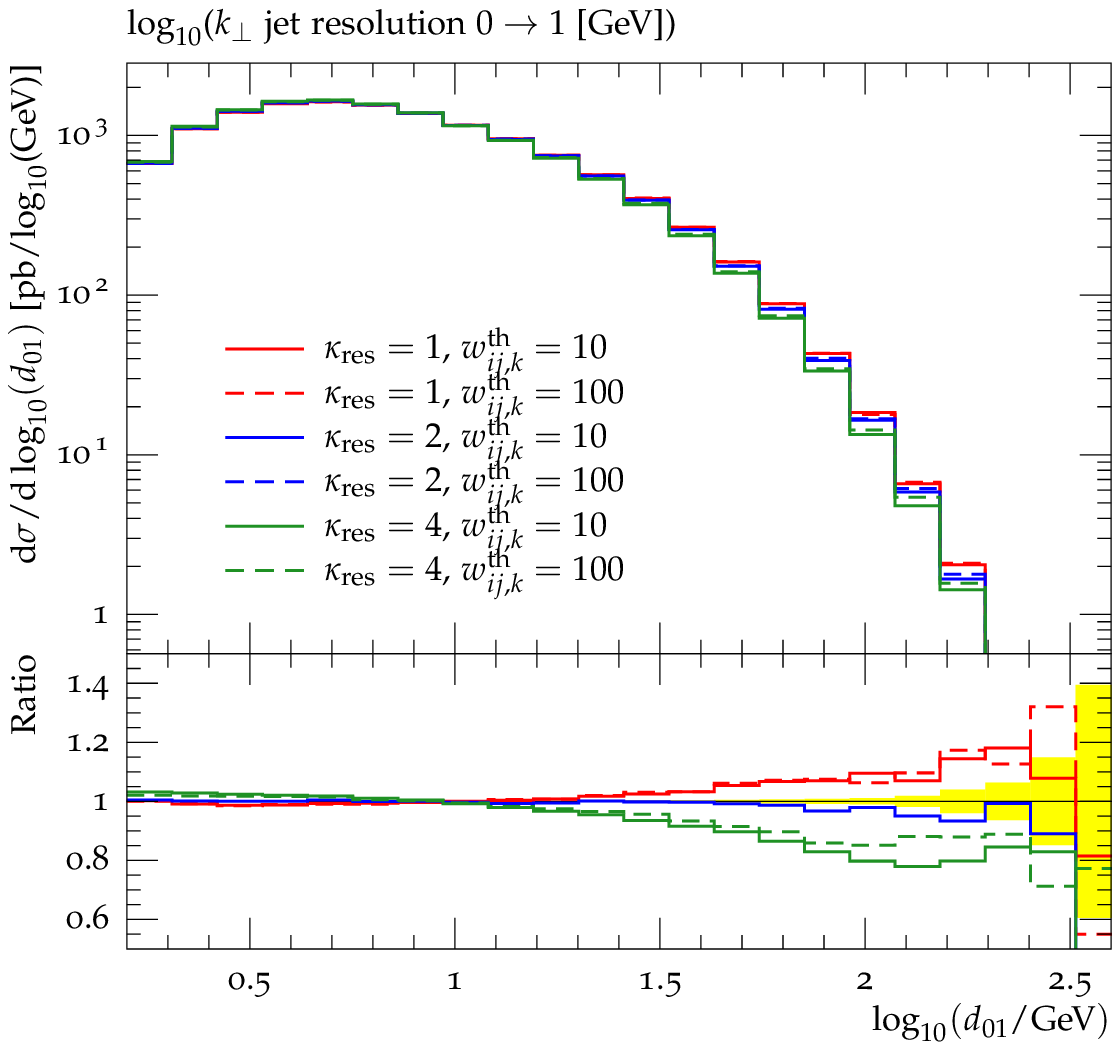}
  \hspace*{0.05\textwidth}
  \includegraphics[width=0.42\textwidth]{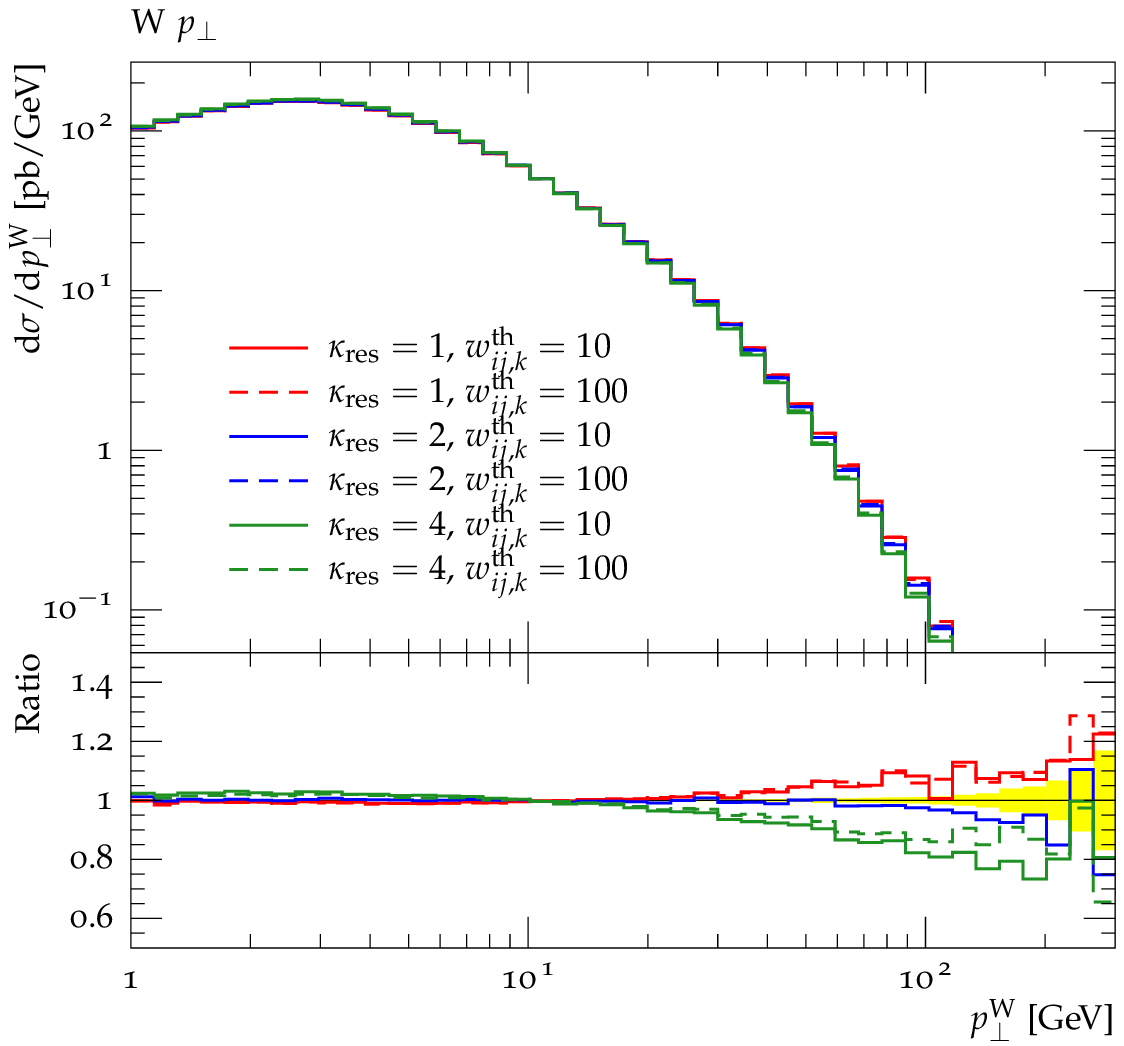}
}
{
  Predictions for $0\to 1$ jet resolution in $k_T$ clustered jets and transverse momentum of 
  the $W$ boson and  in $W$ boson production at the \Tevatron for different 
  settings of the $Z$-$H$-splitting parameters $\kappa_\text{res}$ and 
  $w_{ij,k}^\text{th}$.
  \label{Fig:ZH_splitting}
}

\clearpage

\subsection{Comparison with tree-level matrix-element parton-shower merging}
\label{sec:results:meps}
By comparing \POWHEG{} results to a standard parton shower combined with LO matrix
elements (LO+PS), it can be established whether observables are produced correctly in
regions where the soft/collinear approximations in the parton shower are
equivalent to the $R/B$ ratios in \POWHEG{}. An example is the distribution
of the jet resolution scale $d_{01}$, using the longitudinally invariant $k_T$-algorithm 
in $W/Z$ + jets production. This observable amounts to the $k_T$-scale where a 1-jet 
event is clustered into a 0-jet event. Figure~\ref{fig:wz:log10_d_01} shows that there 
is good agreement between the LO+PS and \POWHEG{} results for $d_{01}<50$ GeV. 
For harder emissions the LO+PS approach fails due to the restricted phase space, 
as explained in the previous section.

In this publication, we regard the \POWHEG method as an advanced matrix-element
reweighting technique for the parton-shower algorithm; the reweighting is 
supplemented with {\em local} $K$-factors to implement full NLO corrections.
It is therefore useful to compare the respective results with matrix-element 
parton-shower merged samples (ME+PS), which are rescaled by a suitably chosen
{\em global} $K$-factor.   Such samples are known to yield approximate 
NLO radiation patterns by effectively implementing higher-order matrix-element 
corrections into the parton shower.  An implementation of one of the most 
advanced ME+PS algorithms to date is available within the \Sherpa 
framework~\cite{Hoeche:2009rj} such that a direct comparison with \POWHEG is 
a straightforward exercise.  However, because of the lack of virtual 
contributions in the LO+PS and ME+PS samples, an agreement on the total rate 
cannot be expected. Thus, in the comparisons below the
following {\it global} $K$-factors were employed:
\begin{itemize} 
\item $K=1.038$ for $e^+e^-\to$ hadrons at LEP energies,
\item $K=1.2$ for $Z/\gamma^*$ and $W$ production at \Tevatron energies,
\item $K=1.2$ for $ZZ$ production at the \LHC (14 TeV),
\item $K=1.34$ for $W^+W^-$ production at the \LHC (14 TeV), and 
\item $K=2.1$ for Higgs production through gluon fusion at the
same \LHC energies.
\end{itemize}

When comparing \POWHEG results to ME+PS results including matrix elements up
to the 1-jet final state one should obtain a very similar radiation pattern.
The observed agreement indeed is very good, as expected. Figure~\ref{fig:wz:log10_d_01}
shows that, for example, the differential one-jet rates in $W/Z$-boson production
agrees on the 20\% level, even for relatively large scales ($d_{01}>50$ GeV). 
The remaining differences can be attributed to the differences in the Sudakov 
form factors: While \POWHEG{} exponentiates $R/B$, the ME+PS method uses standard
Sudakov form factors at the logarithmic accuracy of the parton shower.

Such differences become visible also in the multiplicity distribution of $k_T$
jets with $p_\perp>20$ GeV in Drell-Yan and $W$ production, 
cf.\ Fig.~\ref{fig:wz:jet_multi_exclusive}.
The 0-jet and 1-jet rates agree within 10\% between \POWHEG{} and ME+PS, but for
higher multiplicity final states the \POWHEG{} method predicts significantly
more jets. Here a ME+PS simulation with more jets in the matrix element would
lead to better agreement.

Now focussing on the properties of the leading jet produced in association with
a $W$ or $Z$ boson, the transverse momentum of the leading jet is shown in 
Fig.~\ref{fig:wz:jet_pT_1}.  Here the LO+PS approach fails to describe the 
hard tail of the distribution, again doe to lacking phase space, while the 
\POWHEG{} and ME+PS approaches agree within 20\%.  The separation in $\eta$-$\phi$ space between this jet and
the $W/Z$ boson is displayed in Fig.~\ref{fig:wz:V_jet1_dR}. Clear differences 
in the shape of the distribution comparing the LO+PS approach with both \POWHEG{} 
and ME+PS are found, as expected, since the other hand, parton showers cover only 
a restricted area of the phase space, and, in addition, they do not encode the full 
final-state correlations described by the matrix elements.  
Results from the \POWHEG{} and ME+PS methods agree very well, with differences 
below 10\% only. A similar finding applies to the transverse momentum of the leading jet, 
which is shown in Fig.~\ref{fig:wz:jet_pT_1}. Here the LO+PS approach fails to describe 
the hard tail of the distribution, while the \POWHEG{} and ME+PS approaches agree within 20\%.

The transverse momentum of the Higgs boson and the transverse momentum of
the leading jet displayed in Fig.~\ref{fig:hlhc:pT} give a similar picture as in
vector boson production: All three methods agree very well for low transverse momenta.
In the high $p_\perp$ region the \POWHEG{} and ME+PS approaches agree within 15\%.

Figure~\ref{fig:hlhc:jets} shows that minor differences arise between the LO+PS
and the \POWHEG and ME+PS approaches in the pseudorapidity spectrum of the leading jet. 
This can be understood as a direct consequence of the different transverse momentum 
distributions in the LO+PS method, as harder jets tend to be more central than 
softer ones. The \POWHEG{} and ME+PS approaches agree well in the central rapidity 
region and show up to 10\% difference only in the forward region.
The distribution of $\eta$-$\phi$ separation between the two leading jets proves 
again that the \POWHEG{} and ME+PS predictions are very similar, with deviations
below the 5\% level. Again, the LO+PS prediction shows a slightly different behaviour,
because of the reasons stated above.

Now we turn to look at diboson production at nominal LHC energies of 14 TeV.
Figure~\ref{fig:zzlhc:jets} (left) shows a comparison of the scalar sum $H_T$ of
the transverse momenta of jets and leptons in $Z$-pair production. Deviations of
up to 50\% become visible between the three compared approaches. This is especially 
true in the high-$H_T$ region. It is well understood that the predictions of the LO+PS
approach are softer than either of the two other approaches, due to the
restricted emission phase space. The relatively large differences between the
ME+PS approach and \POWHEG{} are naively not expected, but might stem from using
consistent but somewhat oversimplified scale schemes. This surely should be analysed
in more detail, in a forthcoming publication, where pair production processes, 
including $WH$ and $ZH$ would be studied.
The transverse momentum distributions of the individual $Z$ bosons
(Figure~\ref{fig:zzlhc:jets} right) on the other hand agree
very well in both approaches, while it is again obvious that the LO+PS
sample cannot describe the hard region of this spectrum.

In the azimuthal separation of the two $Z$ bosons, see
Fig.~\ref{fig:zzlhc:angles}, a similar feature as in $H_T$ can be found:
The events are harder in ME+PS than in \POWHEG{}, leading to increased
decorrelation of the boson pair. In Figure~\ref{fig:zzlhc:angles} (right) it can
be seen that the angle between the boson decay planes is predicted very
consistently in all three approaches.

Properties of the leading jet in $W^+W^-$ pair production at LHC energies are
displayed in Figure~\ref{fig:wwlhc:jets}. On the left hand side one can see
the transverse momentum of the leading jet and on the right hand side the
separation between lepton and leading jet.
For both the ME+PS and \POWHEG{} approaches agree well within 20\% and the LO+PS
sample severely underestimates the hardness of the first jet due to its
phase-space restrictions.

Figure~\ref{fig:wwlhc:angles} displays observables related to the two
oppositely charged leptons from the two decays. The pseudorapidity difference
(left) agrees within 20\% for all three approaches, while their
azimuthal decorrelation is significantly lower in the LO+PS sample than in the
ME+PS and \POWHEG{} approaches, which agree very well.

\clearpage

\myfigure{p}{
  \vspace*{-10mm}
  \includegraphics[width=0.45\textwidth]{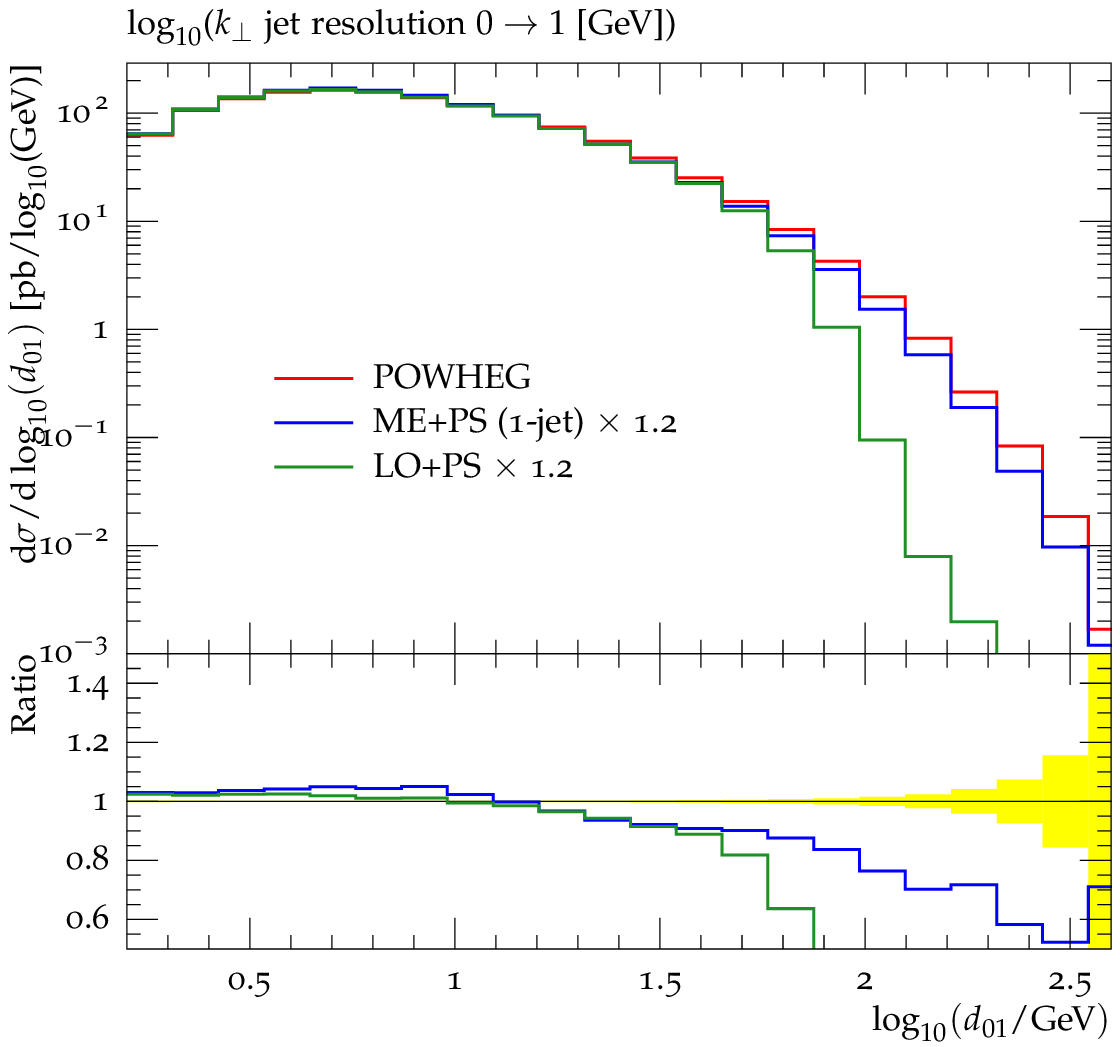}
  \hspace*{0.05\textwidth}
  \includegraphics[width=0.45\textwidth]{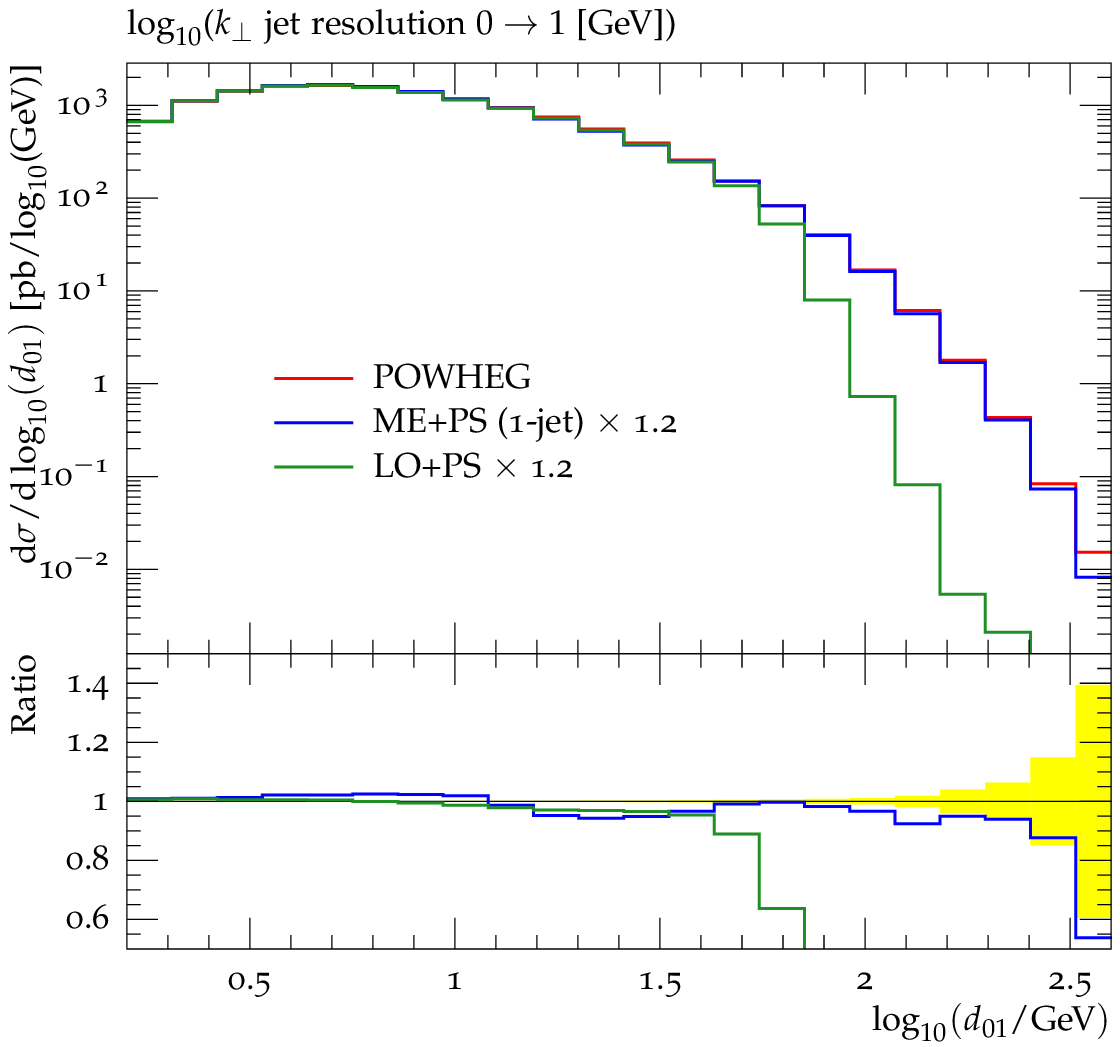}
  \vspace*{-3mm}
}
{
  Predictions for the $0\to 1$ jet resolution in $k_T$ clustered jets
  in $Z/\gamma^*$ (left) and $W$ (right) boson production at the 
  \protect\Tevatron.\label{fig:wz:log10_d_01}
}

\myfigure{p}{
  \vspace*{-10mm}
  \includegraphics[width=0.45\textwidth]{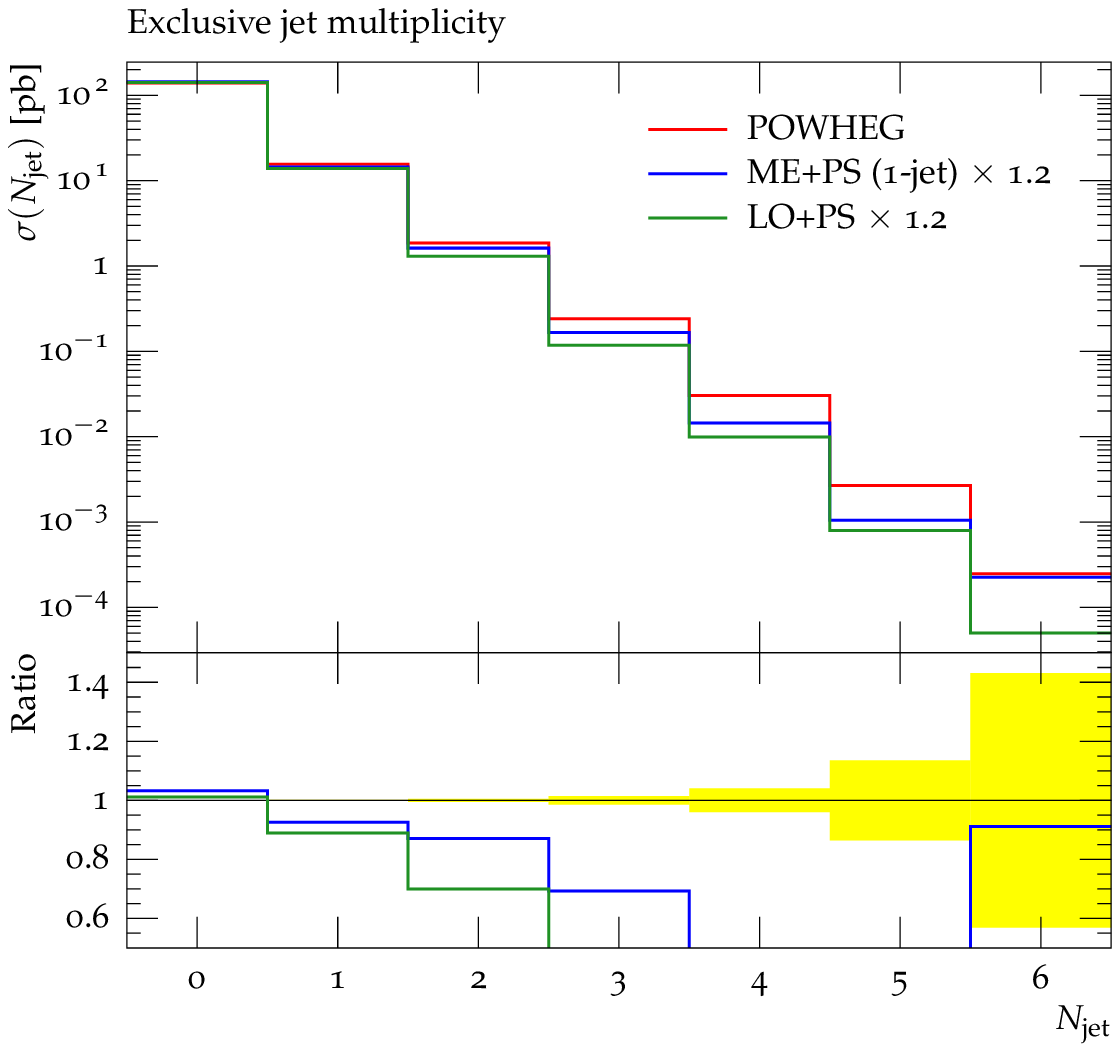}
  \hspace*{0.05\textwidth}
  \includegraphics[width=0.45\textwidth]{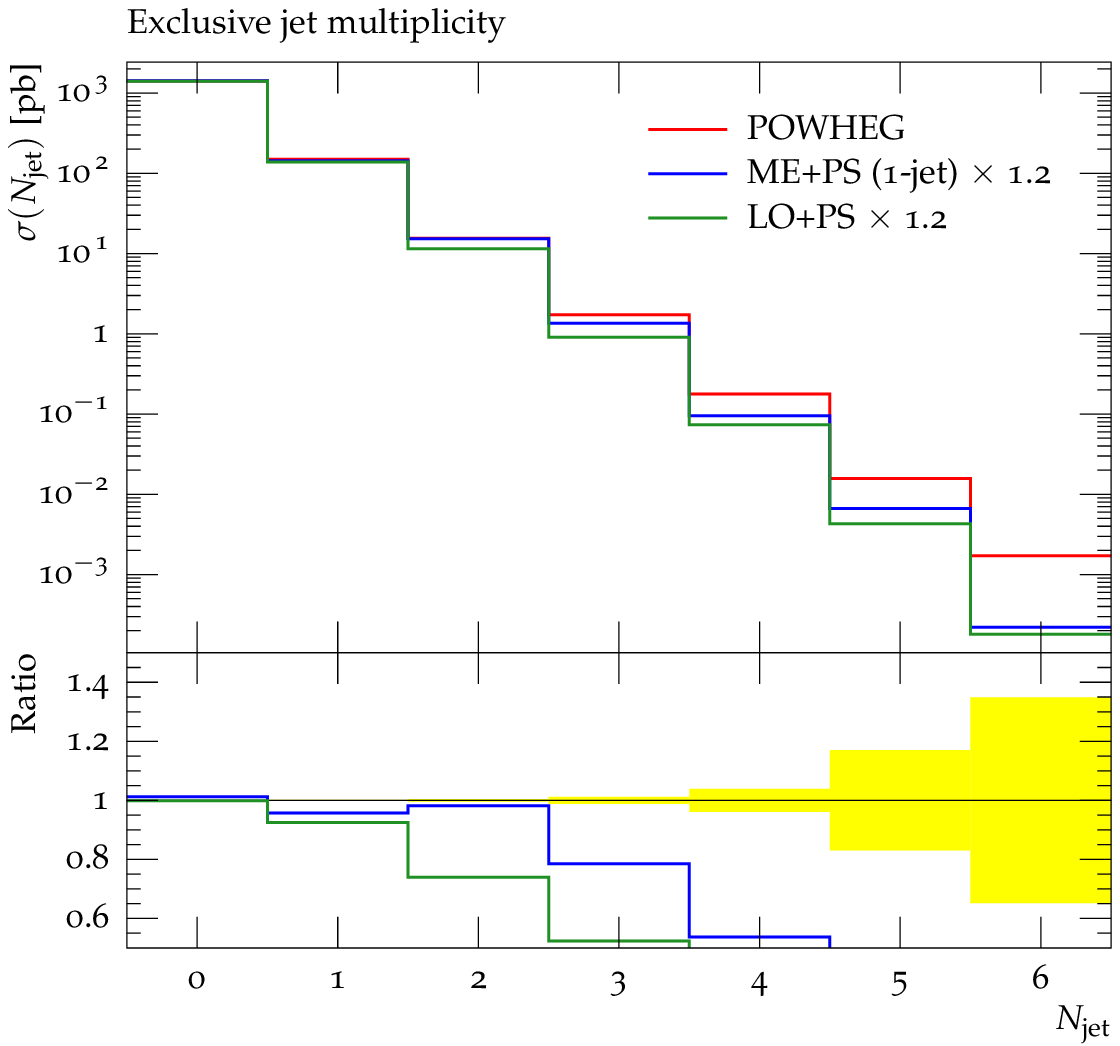}
  \vspace*{-3mm}
}
{
  Exclusive jet multiplicity for jets with $p_\perp>20$ GeV in the $k_T$ algorithm
  in $Z/\gamma^*$ (left) and $W$ (right) boson production at the 
  \protect\Tevatron.\label{fig:wz:jet_multi_exclusive}
}

\myfigure{p}{
  \vspace*{-10mm}
  \includegraphics[width=0.45\textwidth]{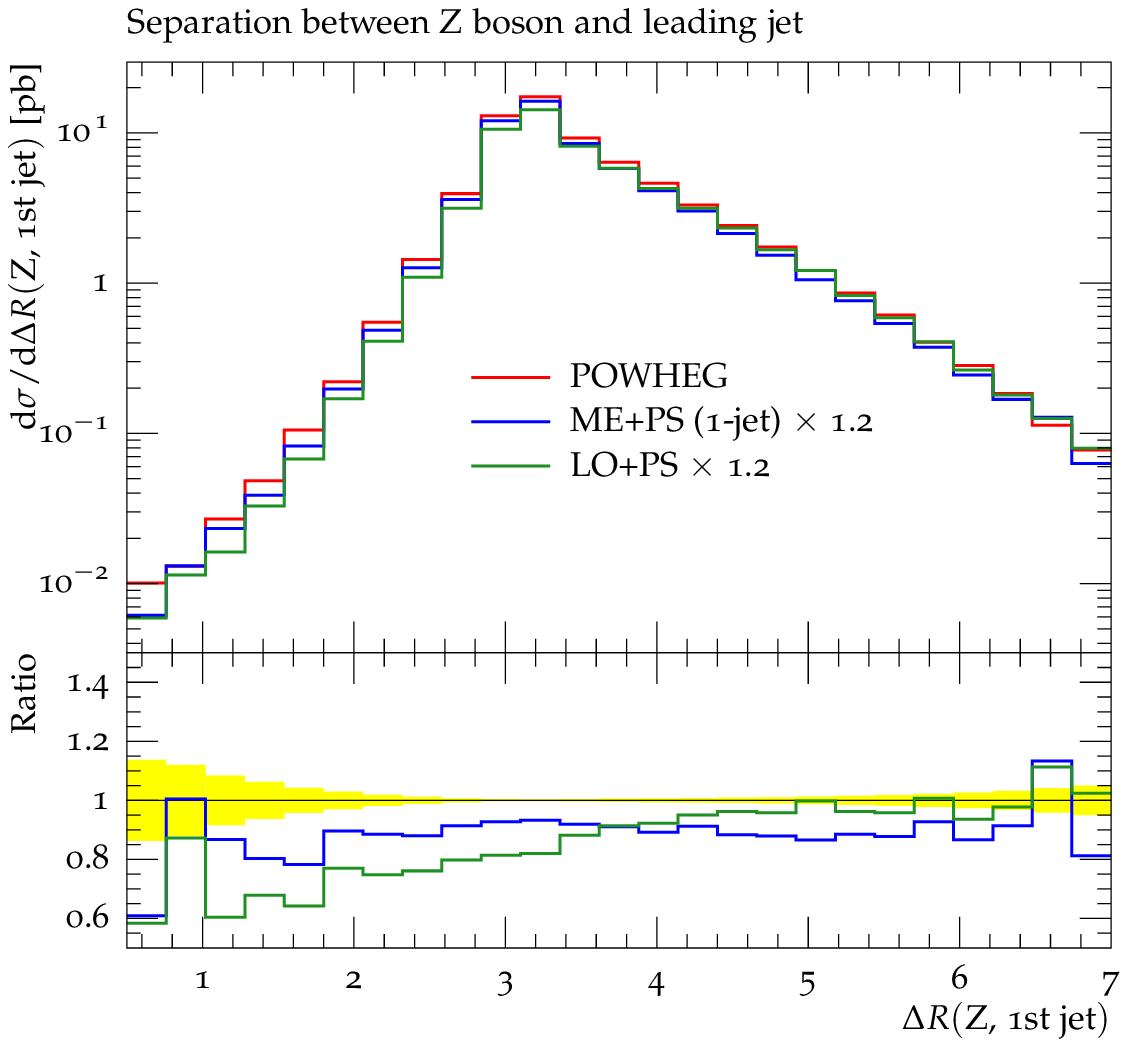}
  \hspace*{0.05\textwidth}
  \includegraphics[width=0.45\textwidth]{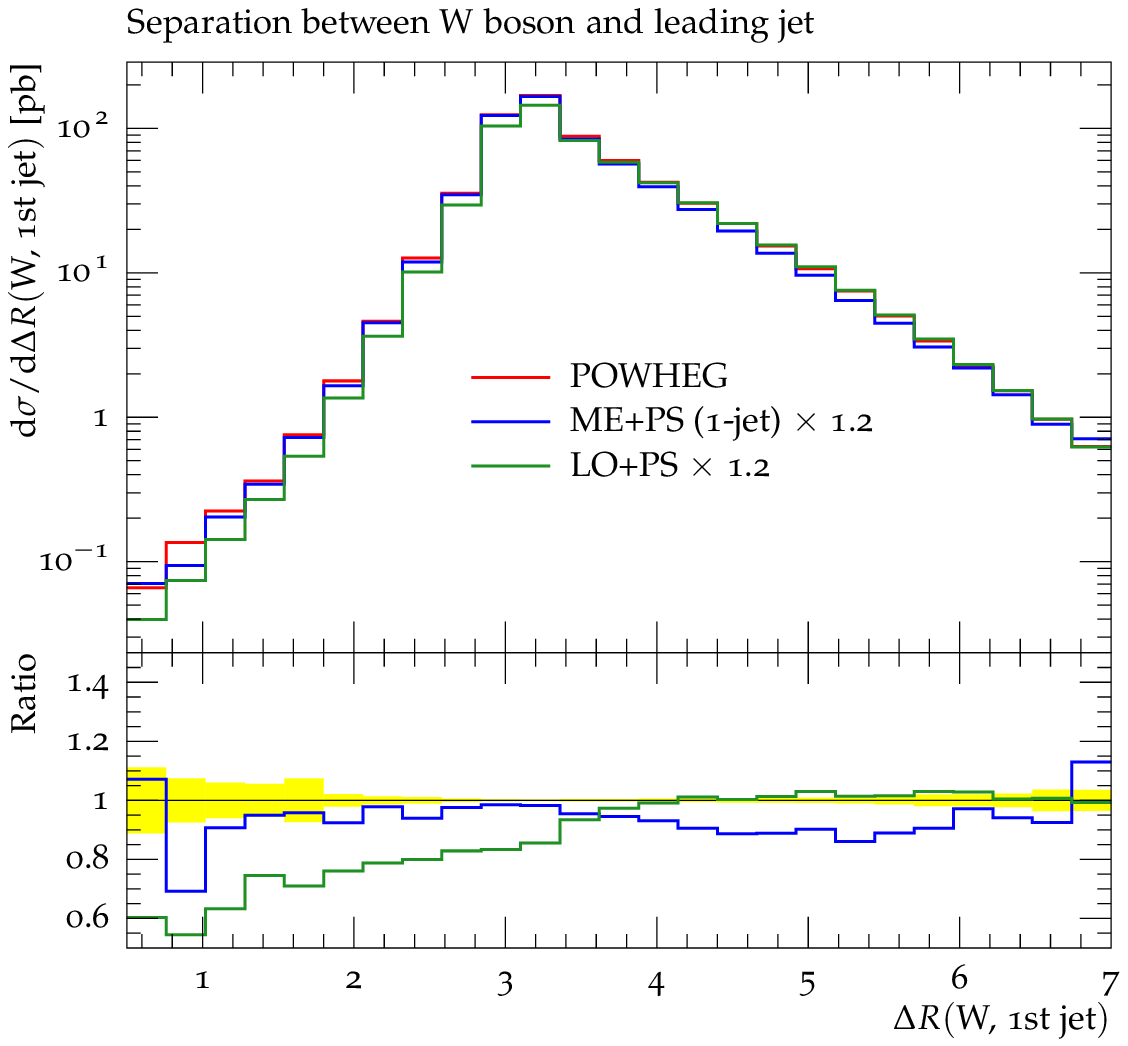}
  \vspace*{-3mm}
}
{
  Separation between vector boson and leading jet
  in $Z/\gamma^*$ (left) and $W$ (right) boson production at the 
  \protect\Tevatron.\label{fig:wz:V_jet1_dR}
}

\myfigure{p}{
  \vspace*{-10mm}
  \includegraphics[width=0.45\textwidth]{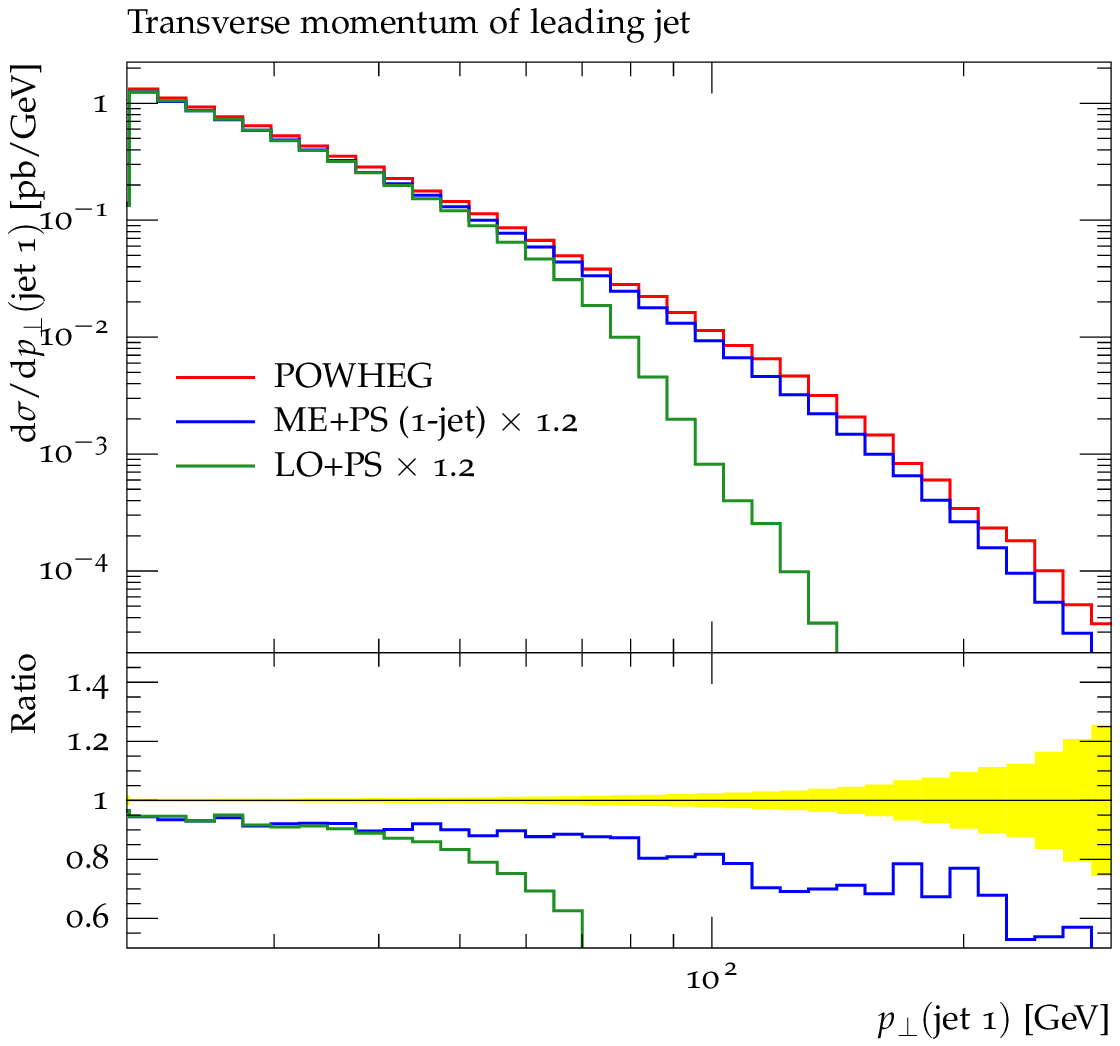}
  \hspace*{0.05\textwidth}
  \includegraphics[width=0.45\textwidth]{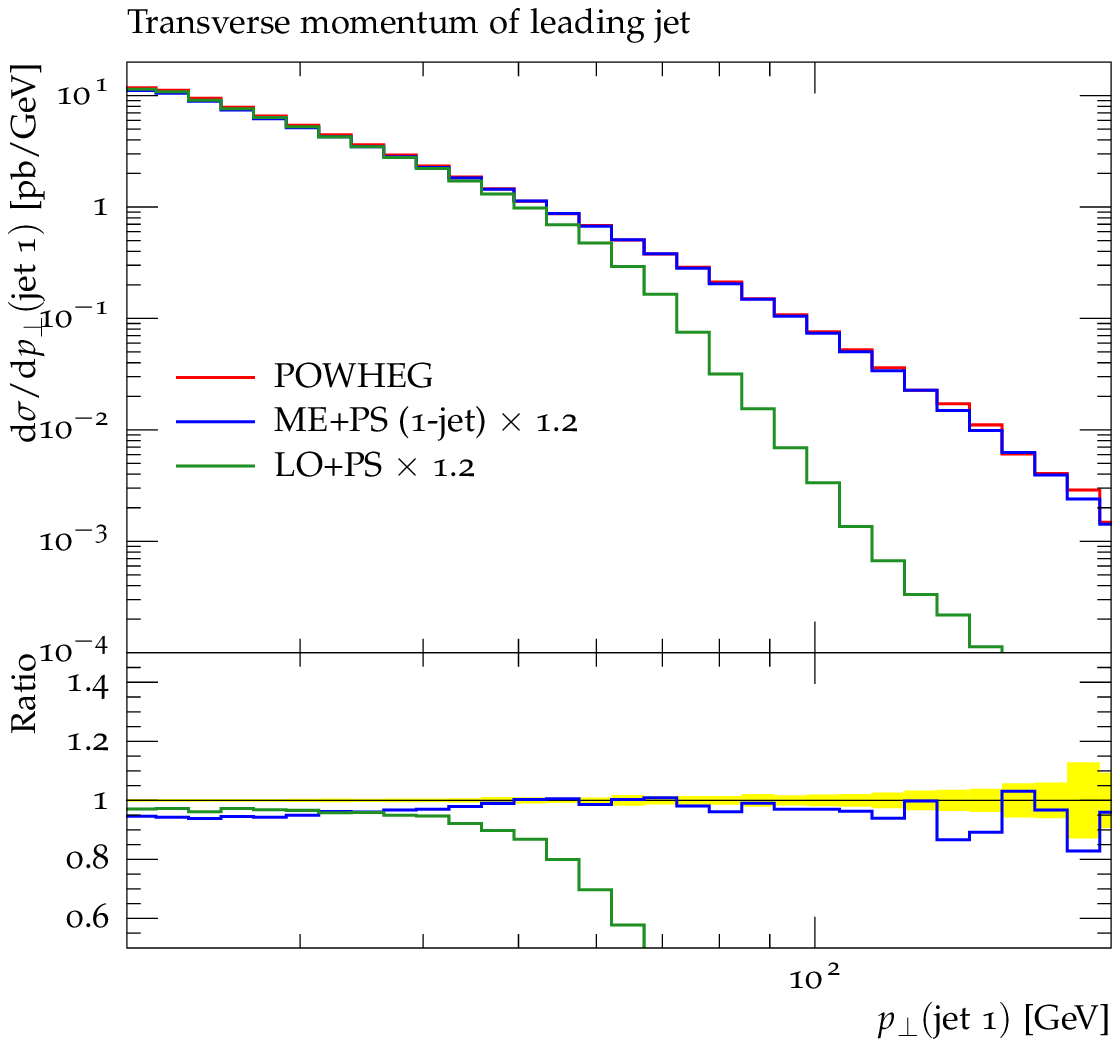}
  \vspace*{-3mm}
}
{
  Transverse momentum of the leading jet
  in $Z/\gamma^*$ (left) and $W$ (right) boson production at the 
  \protect\Tevatron.\label{fig:wz:jet_pT_1}
}

\myfigure{p}{
  \vspace*{-10mm}
  \includegraphics[width=0.45\textwidth]{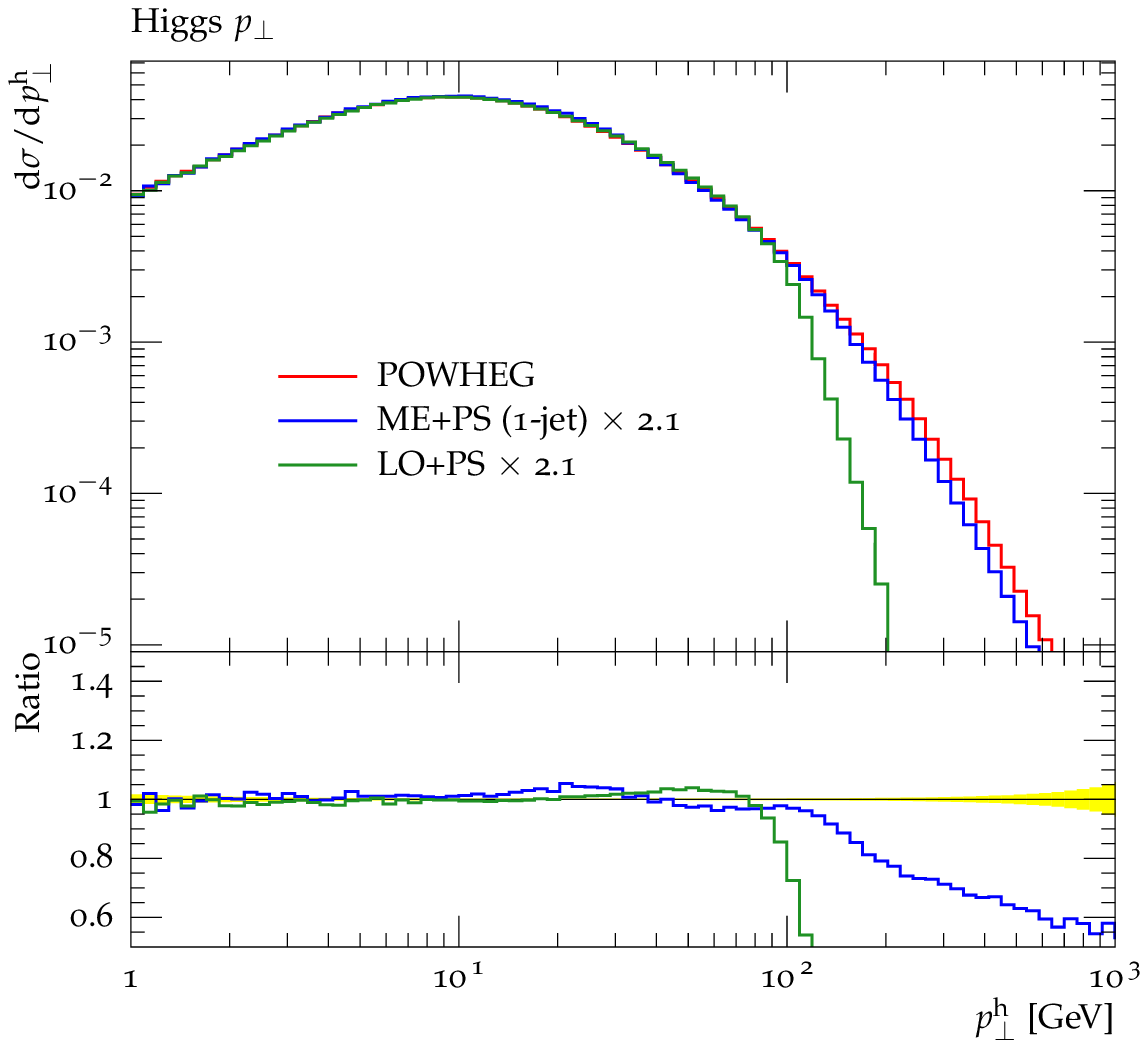}
  \hspace*{0.05\textwidth}
  \includegraphics[width=0.45\textwidth]{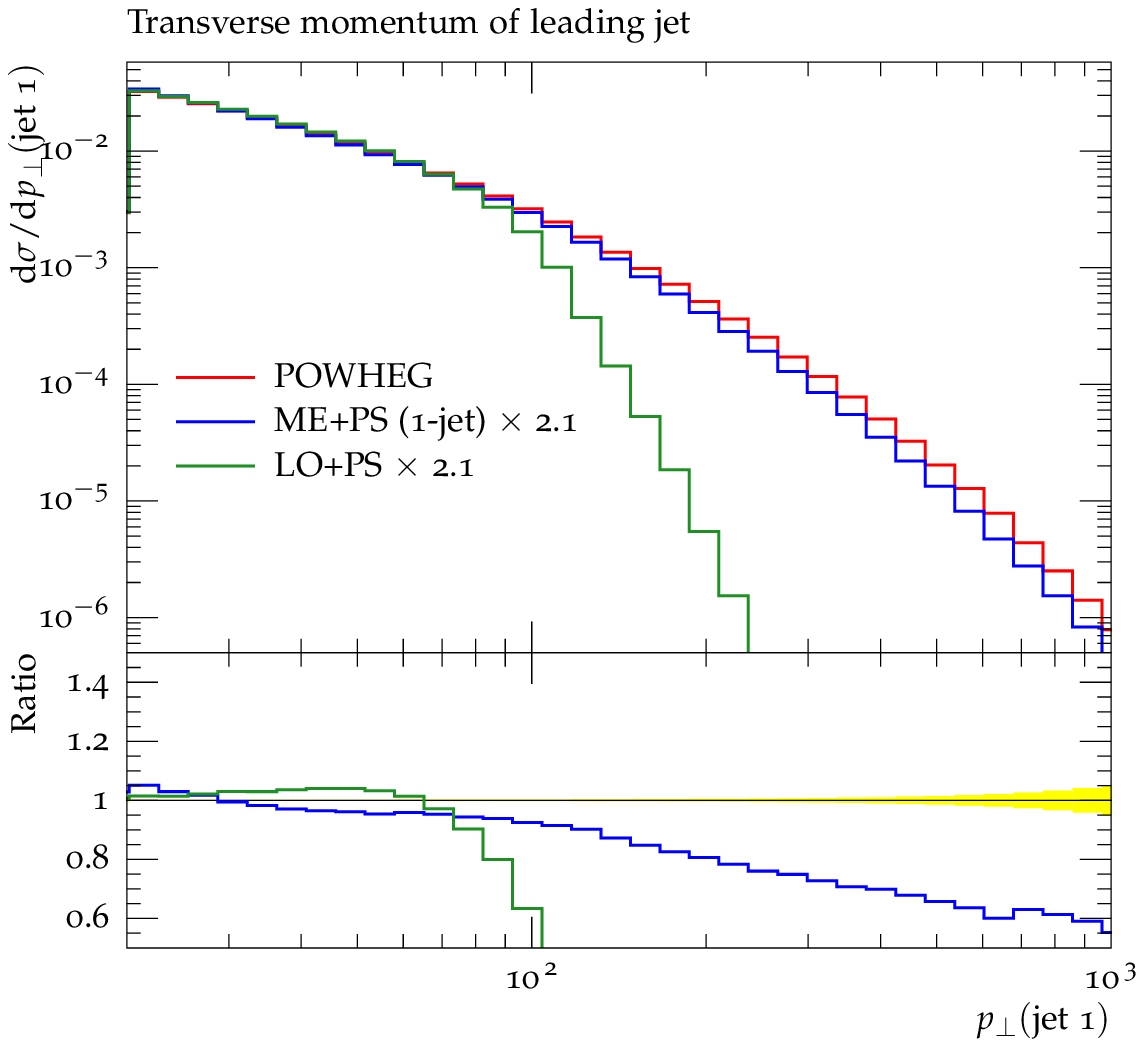}
  \vspace*{-3mm}
}
{
  Transverse momentum of the Higgs boson (left) and leading jet (right)
  in $gg\to h$ fusion at nominal \protect\LHC energies.
  \label{fig:hlhc:pT}
}

\myfigure{p}{
  \vspace*{-10mm}
  \includegraphics[width=0.45\textwidth]{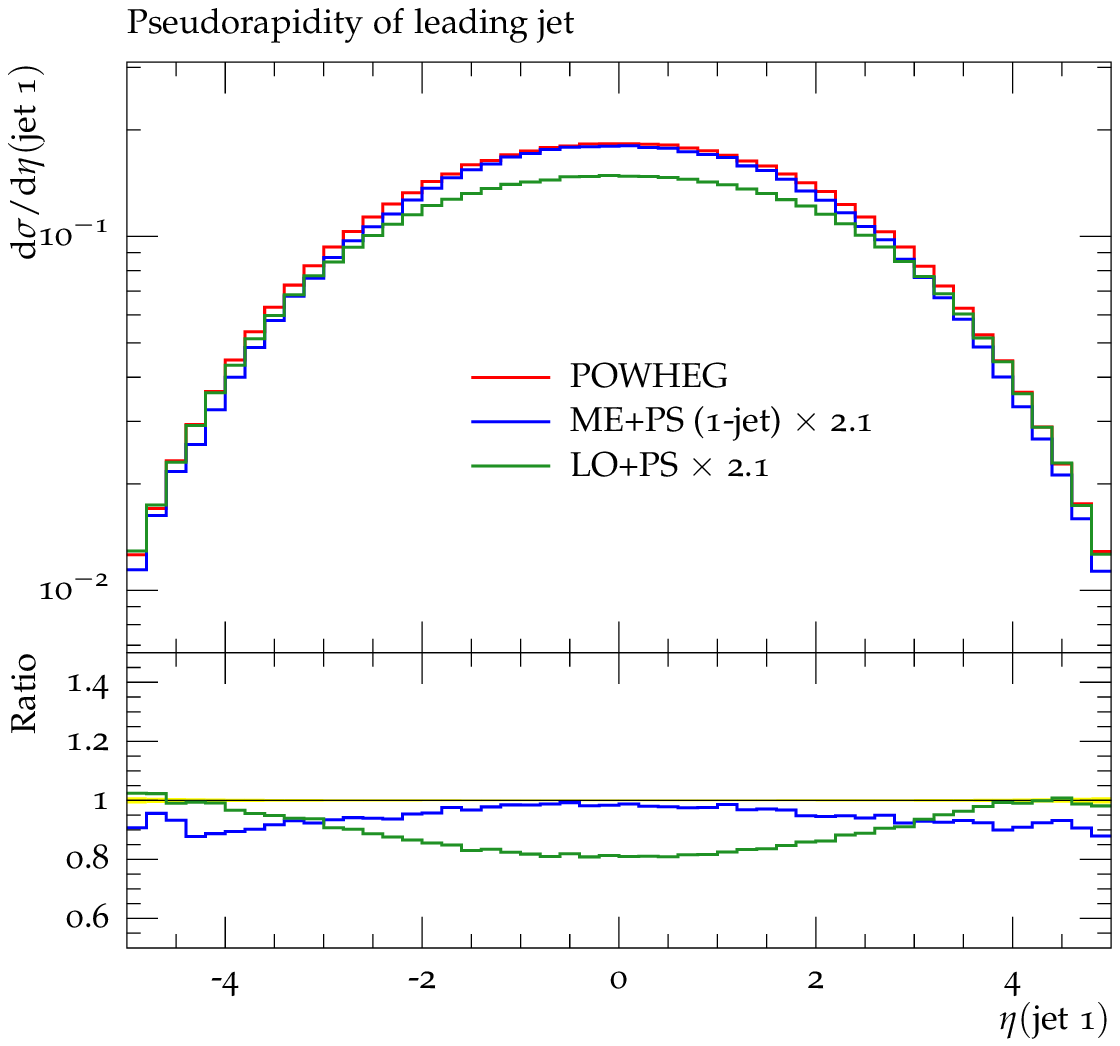}
  \hspace*{0.05\textwidth}
  \includegraphics[width=0.45\textwidth]{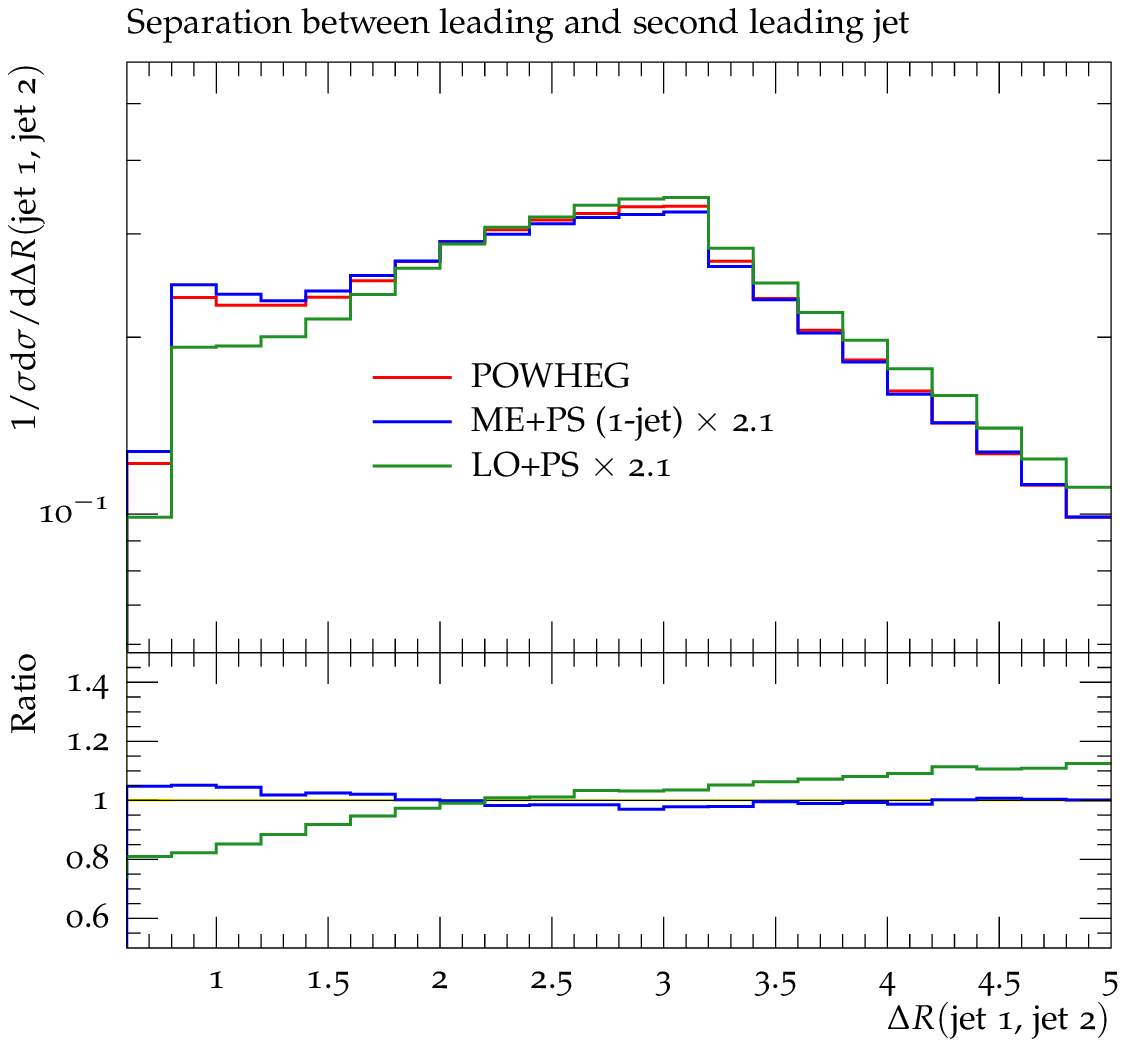}
  \vspace*{-3mm}
}
{
  Rapidity of the leading jet (left) and
  separation of the leading and second-leading jet (right)
  in $gg\to h$ fusion at nominal \protect\LHC energies.
  \label{fig:hlhc:jets}
}

\myfigure{p}{
  \vspace*{-10mm}
  \includegraphics[width=0.45\textwidth]{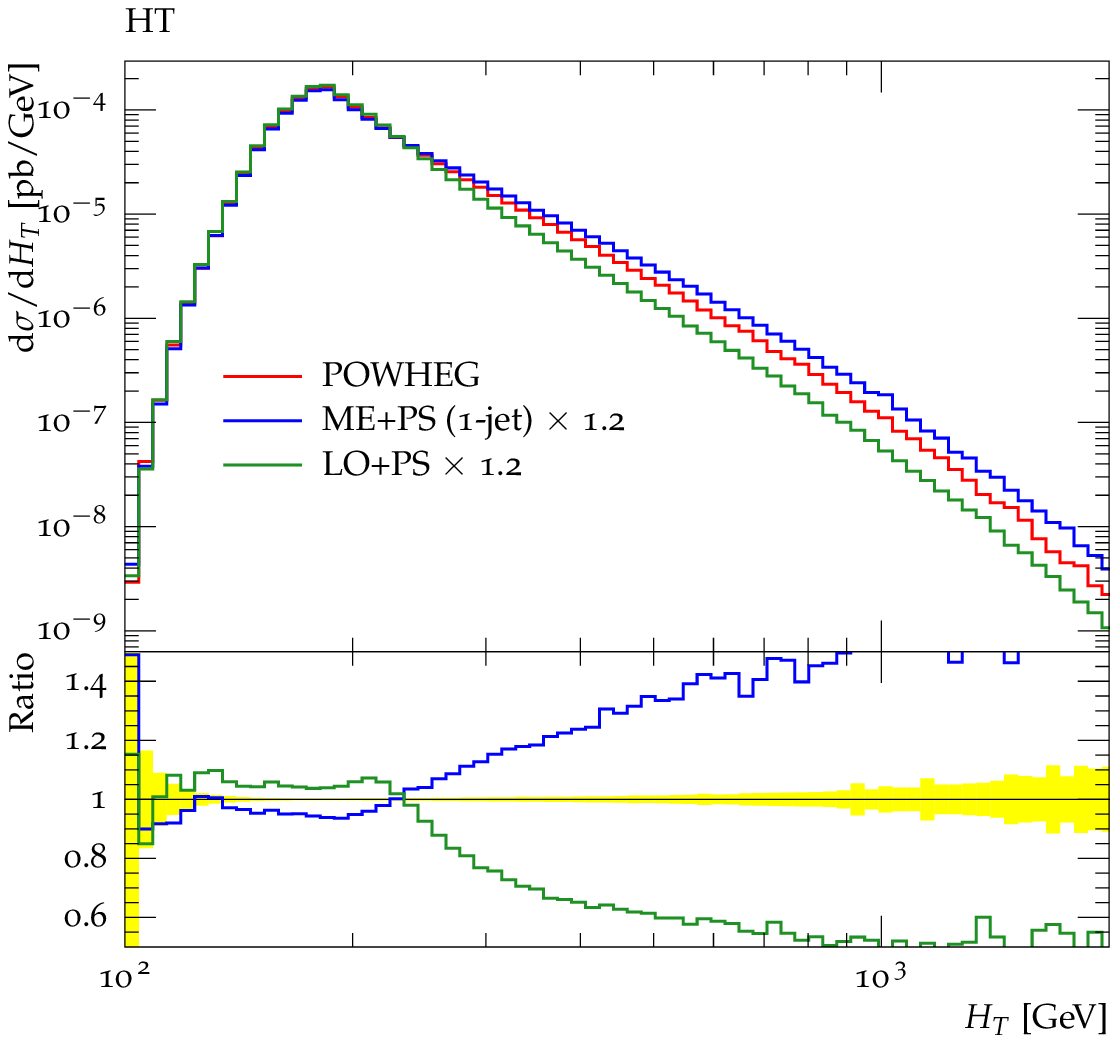}
  \hspace*{0.05\textwidth}
  \includegraphics[width=0.45\textwidth]{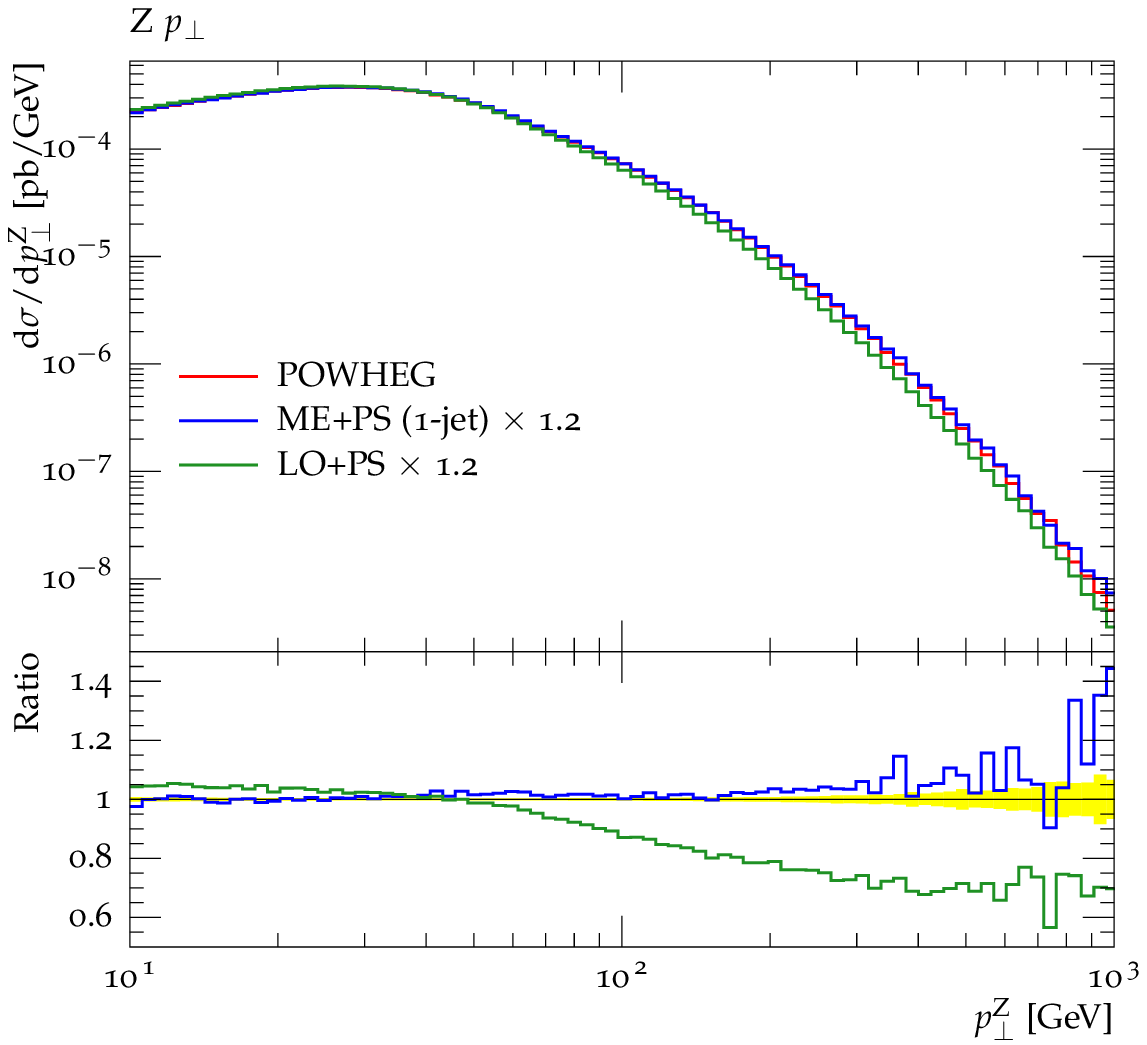}
  \vspace*{-3mm}
}
{
  $H_T$ (left) and transverse momentum of the individual $Z$ bosons (right)
  in $ZZ$ production at nominal \protect\LHC energies.
  \label{fig:zzlhc:jets}
}

\myfigure{p}{
  \vspace*{-10mm}
  \includegraphics[width=0.45\textwidth]{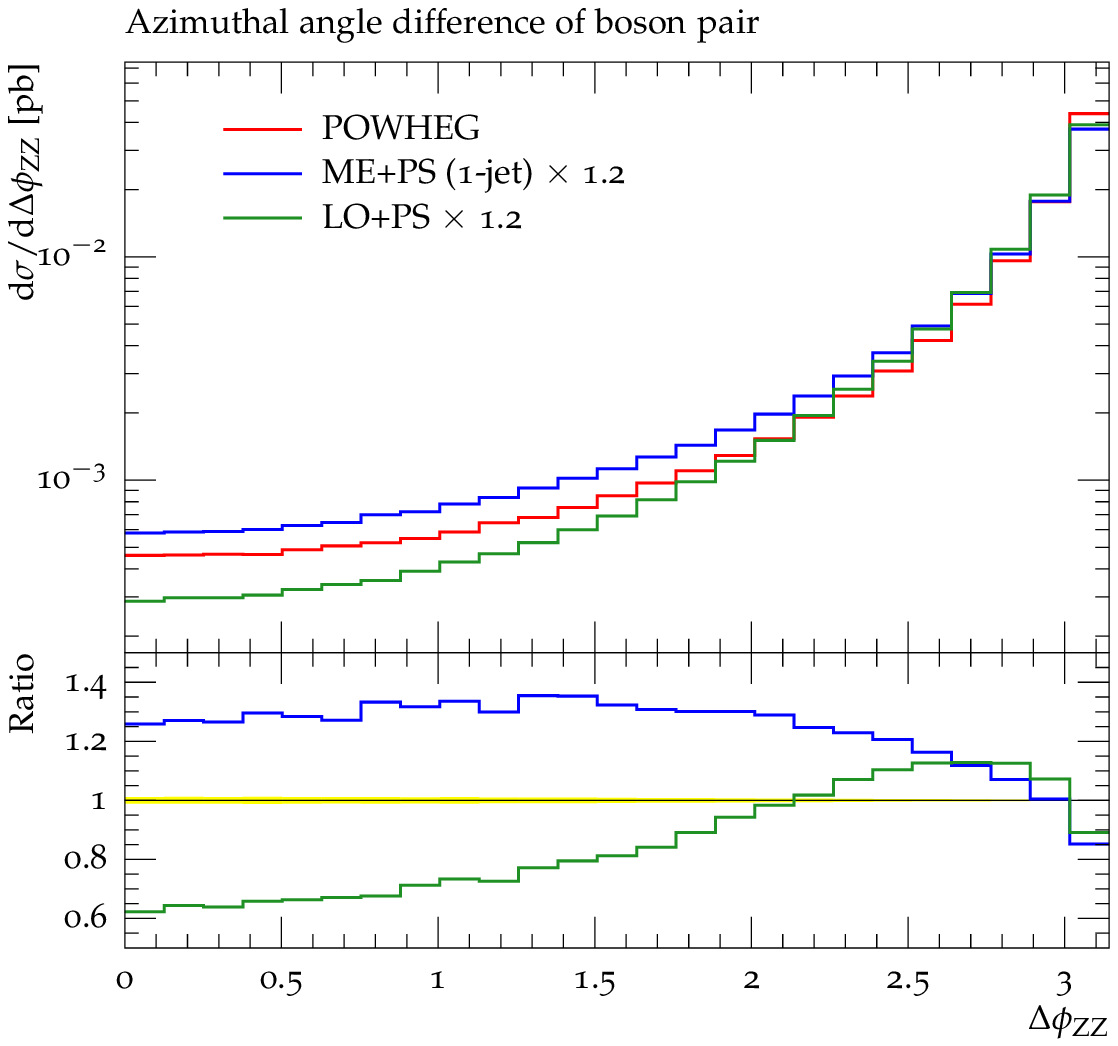}
  \hspace*{0.05\textwidth}
  \includegraphics[width=0.45\textwidth]{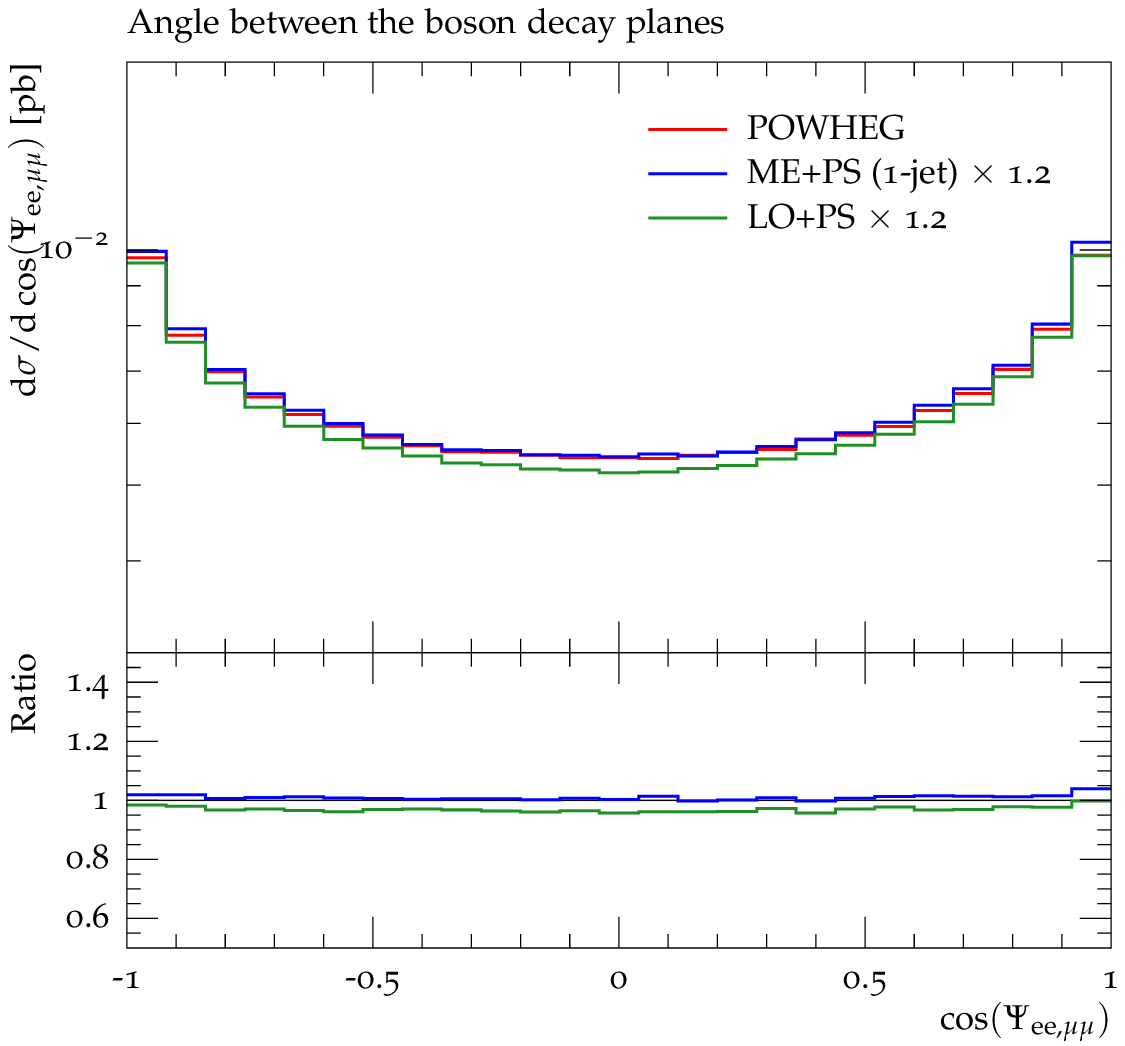}
  \vspace*{-3mm}
}
{
  Azimuthal angle between the two $Z$ bosons (left) and
  angle between the two $Z$ decay planes (right)
  in $ZZ$ production at nominal \protect\LHC energies.
  \label{fig:zzlhc:angles}
}

\myfigure{p}{
  \vspace*{-10mm}
  \includegraphics[width=0.45\textwidth]{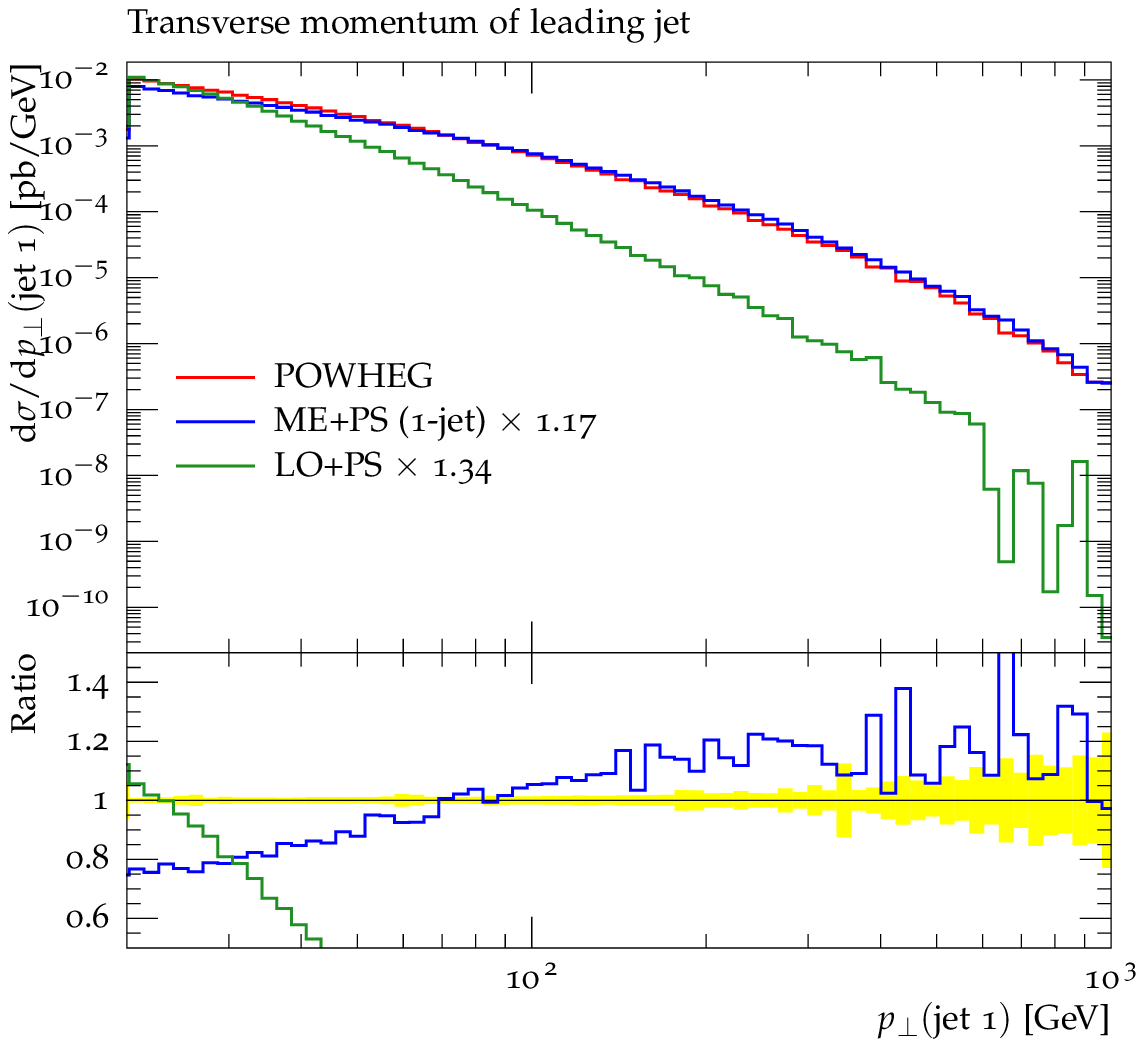}
  \hspace*{0.05\textwidth}
  \includegraphics[width=0.45\textwidth]{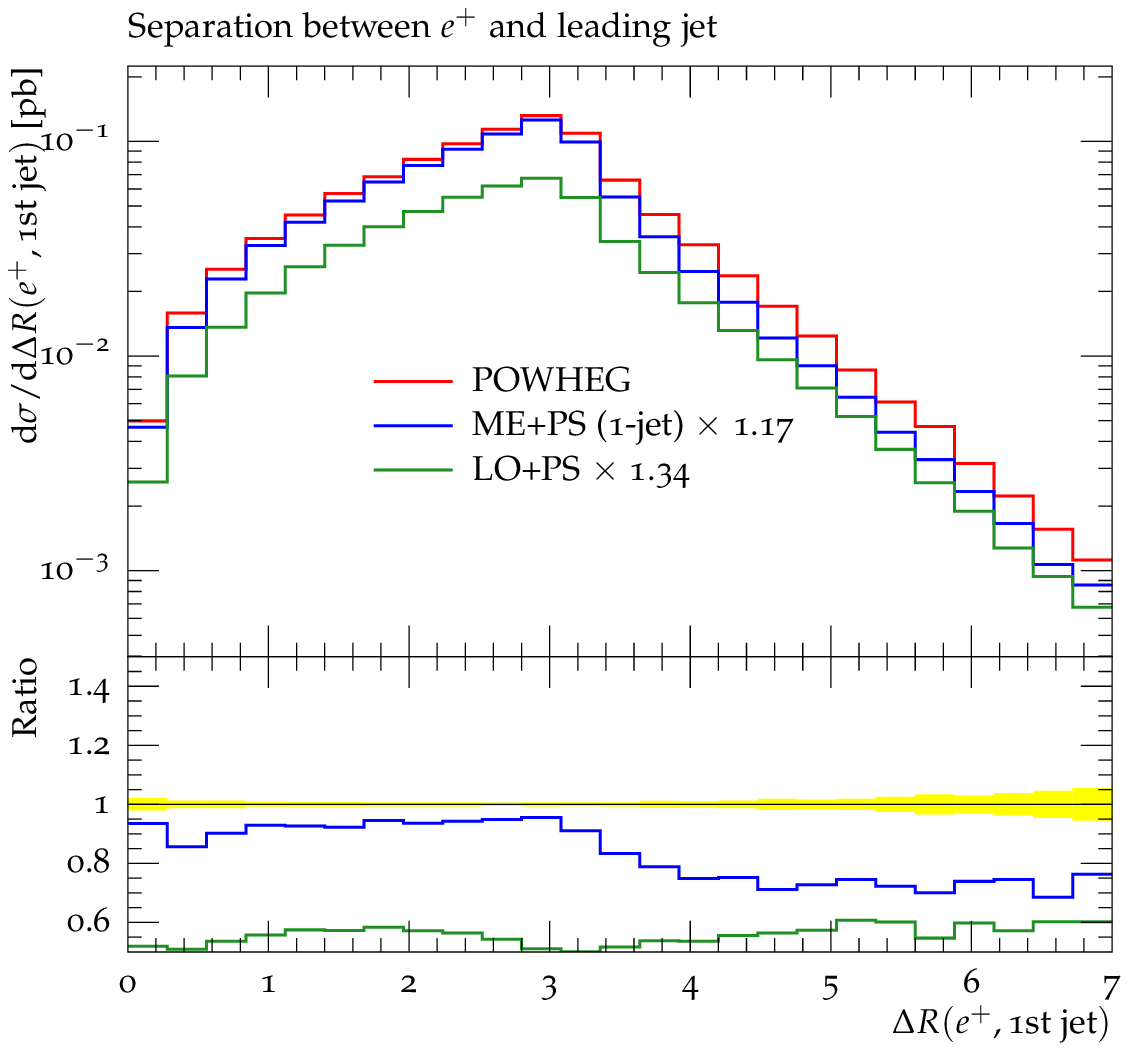}
  \vspace*{-3mm}
}
{
  Transverse momentum of the leading jet (left) and
  separation of the lepton and leading jet (right)
  in $W^+W^-$ production at nominal \protect\LHC energies.
  \label{fig:wwlhc:jets}
}

\clearpage

\subsection{Comparison with experimental data}
\label{sec:results:data}
The remainder of this section is dedicated to a comparison of results from the 
\POWHEG approach with experimental data to assert the improved description
of data, provided by this method.

For the reaction $e^+e^-\to$ hadrons at LEP1 energies the LO+PS and ME+PS
predictions do not show significant differences except in extreme regions of
phase space. The \POWHEG{} prediction confirms that picture. 
This is largely due to the fact that the parton-shower algorithm, which is employed 
in \Sherpa is based on Catani-Seymour subtraction terms and those terms constitute
a rather good approximation to the real-emission matrix element in the process
$e^+e^-\to q\bar{q}g$.

In the distribution of the Durham jet resolution at which 3-jet events are
clustered into 2-jet events (Fig.~\ref{fig:lep:aleph1} left) all three approaches
agree very well with the measurement over large parts of the phase space.
Only in the hard emission region $y_{23}>0.05$ deviations from the LO+PS result can be
seen. It is encouraging, although not surprising to note that both \POWHEG{} and ME+PS 
describe the data better.
Good agreement of all three approaches with each other and with the measurement is also
observed e.g.\ for the thrust distribution (Fig.~\ref{fig:lep:aleph1} right),
the total jet broadening (Fig.~\ref{fig:lep:aleph2} left) and the C-Parameter 
(Fig.~\ref{fig:lep:aleph2} right).

As was discussed in~\cite{Carli:2010cg}, deep-inelastic scattering processes offer 
the opportunity to test perturbative QCD in a region where the factorisation scale
of the inclusive process, $Q^2$, is smaller than the scale of the actual event, which
might be set e.g.\ by the transverse momentum of a hard jet. As measurements can be
performed down to very low values of $Q^2$, many hard jets must usually be included
in the simulation to achieve a good description of data throughout the phase space.
This method cannot be used in the context of this work, as the \POWHEG implementation
in \Sherpa can so far only be employed for the core process $e^\pm q\to e^\pm q$. 
Therefore, results are presented for the high-$Q^2$ region only and the discussion 
of the low-$Q^2$ domain is postponed to a forthcoming publication~\cite{Hoeche:2010xx}, 
where the merging of \POWHEG{} samples with higher-multiplicity matrix elements will be discussed.
Figure~\ref{fig:dis:q2r32} shows reasonable agreement between the \POWHEG results
and experimental data in a measurement of the di-jet cross section performed at the 
H1 experiment~\cite{Adloff:2000tq,Adloff:2001kg}. Deviations from the LO+PS result
are apparent, especially at lower values of $Q^2$, as the phase space of the parton 
shower is severely restricted by the low factorisation scale.
Similar findings apply to the rapidity spectra shown in Fig.~\ref{fig:dis:etaetsum}.

The probably most discussed observable probing the influence of QCD radiation 
in hadron-hadron collisions is the transverse momentum of the
lepton pair in Drell-Yan production, which is displayed in Fig.~\ref{fig:ztev:pTy}.
Very good agreement, compared to a recent measurement, is found for both the 
\POWHEG{} and ME+PS approaches, while the LO+PS method is not able to describe large parts
of the spectrum because of the restricted real-emission phase space. 
The rapidity of the Z boson in Fig.~\ref{fig:ztev:pTy} is described 
very well by all three approaches.

The situation is very similar in $W$-boson production. A comparison of \POWHEG 
predictions with Tevatron data from the \DO experiment~\cite{Abbott:2000xv}
is shown in Fig.~\ref{fig:wtev:pT}, where very good agreement between the Monte-Carlo 
result and the data can be observed. In addition to the direct comparison the 
uncertainties related to a variation of the renormalisation and factorisation 
scales are also shown. Thereby, two different strategies are pursued: While the inner (dark) band
shows the uncertainty related to a variation of the scale in the hard matrix elements
only, the outer (light) band shows the influence of varying the scales also in the
parton-shower evolution. It is rather obvious that the latter approach yields the
larger variations, as it is associated with an uncertainty in the choice of
the strong coupling for the real-emission subprocess. While this process essentially
determines the shape of the transverse momentum distribution in Fig.~\ref{fig:wtev:pT},
it enters at tree-level accuracy only, thus leading to a rather large scale dependence.

\clearpage

\myfigure{p}{
  \vspace*{-10mm}
  \includegraphics[width=0.45\textwidth]{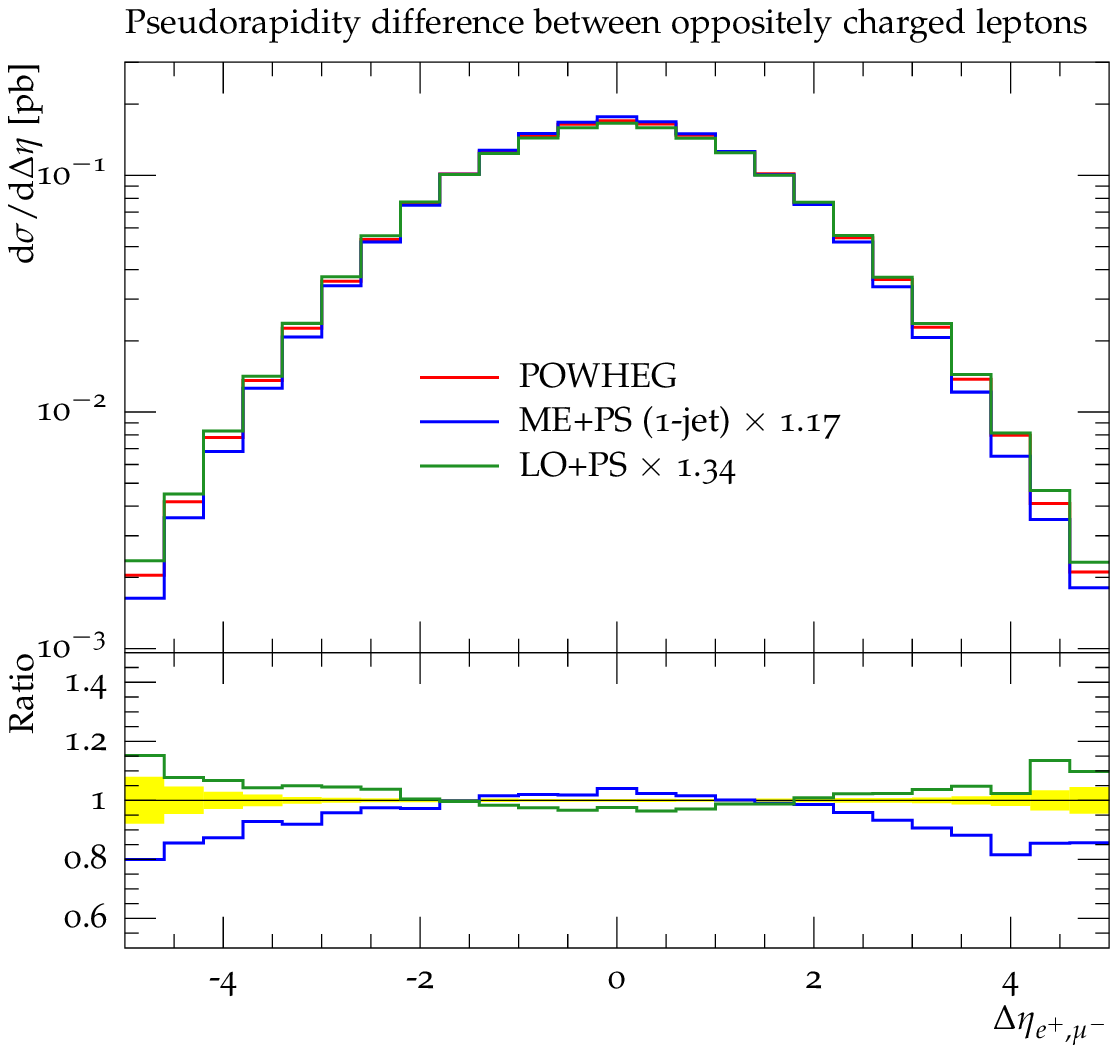}
  \hspace*{0.05\textwidth}
  \includegraphics[width=0.45\textwidth]{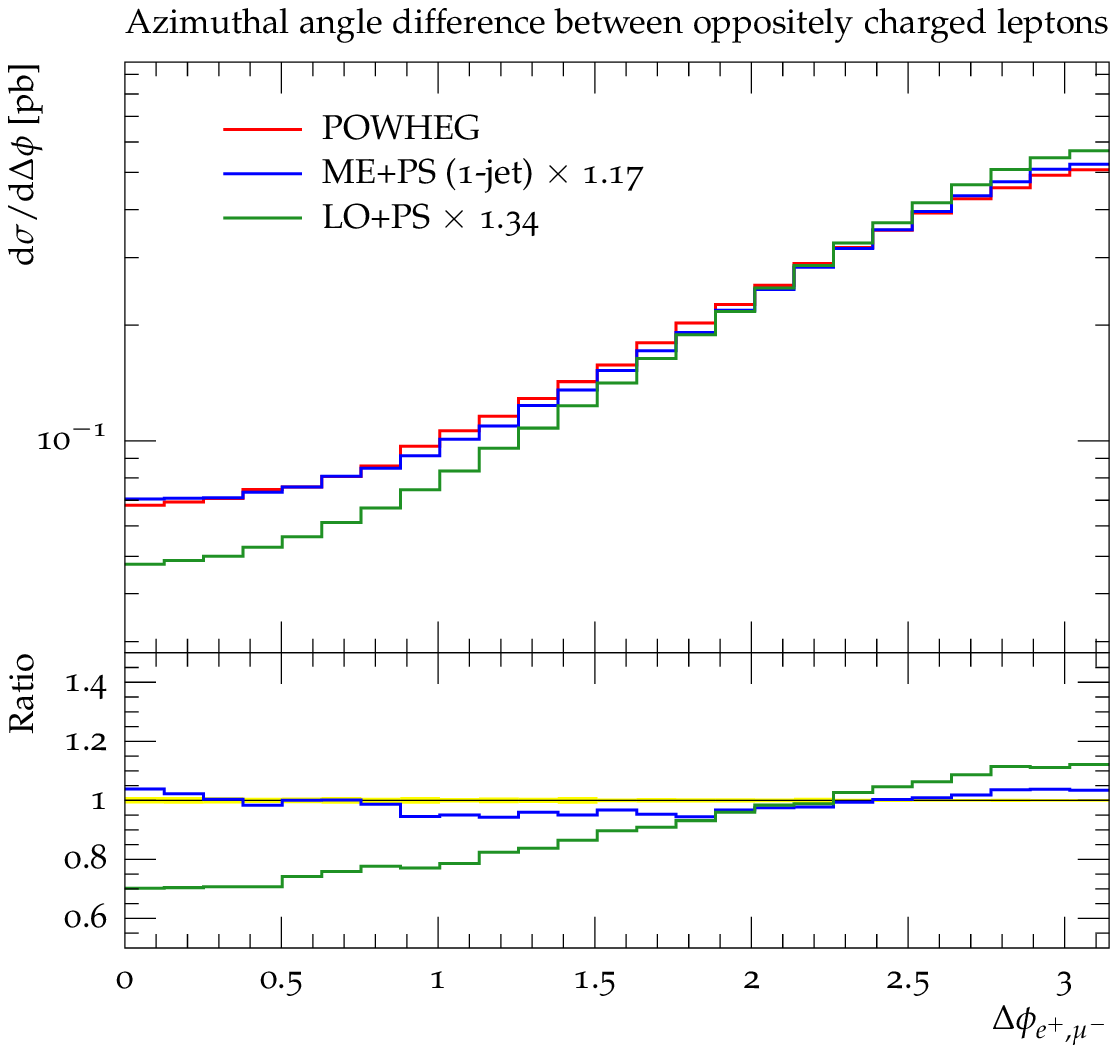}
  \vspace*{-3mm}
}
{
  Pseudorapidity difference (left) and
  azimuthal angle (right) between the two oppositely charged leptons
  in $W^+W^-$ production at nominal \protect\LHC energies.
  \label{fig:wwlhc:angles}
}

\myfigure{p}{
  \vspace*{-10mm}
  \includegraphics[width=0.45\textwidth]{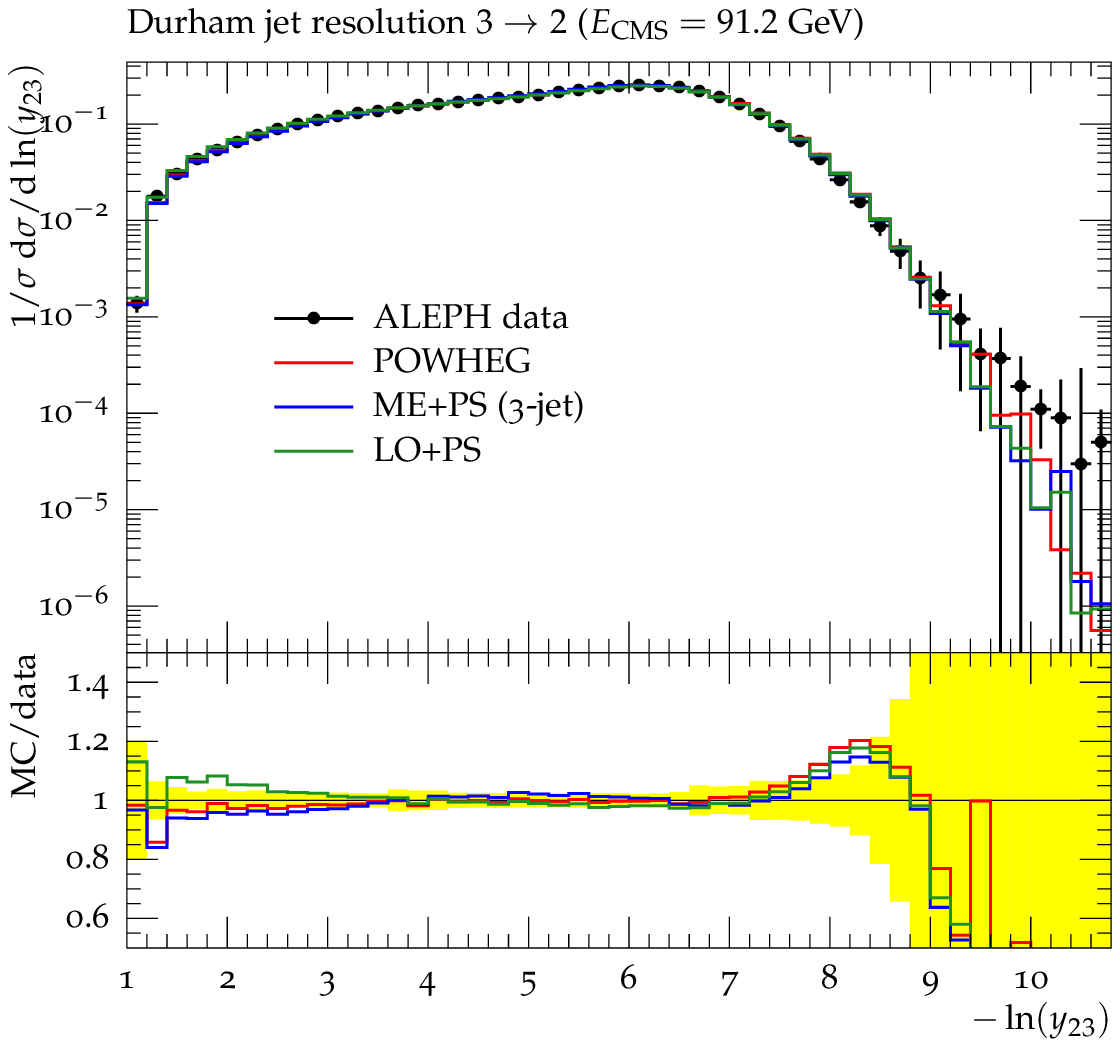}
  \hspace*{0.05\textwidth}
  \includegraphics[width=0.45\textwidth]{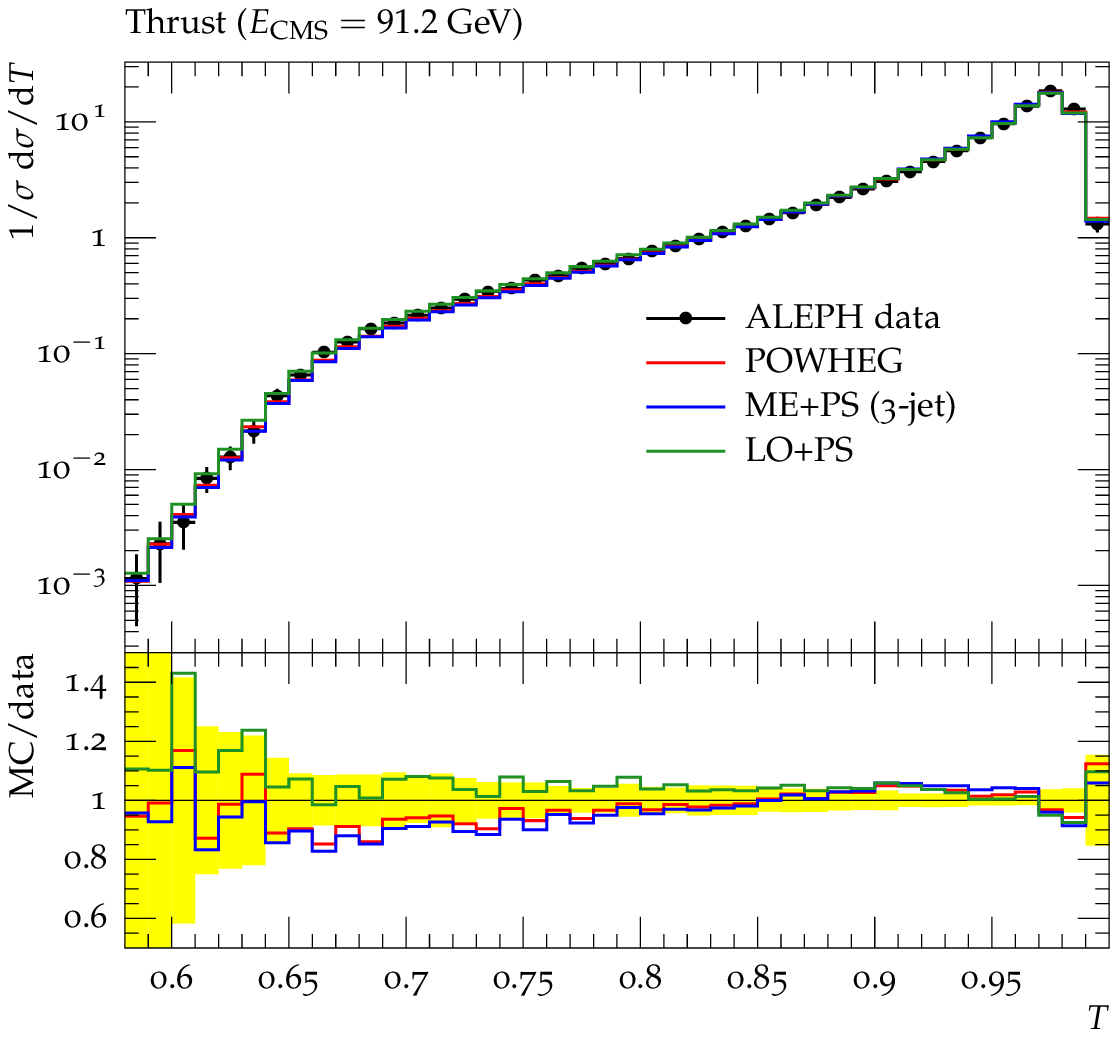}
  \vspace*{-3mm}
}
{
  Durham $2\to 3$ jet resolution (left)
  and thrust distribution (right)
  compared to data from the \protect\Aleph experiment\cite{Heister:2003aj}.
  \label{fig:lep:aleph1}
}

\myfigure{p}{
  \vspace*{-10mm}
  \includegraphics[width=0.45\textwidth]{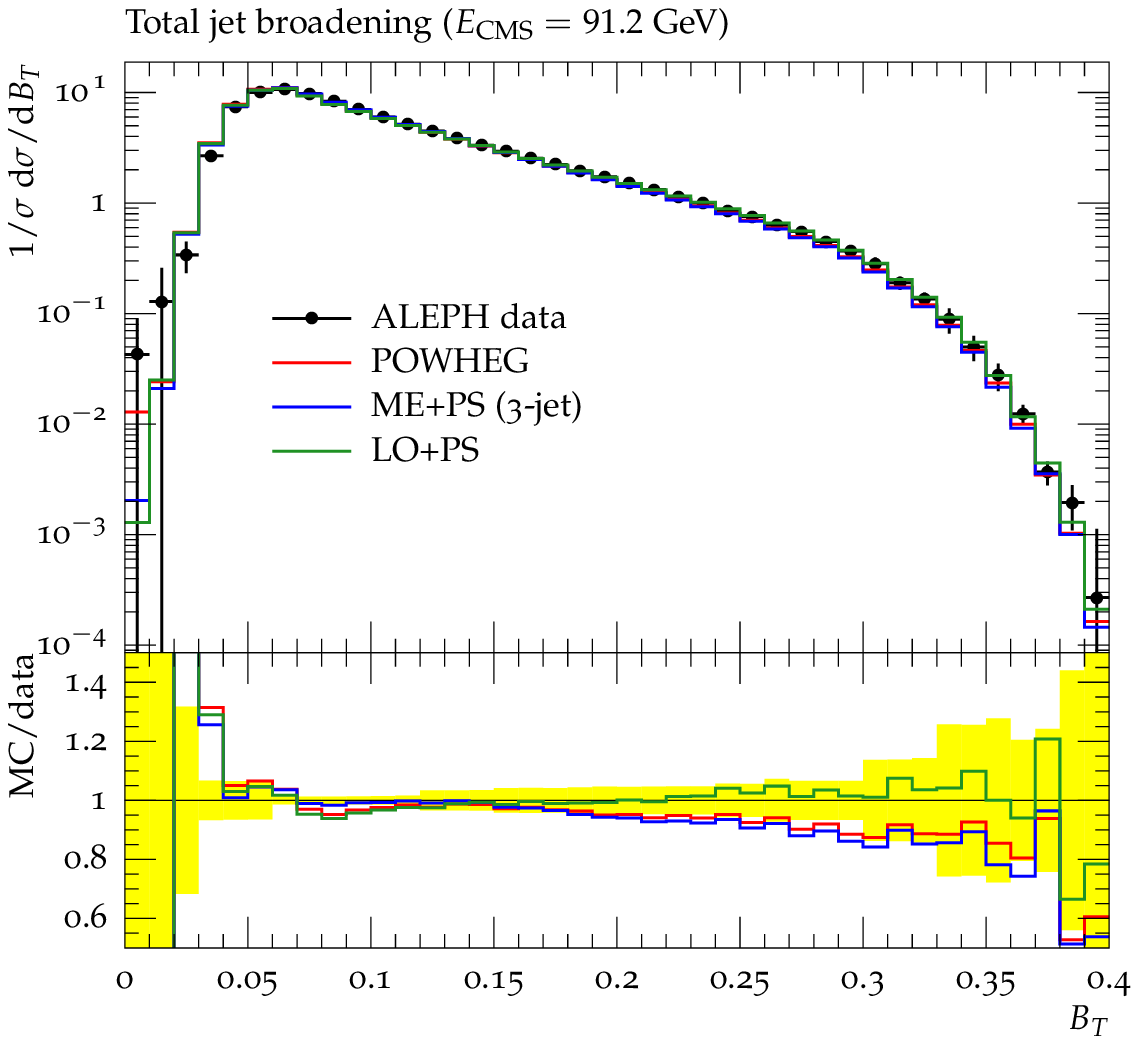}
  \hspace*{0.05\textwidth}
  \includegraphics[width=0.45\textwidth]{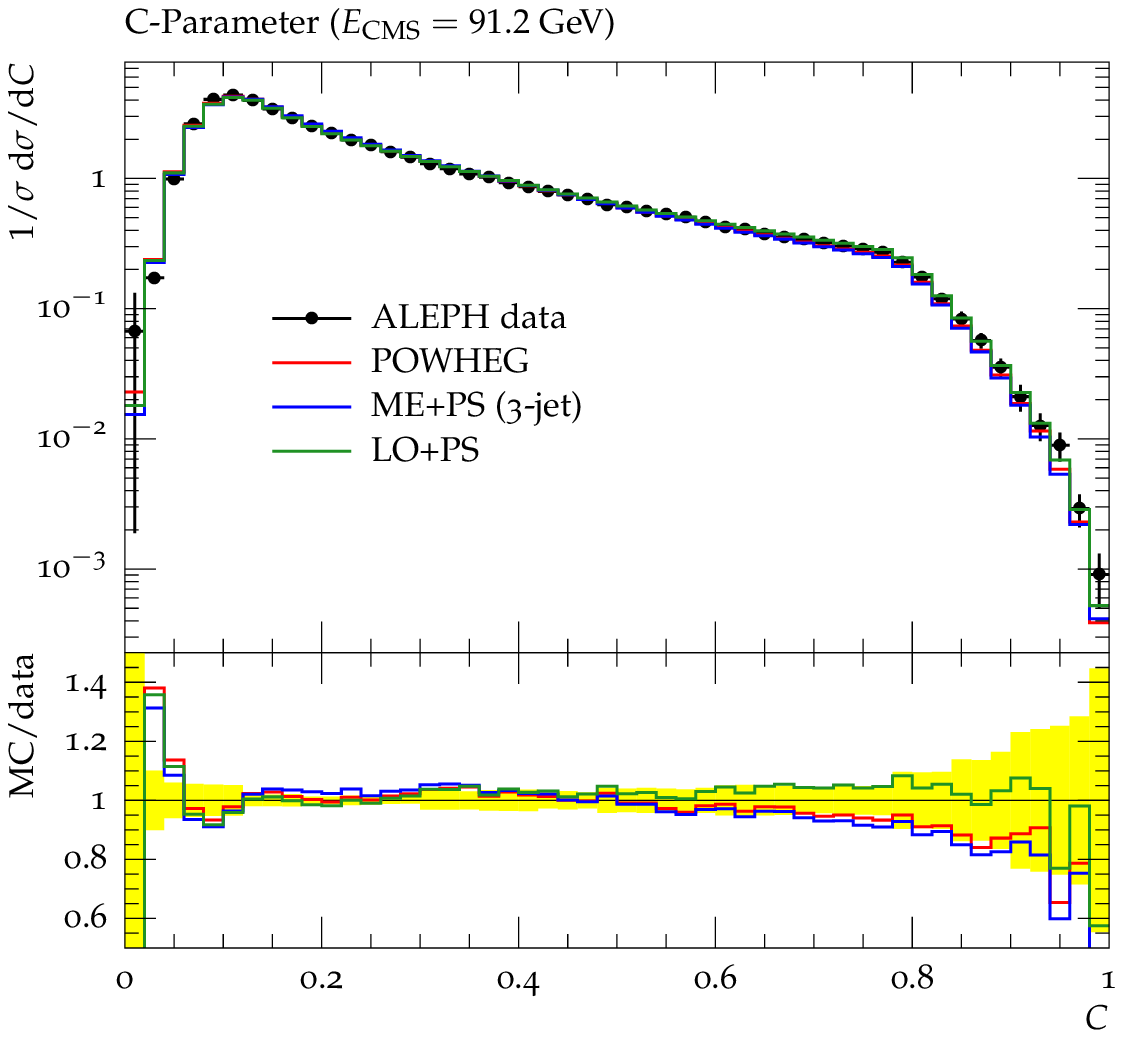}
  \vspace*{-3mm}
}
{
  Total jet broadening (left)
  and C-Parameter (right)
  compared to data from the \protect\Aleph experiment\cite{Heister:2003aj}.
  \label{fig:lep:aleph2}
}

\myfigure{p}{\vspace*{-5mm}
  \begin{minipage}{0.545\textwidth}
  \includegraphics[width=\linewidth]{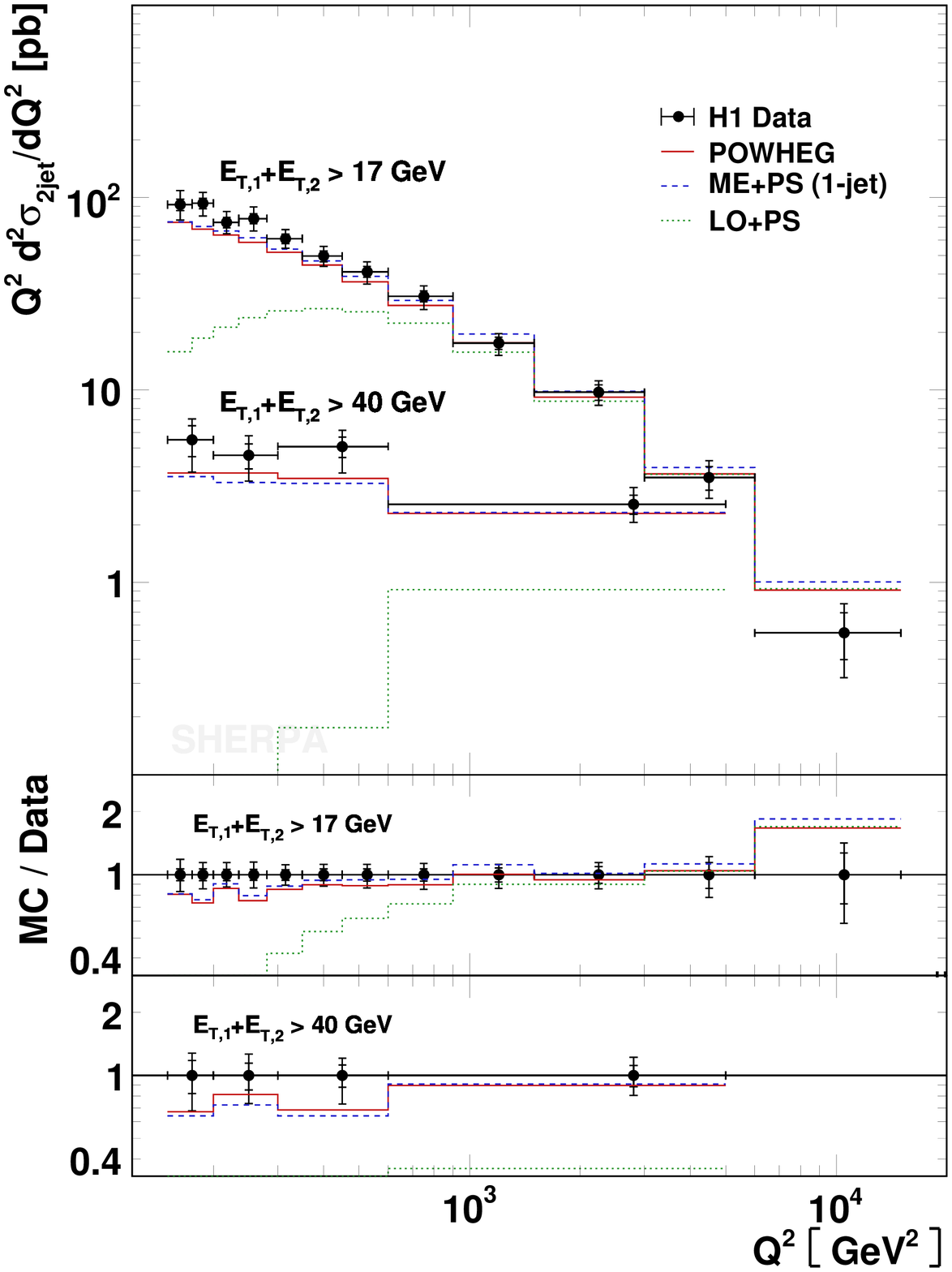}
  \end{minipage}\hfill
  \begin{minipage}{0.4\textwidth}
  \includegraphics[width=\linewidth]{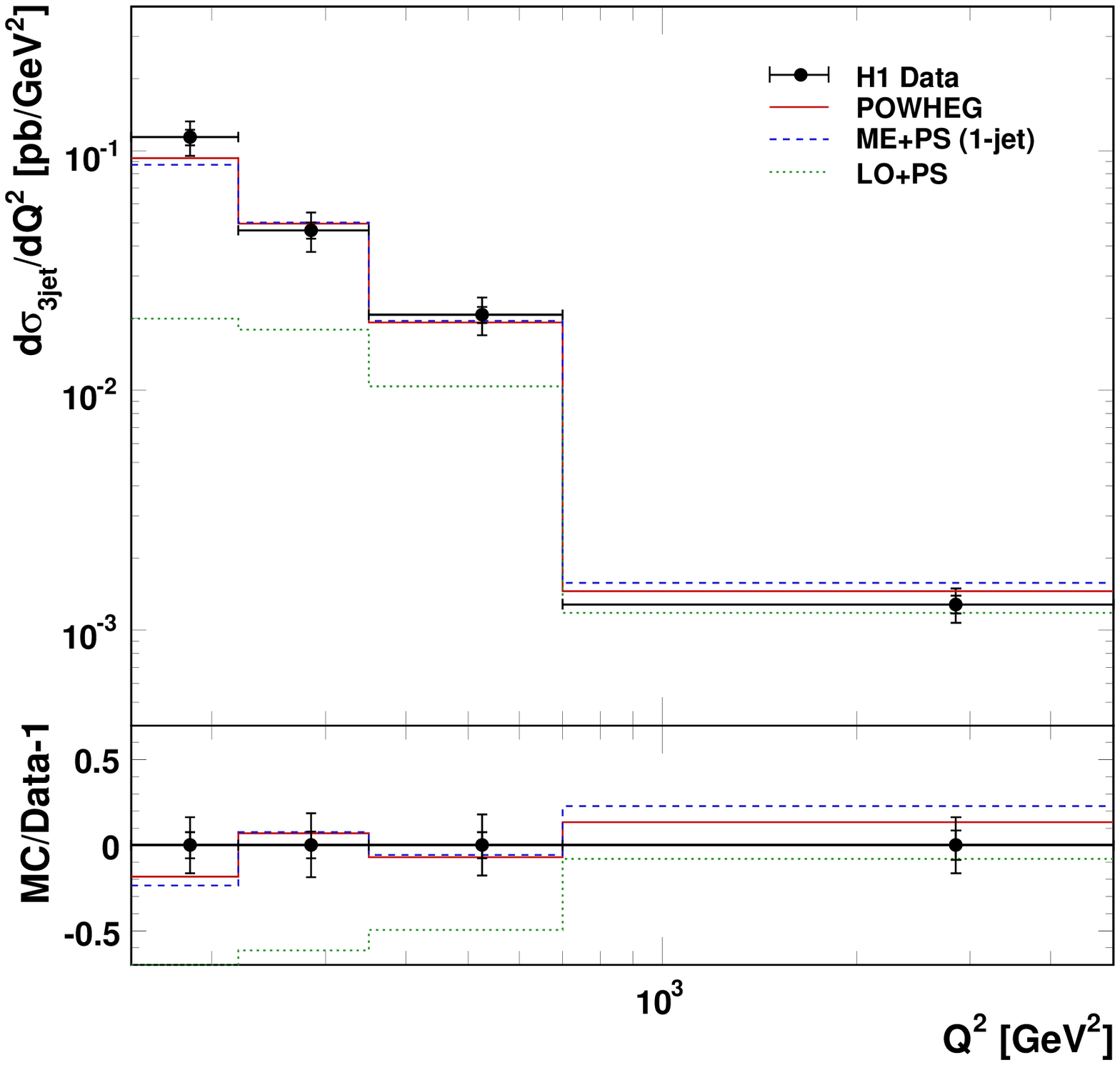}\\[-4mm]
  \includegraphics[width=\linewidth]{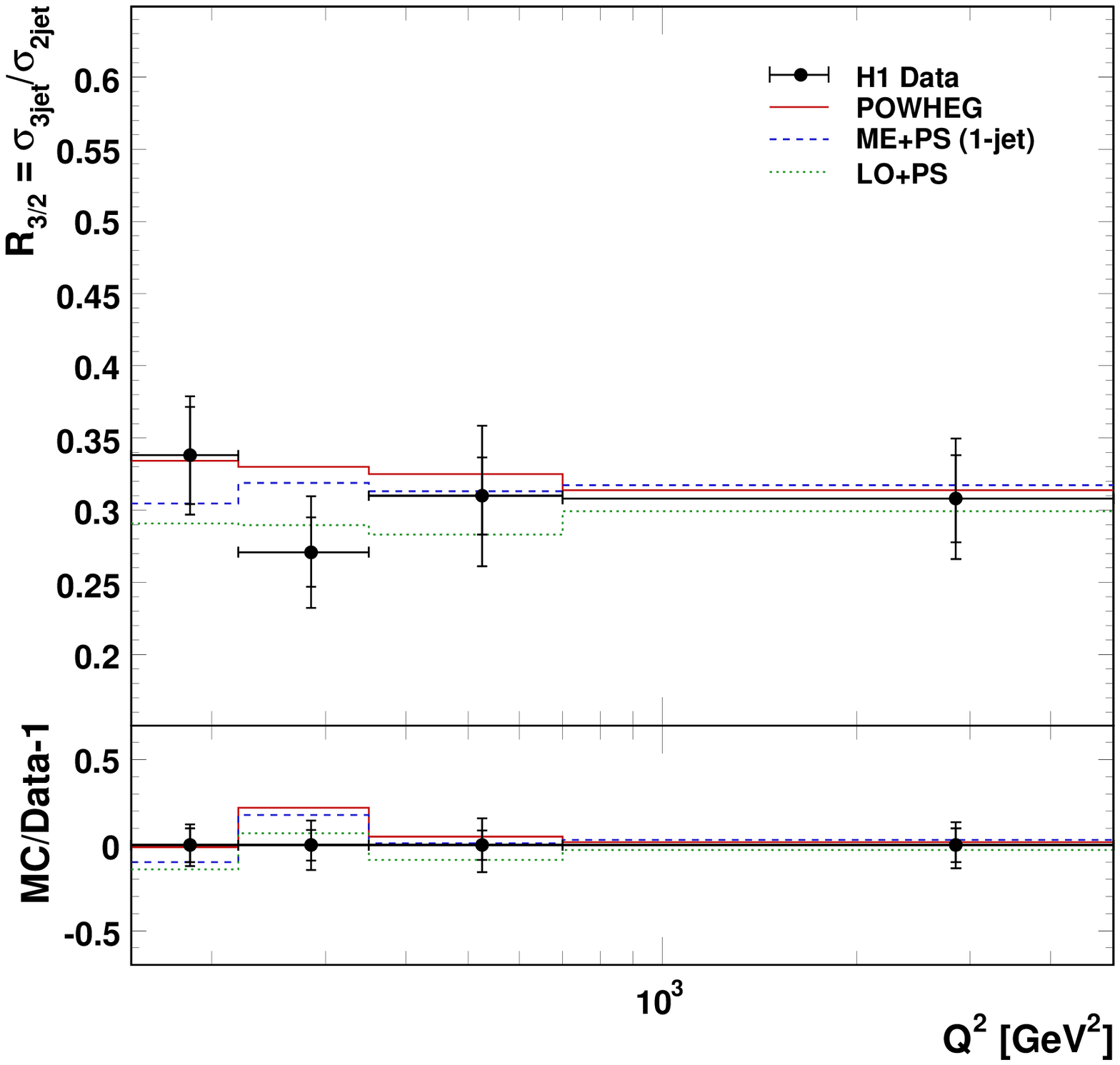}\\[2mm]
  \end{minipage}
}
{
  The di-jet cross section as a function of $Q^2$ in bins of $E_{T,1}+E_{T,2}$ (left),
  the three-jet cross section as a function of $Q^2$ (right top) and the
  ratio of the three- over the two-jet rate as a function of $Q^2$ (right bottom)
  compared to data from the H1 experiment~\protect\cite{Adloff:2000tq,Adloff:2001kg}.
  \label{fig:dis:q2r32}
}

\myfigure{p}{
  \includegraphics[width=0.45\linewidth]{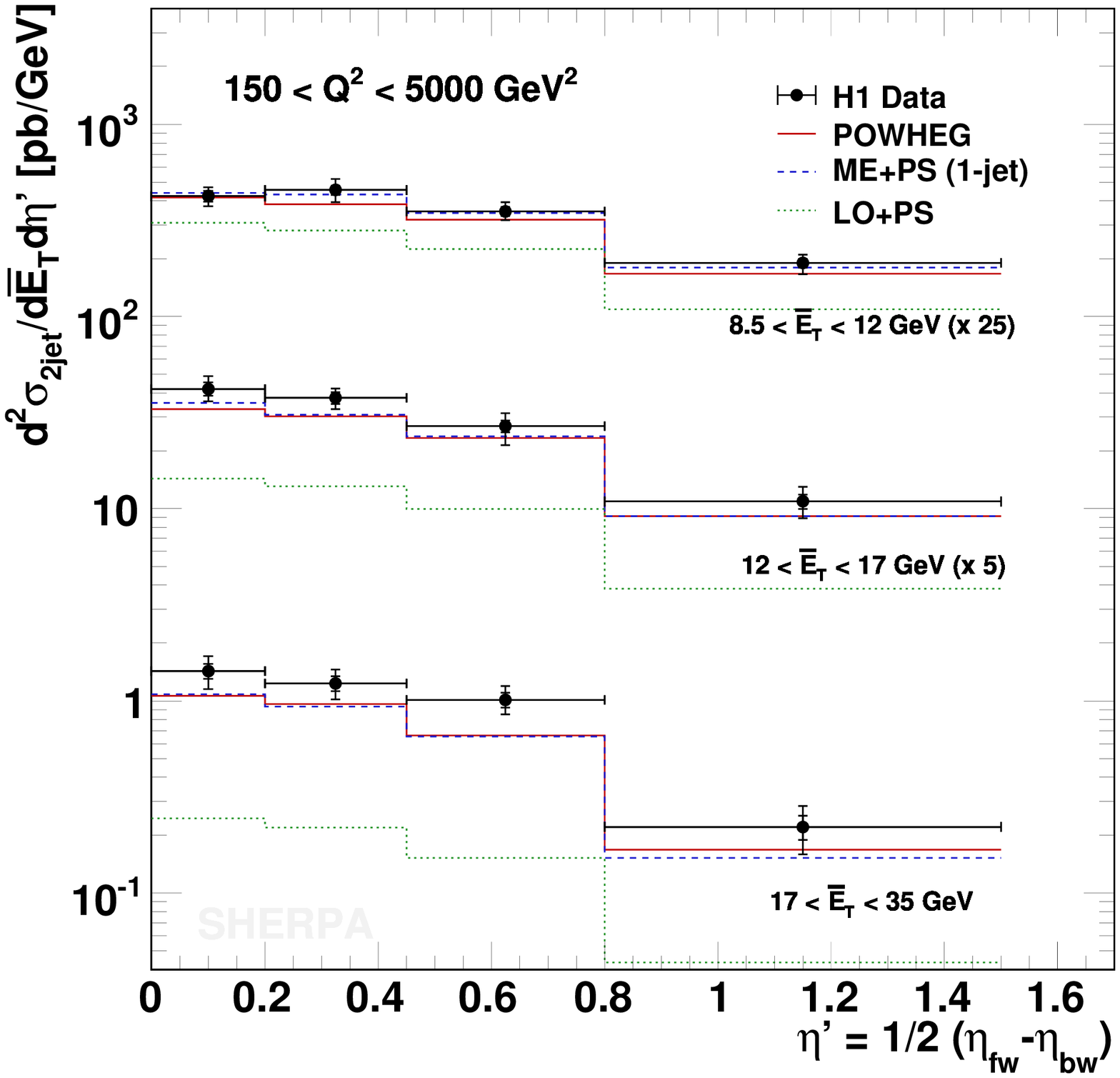}
  \hspace*{0.05\linewidth}
  \includegraphics[width=0.45\linewidth]{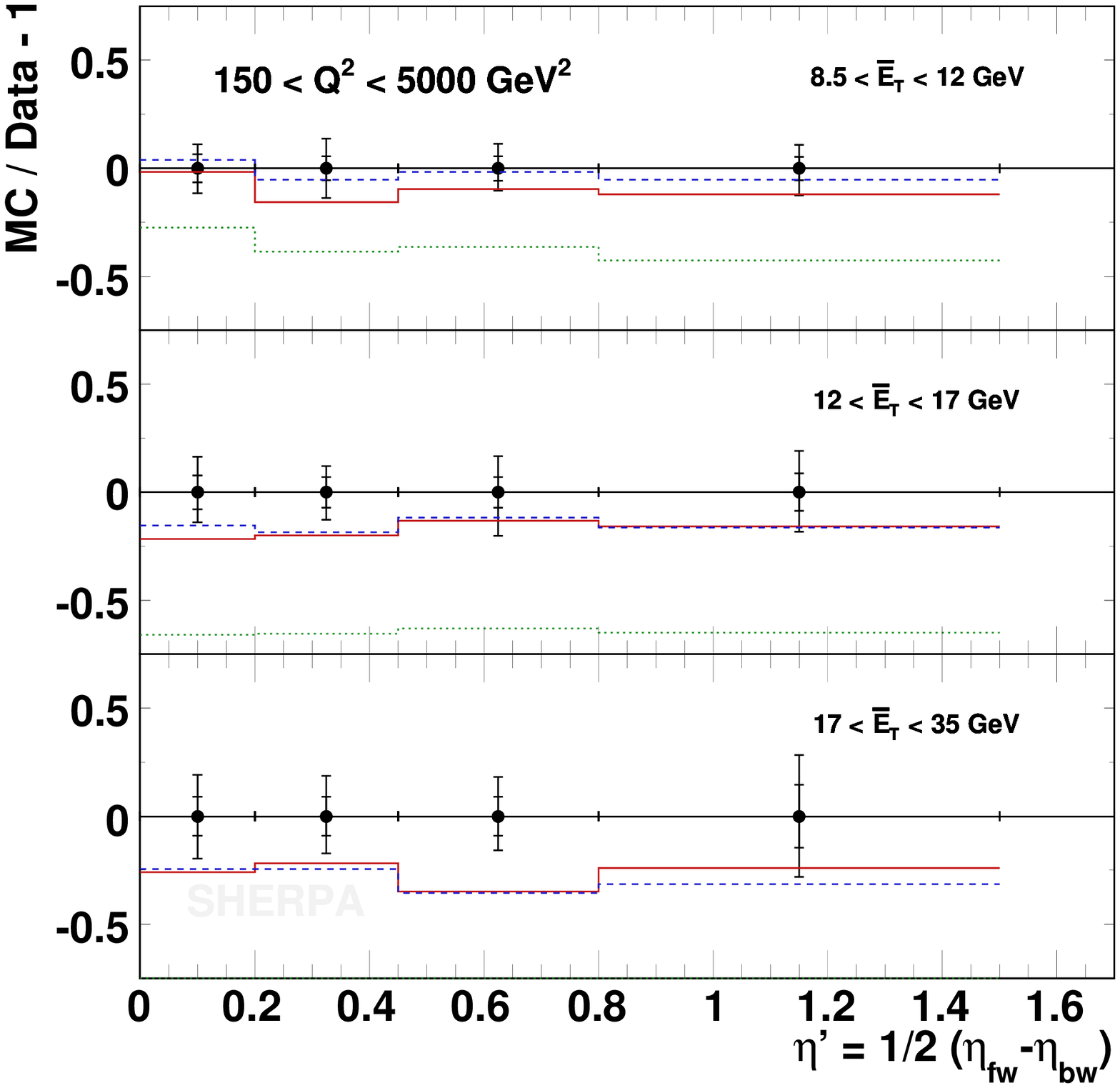}
}
{
  The di-jet cross section as a function of $\eta'$, compared to data 
  from the H1 experiment~\protect\cite{Adloff:2000tq}. $\eta'$ denotes half 
  the rapidity difference of the two leading jets in the Breit frame.
  \label{fig:dis:etaetsum}
}

\myfigure{p}{
  \includegraphics[width=0.45\textwidth]{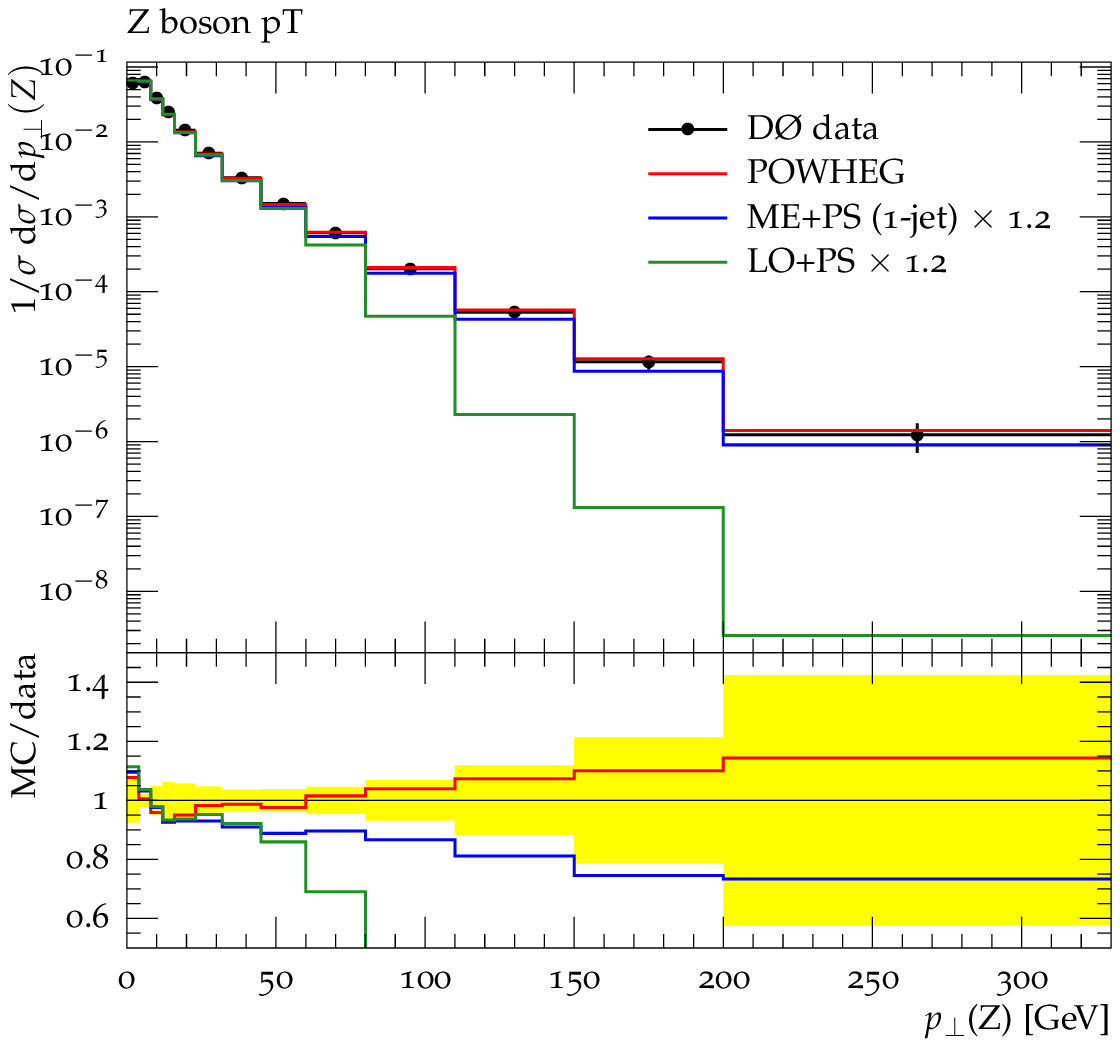}
  \hspace*{0.05\textwidth}
  \includegraphics[width=0.45\textwidth]{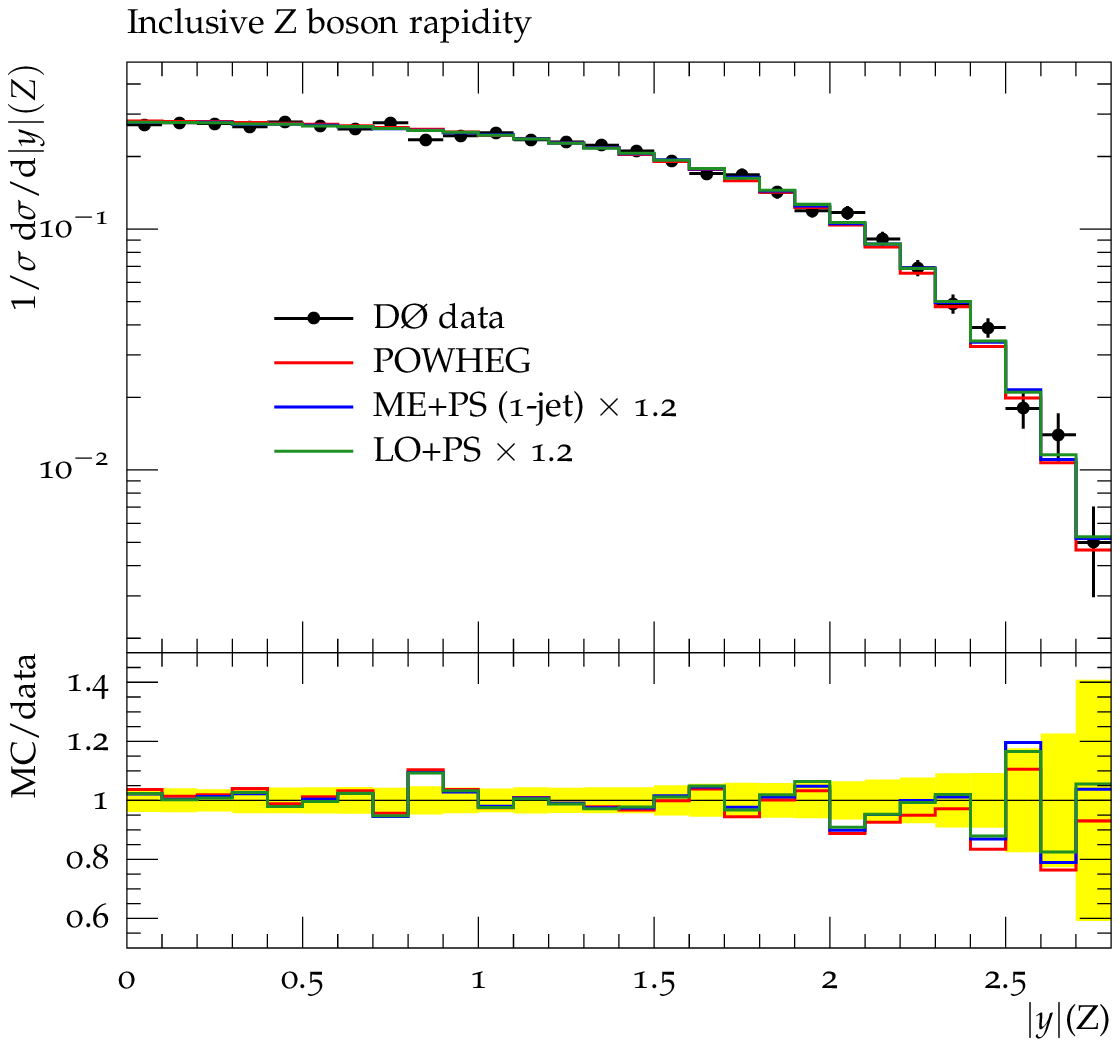}
}
{
  Transverse momentum and rapidity of the $Z$ boson in
  Drell-Yan lepton-pair production at the Tevatron compared to data
  from the \DO experiment\cite{Abazov:2010kn,*Abazov:2007jy}.
  \label{fig:ztev:pTy}
}

\myfigure{p}{
  \includegraphics[width=0.45\textwidth]{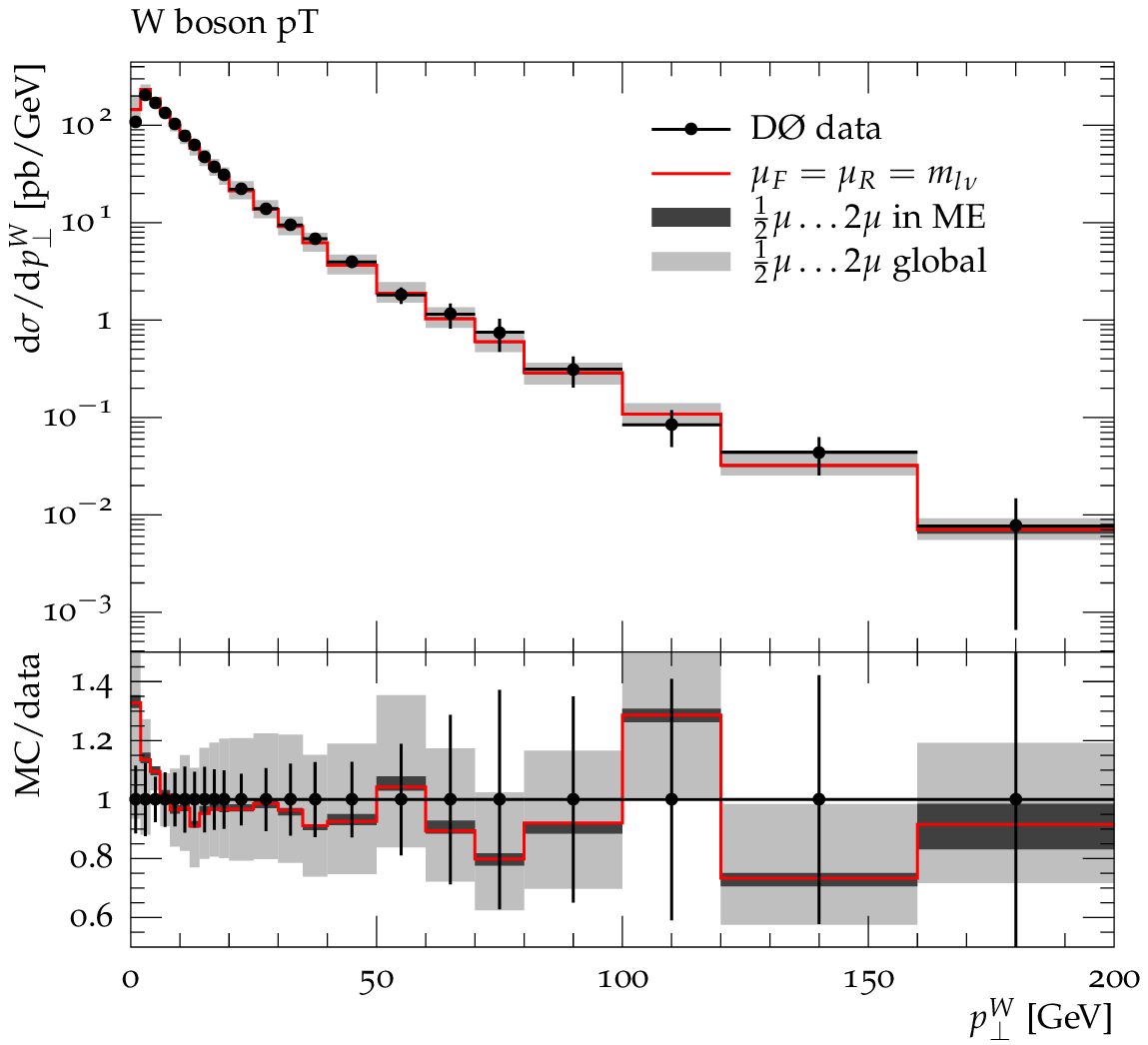}
  \hspace*{0.05\textwidth}
  \includegraphics[width=0.45\textwidth]{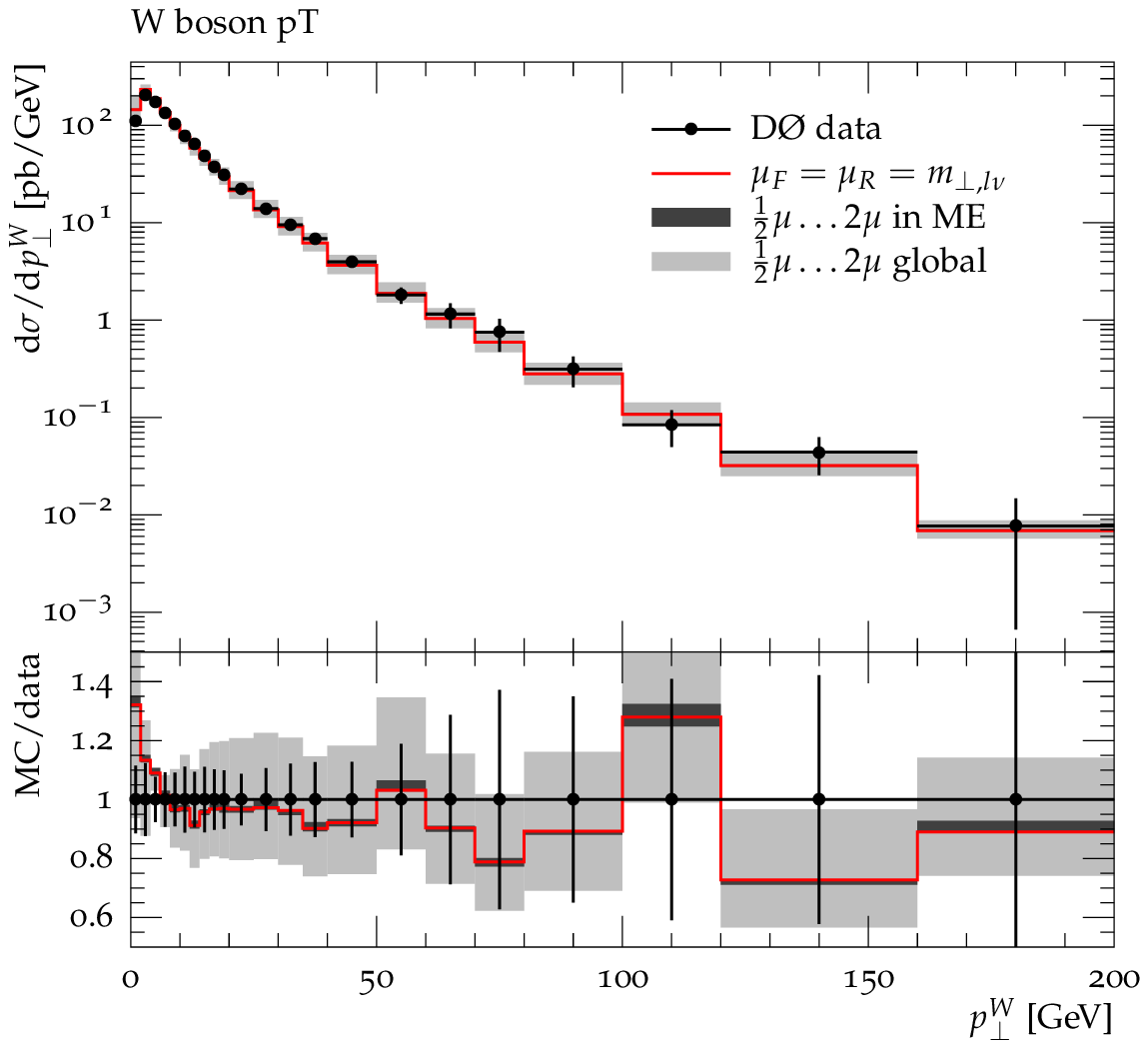}
}
{
  Transverse momentum of the $W$ boson in
  $W$+jets production at the Tevatron compared to data
  from the \DO experiment~\cite{Abbott:2000xv}.
  Scale variations of the \POWHEG{} prediction by factors of $1/2$ and
  $2$ are displayed for two different scale schemes, $\mu_F=\mu_R=m_{l\nu}$ (left)
  and $\mu_F=\mu_R=m_{\perp,l\nu}$ (right). The inner (dark) band displays the
  variations associated with redefining the scales for matrix elements alone, 
  while the outer (light) band also takes variations in the running coupling of
  the parton shower evolution into account.
  \label{fig:wtev:pT}
}

\clearpage

\section{Conclusions and outlook}
\label{sec:conclusions}

In this publication the successful implementation of the \POWHEG
algorithm into the \Sherpa framework was reported.  The program is 
fully automated, relying on \Sherpa's efficient matrix-element 
generation modules, which allow to construct real correction terms
for given processes and their Catani-Seymour dipole subtraction kernels
in both differential and integrated form.  It is worth stressing that 
this is the first time that the \POWHEG method has been applied simultaneously 
to various higher-order calculations using Catani-Seymour dipole terms
for partitioning the real-emission phase space. This implementation
makes a number of processes, computed at NLO
and available in public program libraries accessible for matching with a 
parton shower. Additional processes are easily added in \Sherpa, by 
merely linking the corresponding code for the virtual correction 
terms\footnote{
  A wealth of processes not discussed yet is, for instance, available in 
  the extremely well-developed \MCFM library or in similar programs.}.

The implementation was validated by a number of systematic checks, including
\begin{itemize}
  \item the stability of cross sections, as exhibited in Tables 
    \ref{Tab:xsec_consistency_eeDIS} and \ref{Tab:xsec_consistency_WZh};
  \item the radiation patterns, through comparison with a fake \POWHEG 
    algorithm, based on shower kernels, c.f.\ Figure~\ref{Fig:approxme_Z};
  \item the automated detection of Born zeroes and their stable cure,
    as shown in Figure~\ref{Fig:ZH_RBs} and \ref{Fig:ZH_splitting};
  \item merged LO samples, see
    Figures~\ref{fig:wz:log10_d_01}--\ref{fig:wz:jet_multi_exclusive};
  \item and comparison with a variety of data, in 
    Figures~\ref{fig:lep:aleph1}--\ref{fig:wtev:pT}.
\end{itemize} 
It also included, for the first time, the treatment of $W$-pair production;
results for this, along with some plots for $Z$-pair production are 
displayed in Figures~\ref{fig:zzlhc:jets}--\ref{fig:wwlhc:angles}.  

In the near future, more processes with one coloured line only, such as $WH$
and $ZH$ associated production, will be added to the \Sherpa framework and in a 
future publication we will also discuss the slightly more subtle treatment of 
processes with more than one coloured line.  Furthermore, the methods 
developed in this publication have been used to generate merged samples with
an inclusive cross section at NLO accuracy, according to the method already
presented in \cite{Hamilton:2010wh}; we will report on our parallel development
in a second publication~\cite{Hoeche:2010xx}.  This set of NLO-implementations will be made
available in the future, in a new and extended release of \Sherpa.

\section*{Acknowledgements}
We would like to thank Jennifer Archibald, Tanju Gleisberg, Steffen
Schumann and Jan Winter for many years of collaboration on the \Sherpa project, 
and for interesting discussions.  We are indebted to Daniel Ma{\^i}tre for
support with linking the BlackHat code and for useful conversation.  We thank
Keith Ellis for his help in interfacing the MCFM library to \Sherpa.
Special thanks go to Thomas Gehrmann, Thomas Binoth and Gudrun Heinrich for 
numerous fruitful discussions on NLO calculations. We also thank Thomas Gehrmann
for valuable comments on the manuscript.

SH acknowledges funding by the Swiss National Science Foundation 
(SNF, contract number 200020-126691) and by the University of Zurich 
(Forschungskredit number 57183003).  MS and FS gratefully acknowledge 
financial support by the MCnet Marie Curie Research Training Network 
(contract number MRTN-CT-2006-035606). MS further acknowledges
financial support by the HEPTOOLS Marie Curie Research Training Network 
(contract number MRTN-CT-2006-035505) and funding by the DFG 
Graduate College 1504. FK would like to thank the theory group at CERN, and
MS would like to thank the Institute for Particle Physics Phenomenology in 
Durham, respectively, for their kind hospitality during various stages of 
this project.

%= appendices ======================================
\appendix
%= bibliography ===================================
\bibliographystyle{bib/amsunsrt_modp}  
\bibliography{bib/journal}
%= end ============================================
\end{document}